\documentclass[a4paper,fleqn,usenatbib]{mnras}
\usepackage{mathptmx}
\usepackage[T1]{fontenc}
\usepackage{ae,aecompl}

\usepackage{amssymb,amsmath,amsfonts,graphicx}
\usepackage{natbib,color}

\def\lesssim {<\kern-1.2em\lower1.1ex\hbox{$\sim$}~}   

\newcommand{\de}{{\rm d}}
\newcommand{\tIa}{\tau_{\rm Ia}}
\newcommand{\Msun}{{\rm M_{\odot}}}

\begin{document}

\title[Precision pollution]{
Precision Pollution - The effects of enrichment yields and timing on galactic chemical evolution
}

\author[P.-A. Poulhazan et al.]{Pierre-Antoine Poulhazan$^{1}$, Cecilia Scannapieco$^{1,2,3}$ and Peter Creasey$^{4}$\\
$^1$ Leibniz-Institute for Astrophysics Potsdam (AIP), An der Sternwarte 16, D-14482, Potsdam, Germany\\
$^2$ Universidad de Buenos Aires,  Facultad de Ciencias Exactas y Naturales, Departamento de F\'{\i}sica. Buenos Aires, Argentina\\
$^3$ CONICET. Buenos Aires, Argentina\\
$^4$ Department of Physics and Astronomy, University of California, Riverside, California 92507, USA}

\maketitle

\begin{abstract} 

We present an update to the chemical enrichment component of the smoothed-particle hydrodynamics model for galaxy formation presented in \cite{S05} in order to address the needs of modelling galactic chemical evolution in realistic cosmological environments. Attribution of the galaxy-scale abundance patterns to individual enrichment mechanisms such as the winds from asymptotic giant branch (AGB) stars or the presence of a prompt fraction of Type Ia supernovae is complicated by the interaction between them and gas cooling, subsequent star formation and gas ejection. In this work we address the resulting degeneracies 
by extending our implementation to a suite of mechanisms that encompasses different IMFs, models for yields from the aforementioned stars, models for the prompt component of the delay-time-distribution (DTDs) for Type Ia SNe and metallicity-dependent gas cooling rates, and then applying these to both isolated initial conditions and cosmological hydrodynamical zoom simulations. We find DTDs with a large prompt fraction (such as the bimodal and power-law models) have, at $z=0$, similar abundance patterns  compared to the low-prompt component time distributions (uniform or wide Gaussian models). However, some differences appear, such as the former having systematically higher [X/Fe] ratios  and narrower [O/Fe] distributions compared to the latter, and a distinct evolution of the [Fe/H] abundance.

\end{abstract}



\section{Introduction}

Over the last decades, the accumulation of precise and detailed observations on the chemical properties of the Milky Way and external galaxies 
have enabled substantial improvements in our understanding of cosmic structure formation. Chemical information of the gas and stars in galaxies, as well as in the intergalactic medium, and for different redshifts, are important tools to complement observations of stellar masses, current star formation rates, magnitudes/colors, morphologies. Important physical processes occurring in galaxies leave imprints on their chemical properties, which carry information on their past star formation rates, the occurrence of mergers and accretion, and the amount of dust that obscures their light. In particular, the chemical properties of the stellar component provide information of the galaxies at different cosmic times, enabling the reconstruction of the galaxy's formation history.

In previous decades the importance of chemical evolution in cosmological simulations has often been seen to be subordinate to the greater uncertainties of the feedback and star formation history of the universe
which can modify the stellar populations by an order of magnitude, rather than, for example, 
the changes in Initial Mass Function (IMF), SNIa rates, stellar life-times and chemical yields, 
 IMF changes being the most significant with changes up to a factor 2 in the fraction of massive stars \citep[e.g.][]{Vincenzo_2016}. 
Nevertheless, our knowledge of the 
galactic chemical profile has made significant progress, with large increases in both the 
number of species for which we can measure abundances \citep[e.g. the neutron-capture elements 
of ][]{Sneden_2008} and the number of stars for which $\alpha$-element abundances can be 
measured \citep[e.g. ][]{Gilmore_2012}. Combining this with the kinematical data from the 
European Space Agency's  Gaia satellite will lead to tight constraints on the galactic 
chemical evolution  of our own galaxy. On the other hand, observations of the chemical 
properties of external galaxies, both locally and at higher redshifts, 
now enable study of the evolution of chemical species in galaxies 
and to look for links between their chemical and dynamical properties, allowing a better understanding on their formation histories.

Accompanying the observational evidence has been increasingly detailed models of the stellar populations, their nucleosynthesis, the evolution and subsequent enrichment of the ISM and the mechanisms by which this is mixed on galactic scales. Consequently, the constraints have the potential to distinguish not just between e.g. the choice of a \citet{Salpeter_1955, Kroupa01} or \citet{Chabrier03} IMF by their effect on the fraction of massive stars (e.g. \citealp{Francois_2004}), but on a whole interrelated network of processes that incorporate different models for the SNII yields \citep[e.g.][]{WW95, P98, Limongi_2003}, the AGB ejecta \citep{M01, Marigo_2008}, the type Ia nucleosynthesis \citep{Nomoto_1997,Iwamoto_1999}, the existence of a prompt component for the aforementioned \citep{Mannucci_2006}, the metal dependent life-times of all their progenitors \citep[][P98 hereafter]{P98}, and how rapidly the metals are mixed \citep[e.g.][]{Aumer_2013} or ejected \citep{Creasey_2015} and their effects on gas cooling \citep{W09}.

As observations of chemical properties increase in amount and detail, a theoretical understanding of the building up of the chemical history of galaxies is needed, and in fact over the last decades a great deal of effort has been made in the field of galactic chemical evolution.
The evolution of chemical species in galaxies is intimately linked to the star formation process, the amount of feedback that injects energy into the medium and the mixing and redistribution of chemical elements. Thanks to the advances in numerical techniques and computer processing capabilities we are gradually improving beyond one zone/ordinary differential equation models of \citet{Matteucci_1994, Matteucci_2001, Vincenzo_2017} or hydrodynamical models based on N-body trees \citep{Minchev_2013} or the semi-analytical models of \citet{Nagashima_2005, Arrigoni_2010, Yates2013,DeLucia_2014} to simulations in a full cosmological context of the chemical properties of galaxies. 
These have evolved from the chemical enrichment modules of \cite{Mosconi01, Lia02, Kawata03, Tornatore04, Okamoto05, S05}, and most of the current state-of-the-art simulation codes consider the production and distribution of metals, their mixing and their effects on the gas cooling process \citep[][]{Crain2009, Schaye2015, Illustris}, which have enabled substantial improvements in our understanding of cosmic structure formation and the chemical evolution of the Universe.

Models of chemical enrichment are, however, subject to many uncertainties, as
the shape and universality of the IMF, the chemical yields of SNII, the nature of the
SNIa progenitors and the SNIa delay distribution are still under debate. 
In three-dimensional hydrodynamical simulations there is still no consensus on the precise mechanism by which one should implement feedback to regulate star formation by reheating or preventing the gas from cooling, which results into uncertainties in the level, distribution and timing of the chemical production and mixing. Constraints have, however, been more forthcoming with simulations which restrict the geometry such as \citet{Francois_2004, Matteucci_2009} on the effects of the Type Ia distribution, and \citet{Cote_2017} w.r.t. inflows and outflows.
 The lack of fundamental theories for several complex physical processes has lead to a number of models  which for a same set of initial conditions predict galaxies with different properties \citep{S12}.
Despite these problems, such simulations have proven to be extremely useful for an understanding of the effects of given physical processes (mergers, interactions, mixing, instabilities, accretion) on the properties of galaxies, allowing a better interpretation of observational results both at the present time and at higher redshifts.

In this work we attempt to restrict our focus away from the emphasis on feedback and onto the application of the many advances that have been made in the chemical enrichment mechanisms and related processes since the pioneering works of \citet{Katz_1992,Steinmetz94,Raiteri96} and \citet{Mosconi01}.
 In particular we are interested in the effects of updates to type II SN yields, cooling, AGB enrichment and the time distribution of type Ia SNe. We present an update of the chemical model of \cite[][S05 hereafter]{S05} and evaluate the effects of the different assumptions on the chemical properties of the baryons, using both idealized and cosmological initial conditions.

The paper is organized as follows. In Section~\ref{sec:implem} we present the chemical model; in Section~\ref{sec:isolated} we use idealized initial conditions to isolate and test the effects of the different assumptions for the IMF, chemical yields, delay time distributions of SNIa and cooling. We present  results for cosmological simulations in Section~\ref{sec:cosmo}; and in Section~\ref{sec:conclu} we summarize our conclusions.


\section{Numerical implementation}
\label{sec:implem}

In this section, we describe the numerical implementation of chemical enrichment and cooling, which are an update to the S05 model. The updated model includes a more sophisticated treatment of the chemical yields, the supernova life-times and rates, the additional treatment of enrichment from low- and intermediate-mass stars in the AGB phase, and more realistic model for the gas cooling process. All  simulations presented here use the standard routines for star formation and feedback of S05 and \cite{S06},  which we briefly describe at the end of this section.

Different channels contribute to the chemical enrichment of the interstellar medium in galaxies; chemical enrichment models can be coupled to galaxy simulations provided the following ingredients are specified:
\begin{enumerate}
\item the Initial Mass Function (IMF), which determines the fractional contribution of stars of different mass in a stellar population,
\item the typical life-times of progenitor stars, which determines the time of release of chemical elements synthesized in stellar interiors,
\item the rates of occurrence of SNII, SNIa and AGB stars, which are related to the choice of the IMF,
\item the chemical yields, namely the amount of each chemical element ejected to the interstellar medium.
\end{enumerate}

In the following subsections, we describe our implementation of each of these ingredients for SNII, SNIa and AGB. In the code, each star particle represents a single stellar population (SSP) of the same age and metallicity. At birth, stars inherit the metallicity of their progenitor gas particle and, once formed,  each stellar particle can experience the occurrence of SNII, SNIa and AGB
during its life-time, where metals will be produced and distributed among its nearest gas neighbors either in a single time-step (i.e. in the case of SNII) or
  over an extended time period  (i.e. for AGB   and SNI), as explained below.

\subsection{The stellar initial mass function}

\begin{figure}
	\begin{center}
	\includegraphics[width=\linewidth]{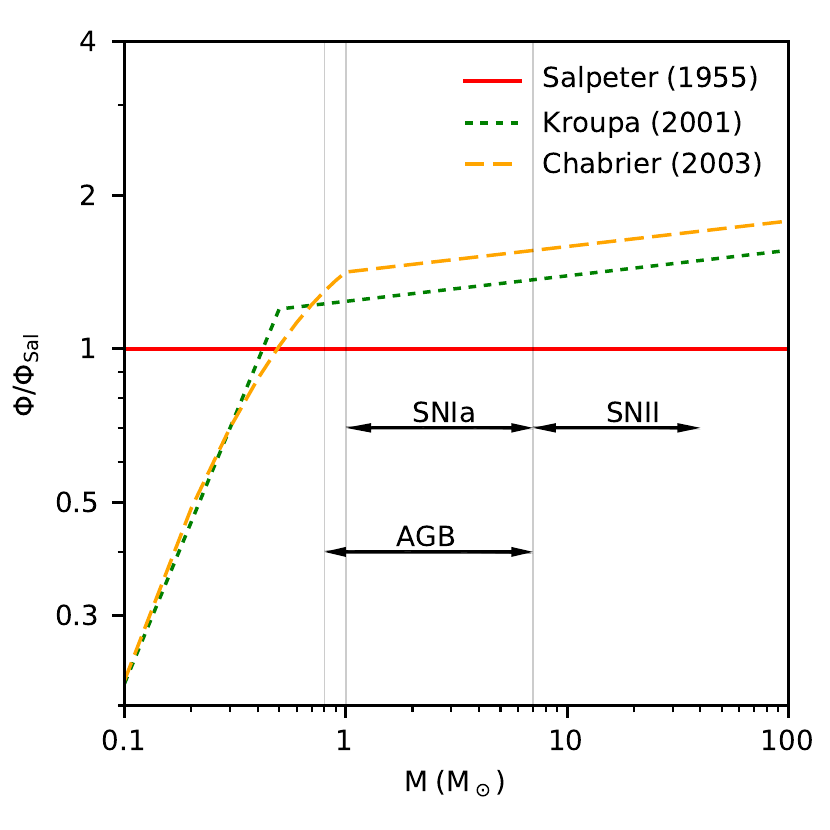}
	\caption{A comparison of the three stellar IMFs by mass, $\Phi (M)$, used in this work. The green dotted and orange dashed lines show results for the \citet{Kroupa01} and \citet{Chabrier03} IMFs, respectively, normalised to that of \citet{Salpeter_1955} (red solid line). The arrows indicate the masses of progenitor stars of SNII, SNIa and AGB events, delimited by the grey vertical lines.}
	\label{fig:IMF}
	\end{center}
\end{figure}

The stellar initial mass function, which gives the fractional distribution of initial masses of a stellar system, is an important ingredient of chemical enrichment models. The choice of the IMF directly affects the chemical enrichment of the interstellar gas, as it determines the relative fraction of long-lived stars with respect to intermediate-mass and massive short-living stars and therefore the relative rates of SNII, SNIa and AGB events. As a consequence, the IMF impacts the relative abundance of elements contributed by different types of enrichment events, and the amount of energy released to the ISM through SN explosions.

Despite its global importance, a full consensus has not yet been reached on the dependencies of the IMF to factors such as time and environment, and as such we restrict ourselves to considering only adjustments to the shape of a time and environment-independent IMF. The IMF used in S05, and still widely used, was the \citet{Salpeter_1955} IMF, $\Phi(M)$, of a single-slope power law:
\begin{equation}
\Phi_{\rm Sal}(M) \equiv {\de N\over \de M} = A_S M^{-2.35}
\end{equation}
where A$_S$ is chosen such that the IMF  is normalised to $1$M$_\odot$ over the mass range considered, and $N$  and $M$ refer, respectively, to the number and mass of the
individual stars.

The primary modification to the Salpeter IMF is to reduce the fraction of stars (i.e. flatten the slope) of mass below $1\;\rm M_\odot$ as proposed by \cite{Kroupa01} and \cite{Chabrier03}.   The Kroupa multi-slope IMF is given by:
\begin{equation}
    \Phi_{\rm Krpa}(M) = \left \{
    \begin{array}{ll}
        {\displaystyle A_K M^{-0.3}} \quad    & 0.01 < M  <  0.08 \; \rm M_\odot \\
        {\displaystyle B_K M^{-1.3}} \quad   & 0.08 \leq  M  <  0.5 \; \rm M_\odot  \\
        {\displaystyle C_K M^{-2.3}} \quad    & M  \geq 0.5 \; \rm M_\odot
    \end{array} \right . \quad
\end{equation}
while the Chabrier IMF is:
\begin{equation}
    \Phi_{\rm Chab}(M) = \left \{
    \begin{array}{ll}
        {\displaystyle \frac{A_C}{M} \, e^{- (\log M - \log M_c)^2 / 2 \sigma^2}} \quad    & M  \leq  1 M_\odot \\
        {\displaystyle B_C \, M^{-2.3}} \quad & M > 1 M_\odot
    \end{array} \right . \quad.
\end{equation}
where M$_c=0.079$M$_\odot$  and $\sigma = 0.69$. The coefficients A$_K$, B$_K$, C$_K$, A$_C$ and B$_C$ are determined by requiring continuity and normalisation over the given mass range.

Our code has the flexibility to adopt any of these three IMFs, and we test the impact of variations of the IMF  in Section~\ref{sec:IMF}. Note that in order to fully specify an IMF, we also need to choose limits for the minimum and maximum masses allowed. In this work, we set these limits to 0.1 and 100M$_\odot$ regardless of our choice of IMF. Fig.~\ref{fig:IMF} shows a comparison of the three IMFs.

\subsection{Stellar life-times} 

\begin{figure}
	\begin{center}
	{\includegraphics[width=\linewidth]{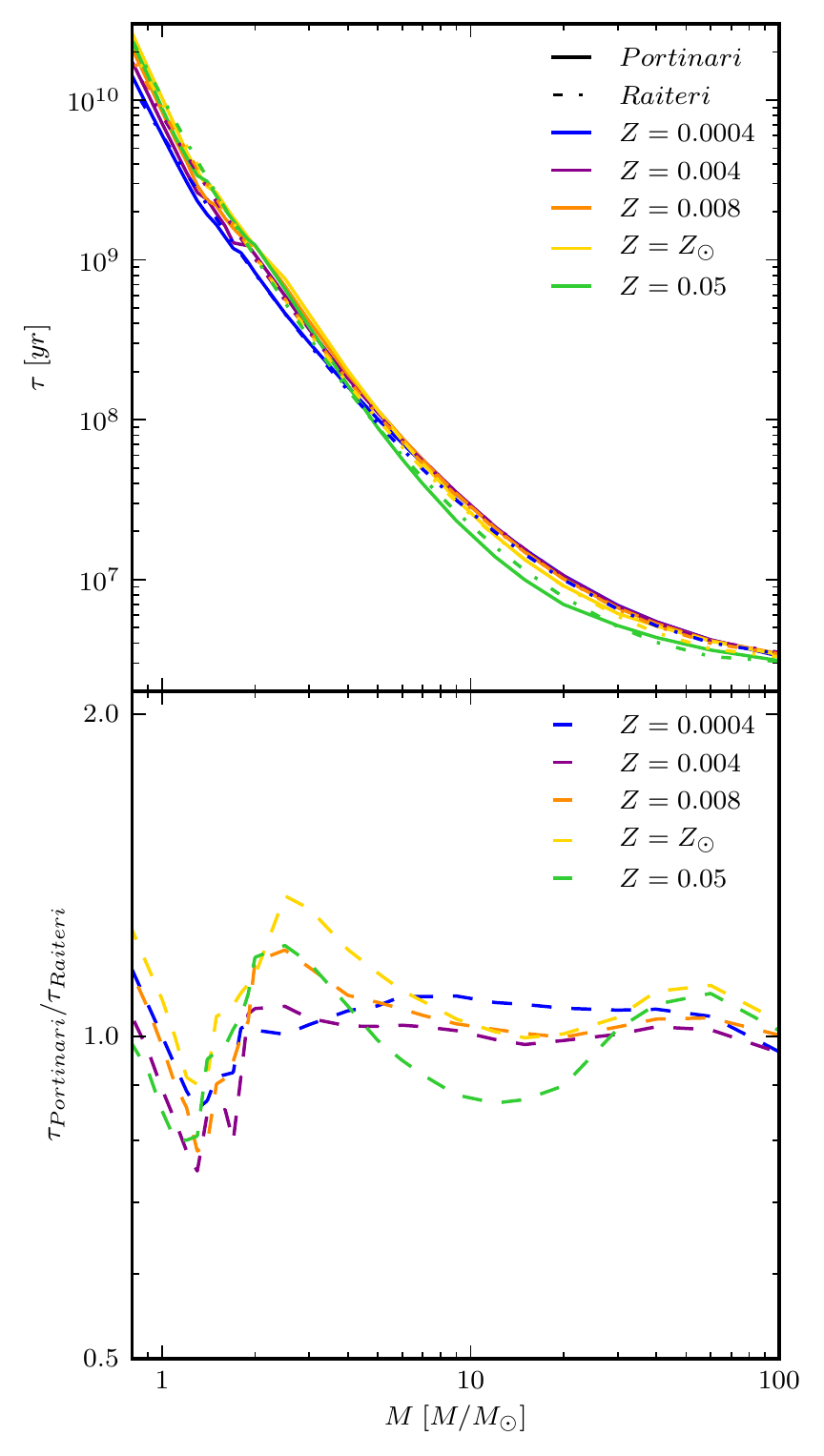}}
	\caption{A comparison of the stellar life-times estimated by \citet{Raiteri96} and \citet{P98} for different metallicities (upper panel), and their ratio (lower panel).  }
	\label{fig:lifetime}
	\end{center}
\end{figure}

Stellar life-times are also an important ingredient of chemical enrichment models as these determine the time of release of the chemical elements that are synthesized in the stellar interiors and in supernova explosions. The life-times depend primarily on the stellar mass with a modest increase/decrease with metallicity below/above $\approx 4\;\rm M_\odot$. In order to be consistent with the various sets of chemical yields that we use (see next subsection), we use two different sets of life-times, as we describe below.

The first set of stellar life-times ($\tau$, in years) are given by \cite{Raiteri96}:
\begin{equation}
\log \tau = a_0(Z) + a_1(Z) \log M +a_2(Z) \log^2 M
\end{equation}
with $M$ being the initial mass of a star particle (in units of M$_\odot$) and $Z$  the metallicity (in the range 0.0004-0.05). The coefficients $a_0$, $a_1$ and $a_2$ are given by:
\begin{eqnarray}
a_0(Z) =& 10.13  +0.07547 \log Z  -0.008084 \log^2 Z &\\
a_1(Z) =& -4.424 -0.7939 \log Z   -0.1187 \log^2 Z  &\\
a_2(Z) =& 1.262   +0.3385 \log Z   +0.05417 \log^2 Z &.
\end{eqnarray}

The second choice for the stellar life-times is given by \cite[][P98 hereafter]{P98} in the form of five tables corresponding to different metallicities (Z=$ 0.0004, 0.004, 0.008, 0.02$ and $0.05$). Both of these sets of life-times are metal-dependent approximations based on the theoretical stellar tracks of the Padova group \citep{Bertelli94}, and agree very well for all masses and metallicities, as shown in Fig.~\ref{fig:lifetime}. (Note that in the case of the Raiteri estimations, we show the life-times which correspond to the same 5 metallicities for which the Portinari life-times are given.)

Note that stars that are progenitors of SNII  (i.e. in our model $M\in [7,40] \; \rm M_\odot$, see next sections) have life-times that vary between $\sim 3\times 10^{6}$ and $\sim 4\times 10^{7}$ yr, unlike less massive stars whose variations in life-times are several orders of magnitude. For this reason, in our model we assume that  the chemical enrichment via SNII occurs in one single episode,  at the mean life-time of the SNII progenitor stars of different masses sampled by a stellar particle, which varies between  $1.5$ and $2.1\times 10^7$ yrs
 (note that this is almost completely insensitive to metallicity). In contrast, as SNIa originate in binary systems the relevant time-scales are given by the evolution of the binary systems. In this case, as well as for AGB stars,  the chemical elements are released over longer time-periods; we model such effect in our code as described in the next subsection.

\subsection{SNII, SNIa and AGB rates}\label{sec:rates}

\begin{figure}
	\begin{center}
	\includegraphics[width=\linewidth]{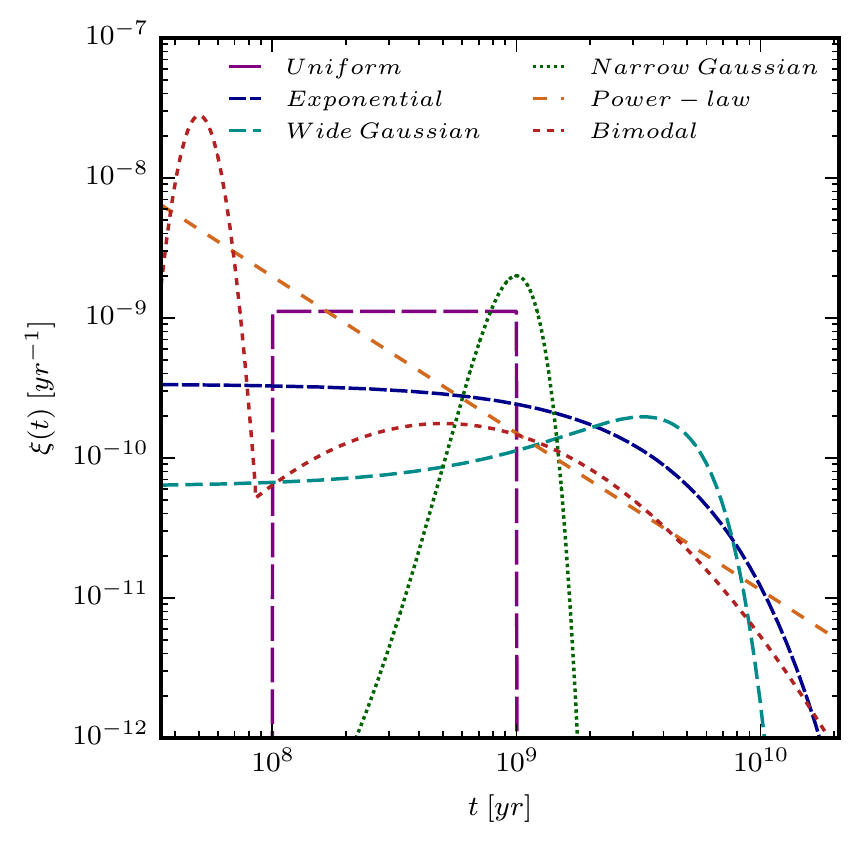}
	\caption{ The five SNIa DTDs considered in our study. The purple dashed line corresponds to the uniform DTD used in S05. The dark blue dashed line corresponds to the exponential DTD given by Eq.\ref{eq:exp}. The dashed light blue  and dotted green lines correspond to the 'wide' and 'narrow' DTDs respectively, given by Eq.\ref{eq:gauss}. The dashed orange line represents the power-law DTD given by Eq.\ref{eq:pow}. Finally the dotted red line corresponds to the bimodal DTD given by Eq.\ref{eq:bimod}.
}
	\label{fig:DTD}
	\end{center}
\end{figure}

Having chosen a model for the life-times of our stellar components we now proceed to identify the fractions that evolve to form SNII, SNIa and AGB stars and determine the rates of the associated mass-loss phases.

SNII are the result of the core-collapse of massive stars, whose short life-times guarantee that their occurrence closely traces the star formation rate of a galaxy, and as a result they are primarily observed in star-forming late-type galaxies.
In our model, we  assume that all stars more massive than $7$ M $_\odot$ and below $40$ M$_\odot$  are progenitors of SNII, and therefore the SNII rate is directly determined by the assumed IMF and life-times of stars.

The rate of AGB stars can be estimated using the same approach, from the number and life-times of low- and intermediate-mass stars. In this case, we use minimum and maximum masses of $0.8$ M$_\odot$ and $7$ M$_\odot$, which correspond to life-times between approximately 53~Myr and $26$~Gyr (with some adjustment for metallicity). As the release of elements via AGB stars is long, we model AGB enrichment assuming three enrichment episodes per particle, that occur at $\sim 100$~Myr, $\sim 1$~Gyr and $\sim 8$~Gyr and contribute to $\sim $25\%, $\sim $55\% and $\sim $20\% of the total mass ejecta respectively.

The nature of the progenitors of type Ia supernova is far less well understood. Observations suggest that SNIa originate from the thermonuclear detonation of CO white dwarfs (WD) near the Chandrasekhar mass, but the mass accretion mechanism is still a topic of considerable debate, and two main approaches have been proposed. First,  the single degenerate scenario where a WD accretes mass from a non-degenerate companion star until it crosses the Chandrasekhar mass limit and explodes;  and second,  the double degenerate scenario which involves the merger of two WDs with a mass of the combination being over the Chandrasekhar mass. Due to this large uncertainty we adopt in our model an empirical approach to describe the SNIa rates similar to that of \cite{W09b} (see also S05 and \citealp{Yates2013}).  In this approach, the SNIa rate is given by the product of $f_{\rm WD}$, the expected cumulative number of WDs per unit stellar mass since star formacion occurred, and the {\it Delay Time Distribution} (DTD) $\xi(t)$, a positive function to determine the rate at which WDs turn into SNIa, s.t. the number of SNIa events in a simulation time-step $\Delta$t  per unit stellar mass is given by 
\begin{equation}
N_{SNIa}(t,t+\Delta t) = A \int_{t}^{t+\Delta t} f_{WD}(t') \xi (t')dt'
\label{eq:number}
\end{equation}
where $A$ represents the fraction of objects in an SSP in the mass range [$1$ M$_\odot$, $7$ M$_\odot$] that are progenitors of SNIa. In this work A is set to 0.01 as in \cite{W09b}. For simplicity we normalise all our DTDs over the interval 29~Myr to 20~Gyr, i.e. 
\begin{equation}
\int_{29\;\rm Myr}^{20\; \rm Gyr} \xi(t) \de t = 1
\end{equation}
 where $ t $ is the time elapsed since the starburst (delay time) and the limits have been chosen conservatively to include all lifetime models for $1$-$7\;\Msun$ stars.

The value of $A$ here is selected in \cite{W09b} as to approximately fit the observed (although not yet fully constrained) cosmic SNIa rate. In this work, we fixed the value of $A$ (except in the case of a uniform DTD, as explained below) and focus on the differences produced by the choice of various DTDs, deferring a detailed study on input parameters for future work.  Notably this is per star rather than per binary, and as such is smaller than those given in other works \citep[e.g.][]{Yates2013}. 
This formalism also introduces some  `double counting' of type Ia and II SNe, in that secondary stars $>1$~M$_\odot$ (in a binary) are assumed to have a white dwarf companion, when in fact some of which will have a higher mass primary that exploded as a type II. 
One can see this is a relatively small fraction by the following argument: If one considers a normalised mass function of binaries in the domain M$_{\rm binary} > 1$~M$_\odot$, then the IMF is approximately a power-law. Taking the mass fraction of secondary 
as $f(\mu) = 2^{1+\gamma} (1+\gamma) \mu^\gamma$ \citep{Greggio_1983}, then $\mu \in [0,0.5]$ implies a mass function of \emph{individual} stars that looks extremely similar (in the $>1$~M$_\odot$ range) up to normalisation \citep[e.g.][]{Malkov_2001} for realistic $\gamma$ values (e.g. 2). Using an IMF exponent of $-2.35$ (Salpeter) gives a fraction of secondaries in the $1$-$7$~M$_\odot$ range with $>7$~M$\odot$ primaries as
around 6\%, i.e. 94\% do indeed have relevant white-dwarf companion for the single-degenerate scenario.

We consider in our study five different DTDs, shown in Fig.~\ref{fig:DTD}, as follows:

{\it i)} A simple power-law DTD with a slope of $-1.12$:
\begin{equation}
\xi_{\rm pl}(\tau) = a \,  \tau^{-1.12},
\label{eq:pow}
\end{equation}
introduced by \cite{Maoz2012} in order to fit the SNIa rate profile derived from the Sloan Digital Sky Survey II, with $a$ the normalization constant.

{\it ii)} An exponential or e-folding delay function  \citep{SBinney09} of the form:
\begin{equation}
\xi_{\rm exp}(\tau) = b\  \frac{e^{{-\tau / \tIa}}} {\tIa}
\label{eq:exp}
\end{equation}
where $\tIa$ is the characteristic delay time, that we assume to be 3 Gyr, and $b$ sets the normalization.

{\it iii)} A bimodal DTD, given by
\begin{equation}
\log \xi_{\rm BM} = c\  \left \{
\begin{array}{ll}
{\displaystyle 1.4-50(\log(\tau)-7.7)^2} \quad    &  \tau  <  \tIa \\
{\displaystyle -0.8-0.9(\log(\tau)-8.7)^2} \quad    & \tau \geq \tIa
\end{array} \right . \quad
\label{eq:bimod}
\end{equation}
proposed by \cite{Mannucci_2006} in order to simultaneously fit the observed SNIa rate and the distribution of SNIa with galaxy B-K color and radio flux, for a redshift range of [0,1.6]. This DTD assumes that 50 percent of the SNIa explode within the first $10^8$yr after the star formation episode while the remaining fraction  have delay times up to 20~Gyr, and $c$ is the normalization parameter.

{\it iv)} `Wide' and `narrow' Gaussian DTD functions, defined as:
\begin{equation}
\xi_{\rm G}(\tau) = d_{W,N}\  \frac{1}{\sqrt{2 \pi \sigma_{W,N}^2}} e^{-\frac{(\tau-\tau_{W,N})^2}{2 \sigma_{W,N}^2}}
\label{eq:gauss}
\end{equation}
with $\tau_W = 3\; \rm Gyr$ and $\sigma_W = 2\; \rm Gyr$ (consistent with the the wide Gaussian model of \citealp{Strolger2004} for Hubble observations and the $\tIa$ of the exponential) and a prompt narrow model of $\tau_N=1\; \rm Gyr$, $\sigma_N = 200\; \rm Myr$ (such a prompt distribution must still be considered, see \citealp{Foerster_2006}), and $d_N$, $d_W$ are the normalization parameters. 

{\it v)} A uniform DTD function (i.e. $\xi(t)=$constant) as in S05 is also implemented in our code for comparison with previous work. Note that in this case, the total number of SNIa events is given by the rate of SNIa which is assumed to be a given (fixed) factor of the SNII rate.

 We emphasise that in Fig.~\ref{fig:DTD} we only plot the different $\xi(t)$ functions, the final SNIa rate depends also on the star formation and lifetime models via the $f_{WD}(t)$ factor in Eqn.~(\ref{eq:number}). As such the total number of SNIa events per solar mass of stars formed is not an explicit parameter of the \citet{W09} model, but is implicit from all these factors.

We implement the enrichment via SNIa  in a discrete number of episodes equally spaced in log$(t)$, during the corresponding time interval as determined by the DTD. In this work, we have used 10 enrichment episodes per SNIa. This number of enrichment steps was chosen as a compromise between computational efficiency and matching of chemical properties in our simulations. As shown in Appendix~\ref{sec:n_SNIa_episodes},  our fiducial choice of $10$ enrichment episodes does not introduce any artificial effect and seems a good option to prevent additional computational cost.

\subsection{Stellar Yields}

The final ingredient of our chemical  model is the selection of chemical yields for SNII, SNIa and AGB stars. We follow the time-release of 11 elements, namely hydrogen, helium, carbon, calcium, nitrogen, oxygen, neon, magnesium, silicon, sulfur and iron. In the following subsections we describe our choices and updates to the chemical yields.

\subsubsection{SNII Yields} 

\begin{figure*}
	\begin{center}
	{\includegraphics{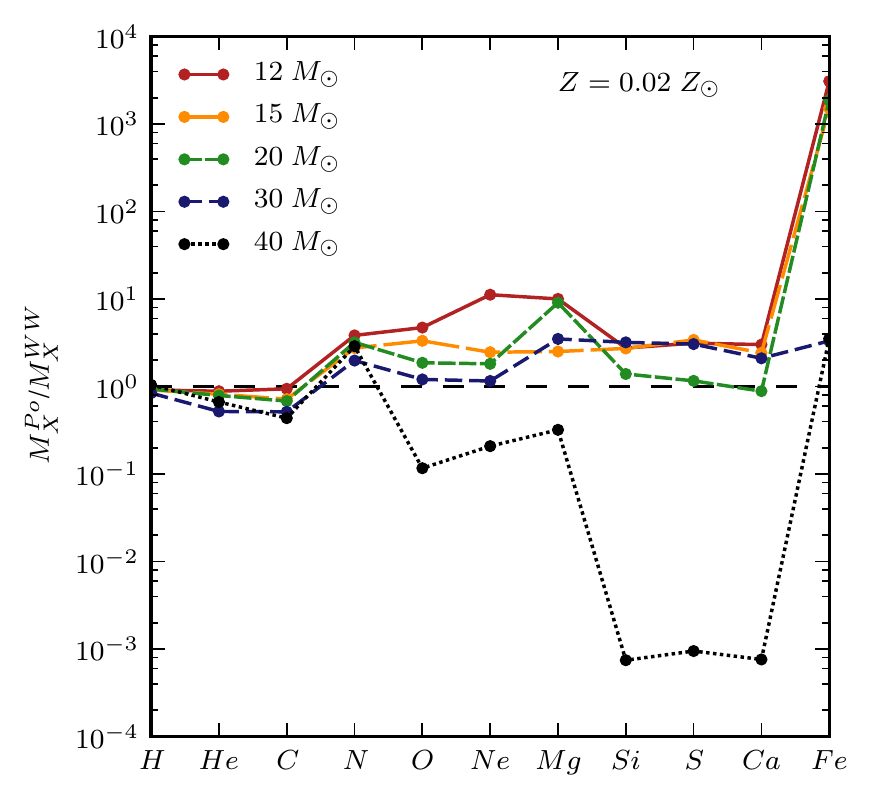}}{\includegraphics{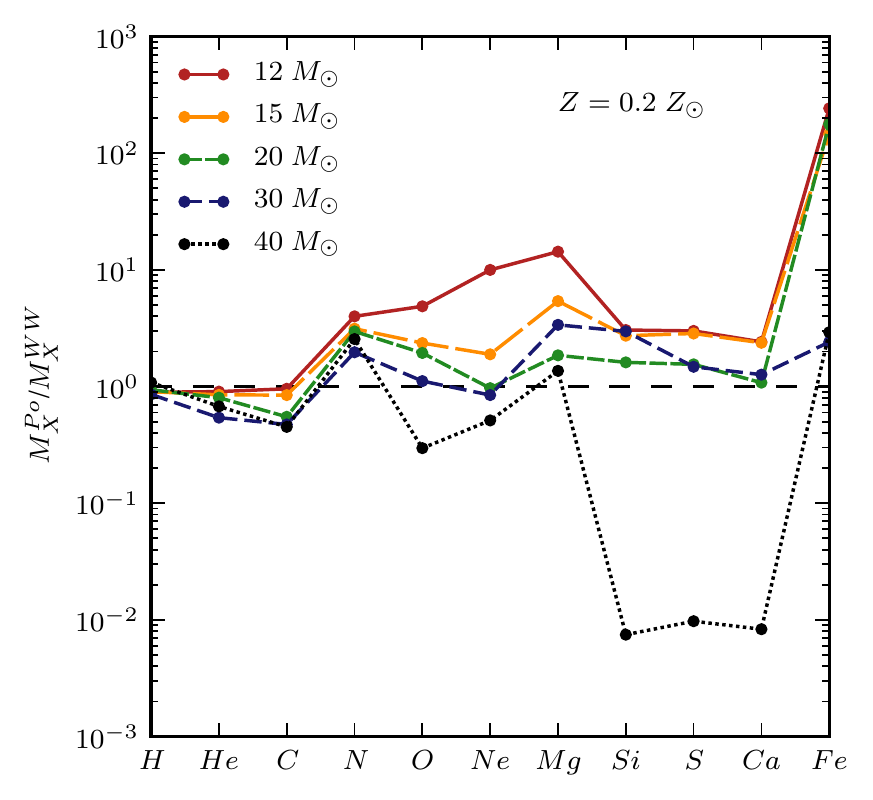}}\\
        {\includegraphics{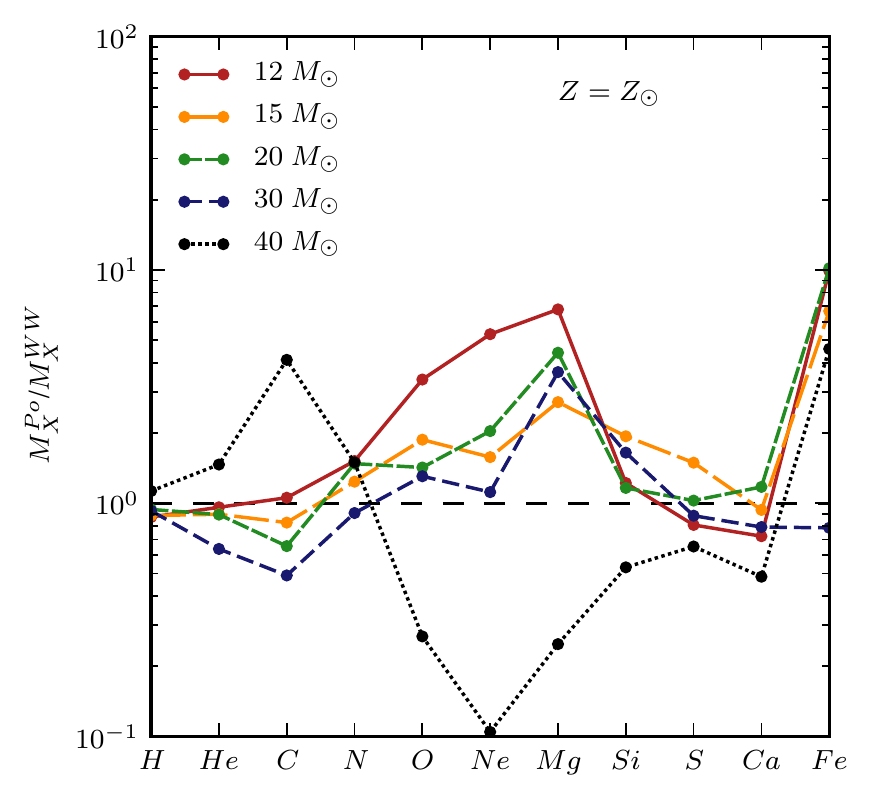}}{\includegraphics{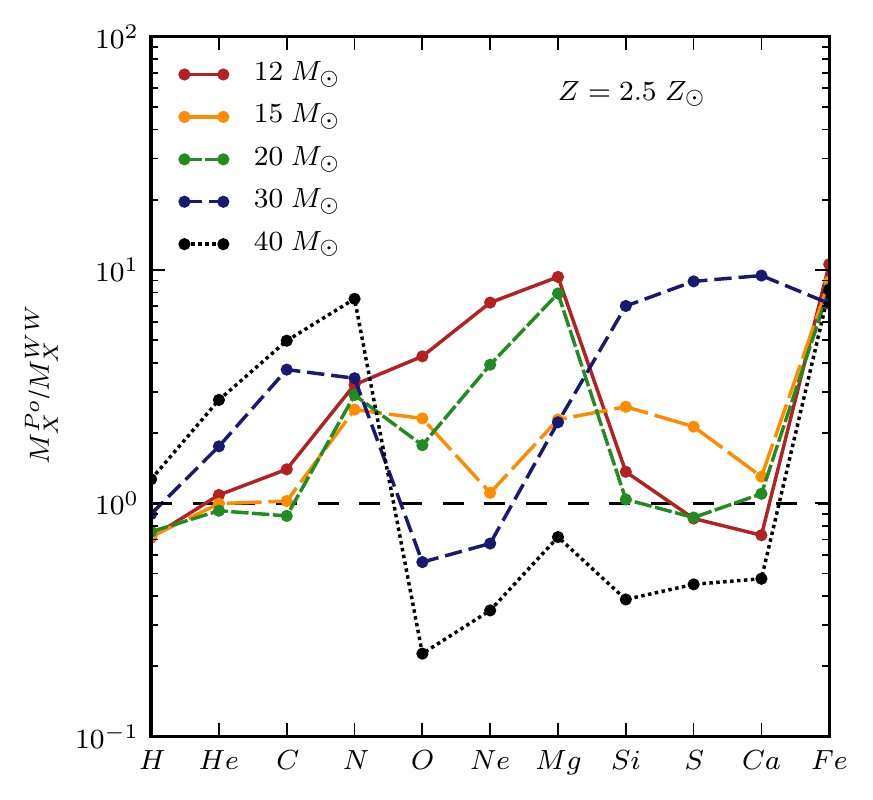}}
	\caption{Comparison of stellar yields of SNII. The figures show the ratio of the masses produced of each element $X$ by a Simple Stellar Population (SSP) in the P98 and WW95 models. Different line-styles indicate different values of the initial mass of the SSP. The results correspond to initial metallicities of $Z = 0.02\;Z_{\odot}$ (upper right panel), $Z =0.2\; Z_{\odot}$ (upper left panel), $Z = Z_{\odot}$ (lower right panel) and $Z =2.5\; Z_{\odot}$ (lower right panel).}
	\label{fig:yields}
	\end{center}
\end{figure*}

SNII produce large quantities of the $\alpha$-elements; in particular, they are the primary producers of oxygen and magnesium in the Universe. Our standard model S05 used the metallicity-dependent yields of \cite[][hereafter WW95]{WW95}\footnote{Note that 
the radioactive nickel yield has been added to the iron yield, and we have used half of the iron yield of WW95 (\citealt{Arrigoni12,Timmes95}).}. In our updated version, we choose instead the metallicity-dependent yields given by \cite{P98} (P98), who tabulate the ejecta from stars with metallicities between 0.0004 and 0.05 and with initial masses between 6 and 1000~$\Msun$. 
In our study we consider stars in the range 7 to $40\;\Msun$ as SNII progenitors. We note that there is no simple cut above which high mass stars form black holes \citep[e.g.][]{Fryer_1999, Smartt_2009}. To give some intuition of the effects of this choice, increasing this limit to $100\;\Msun$ would increase the stellar mass in SNII progenitors by around 5\%.

The P98 yields are based on the explosive nucleosynthetic calculations of WW95. The latter, despite their wide use, are however calculated only up to 40~$\Msun$ and neglect the pre-SN mass loss through winds, which has encouraged the adoption of the P98 yields \citep[e.g.][]{W09b}. P98 also accounts for the decay of nickel into iron shortly after the SN event, unlike WW98. Note that, as in the works of \cite{Portinari2004}, \cite{W09b} and \cite{Yates2013},  we corrected the P98 yields by doubling the Mg and halving the C and Fe ejecta.

In Fig.~\ref{fig:yields} we compare the SNII yields of WW95 and P98 for stars with masses between 12 and 40 M$_{\odot}$ and metallicities between 0.02 and 2.5 Z$_{\odot}$\footnote{For super-solar metallicity (bottom left panel in Fig.~\ref{fig:yields}), the comparison is done using the yield tables of the highest available metallicities: Z=1$Z_\odot$ for WW95 and Z=2.5$Z_\odot$ for P98.}.
In general, the two sets of yields  agree well for hydrogen, helium, carbon and nitrogen; the largest differences appear in the case of the most massive stars (i.e. 40 M$_\odot$) and increase with  metallicity. In the case of heavier elements, the discrepancies between the two models are larger. The largest variations between WW95 and P98 appear in the iron yields: the P98 yield values are between 1 and 3 orders of magnitude higher than those of WW95 depending on the metallicity. Significant differences are also found for silicon, sulfur and calcium, particularly for low metallicities,
with P98 predicting lower yield values compared to WW95,
 likely because these heavier elements remaining locked in the stellar remnant due to the inclusion of stellar winds in the models (P98). 
 Note that the P98 yields (unlike WW95) include the mass loss through winds prior to the SN explosion, explaining the general trend of higher yields in P98 compared to WW95.

\subsubsection{AGB yields}

AGB stars are major producers of carbon and nitrogen. To describe the effects of stars in the AGB phase on the chemical production, we use the metallicity-dependent yield tables of \citet{M01} for the mass range $0.8$-$5$ M$_\odot$ and the tables of P98 for the mass range $5$-$7$ M$_\odot$. Note that both the Marigo yields and those of P98, which we also used for the SNII chemical enrichment contribution, are based on the Padova evolutionary tracks, and together  provide a complete set of yields for the whole mass range $0.8$-$120$ M$_\odot$.

\subsubsection{SNIa yields} 

Type Ia SNe are the primary producers of iron and nickel; however, as explained above,  there are many uncertainties on their chemical yields, which follow the uncertainties on their progenitors and delay-time distribution. As in our previous implementation (S05), in our updated model we adopt the widely-used SNIa  yields of \cite{T2003}, which are based on the spherically symmetric W7 model for the progenitor systems of \cite{Nomoto_1984}. 

\subsection{Relative contribution of SNII, SNIa and AGB stars to the chemical enrichment}  

\begin{figure*}
	\begin{center}
	\includegraphics[height=8cm]{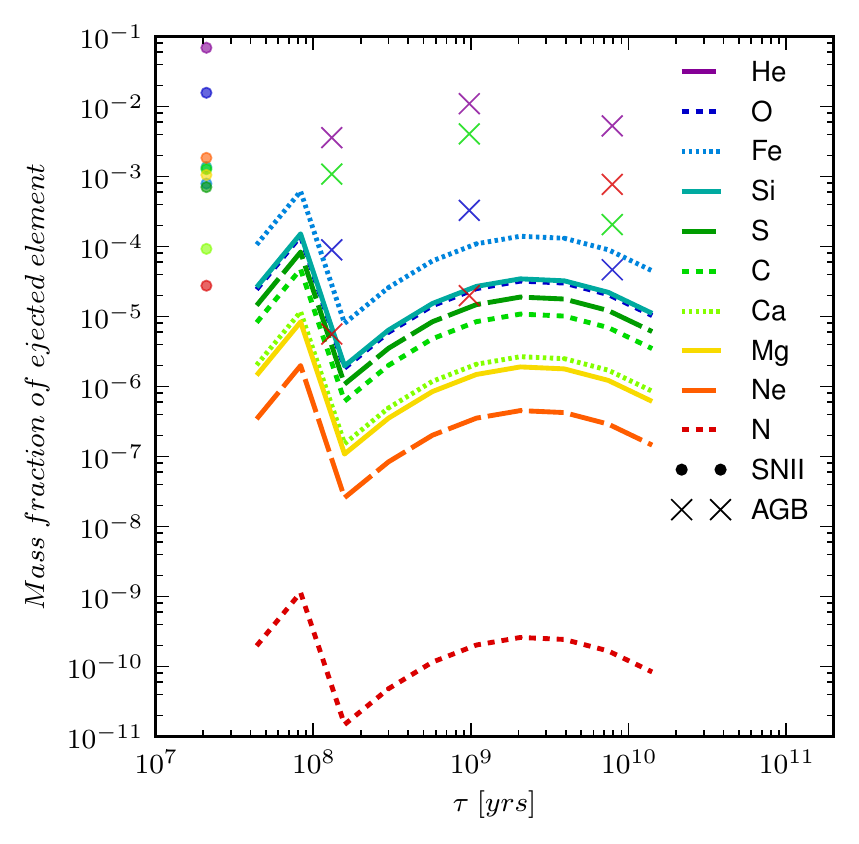}{\includegraphics[height=7.9cm]{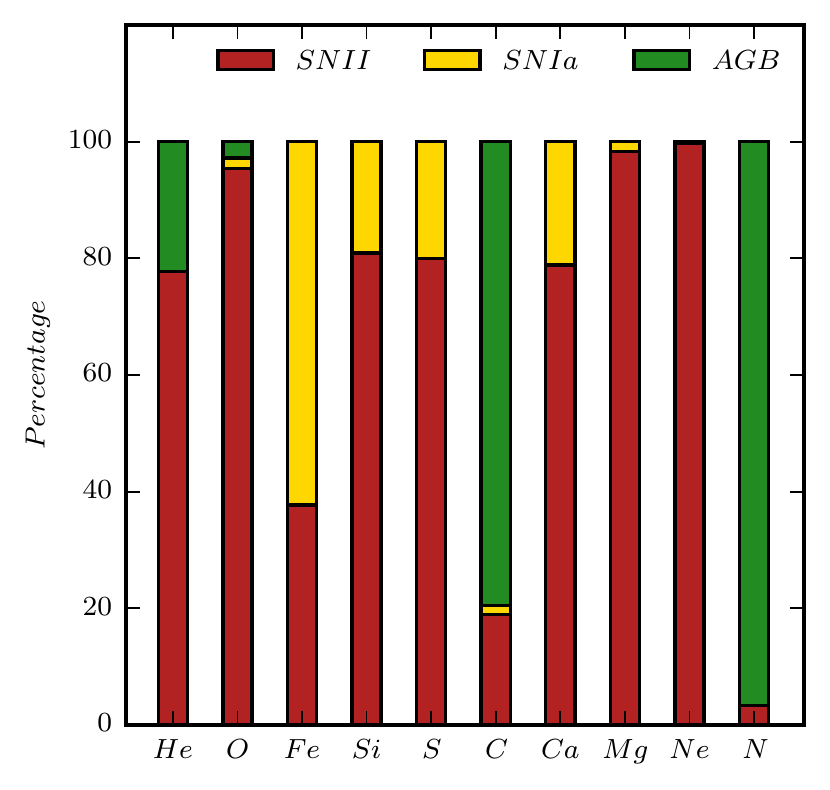}}
	\caption{ {\it Left panel}: Mass fraction released by SNII (solid circles), SNIa (lines) and AGB winds (crosses) according to the tabulations of \citet{M01}. These results are for a SSP of 1 M$_\odot$ with solar initial metallicity and assuming a bimodal DTD, a Chabrier IMF and using the P98, M01 and \citet{T2003} chemical yields. Different colors and line-styles apply for the different elements considered in our study. {\it Right panel}: Comparison of the mass ejected in the different chemical elements by a SSP with  Z$_\odot$ via SNII (red), SNIa (yellow) and AGB (green) winds.}
	\label{fig:contribution}
	\end{center}
\end{figure*}

In this subsection we compare the relative contribution of SNII, SNIa and AGB stars
to the various chemical elements considered in our study. Note that, 
the relative contribution of each of these to the chemical enrichment
of a galaxy will depend not only on the choice of IMF
and chemical yields, but also on its particular star formation and chemical
history which is affected by environment (e.g. amount of un-enriched material
from infall), merger activity (e.g. amount of gas/metals contributed by satellites)
and
formation time of the stars (that determines, at a given time, how much
of the total contribution to metals through the different channels has been 
completed).

In Fig.~\ref{fig:contribution} we show the element mass ejected
via the different enrichment events
by a SSP of 1 M$_\odot$, assuming a bimodal DTD and a Chabrier IMF (left-hand panel)
 and the corresponding relative contributions (right-hand panel) in the limit $t\to \infty$. 
We have ignored hydrogen for brevity, although it forms a considerable fraction of the ejecta.
AGB stars are clearly the major producers of C (97$\%$) and N (80$\%$), and 
SNIa contributes about $60\%$ of the the total amount of Fe with the remaining fraction
coming from SNII. 
O, Mg, Ne are almost exclusively produced via SNII, that also contribute
about $80\%$ of the total amount of Si, S and Ca.

\subsection{Cooling}

The original S05 model included metal-dependent cooling as given by
\cite{SD93}, which assumed collisional ionization equilibrium (CIE).
A number of astrophysical environments require significant
corrections for photo-ionisation due to the UV background.
This can suppress the cooling rates by a large factor, particularly
for the temperature range  $10^4-10^5$~K in the case of primordial gas
\citep{E92} and up to $10^6$~K for enriched gas \citep[e.g.][]{W09}.
In addition the presence of photo-ionization introduces a much stronger
dependence on gas density.

In our updated model, we use pre-computed cooling tables
for an optically thin
and dust-free gas in ionization equilibrium 
(ignoring the depletion of metals onto dust grains, e.g. as in \citealp{Whittet_2010})
in the presence of the cosmic
microwave background (CMB) radiation and an uniform redshift-dependent
UV background radiation field from galaxies \citep{HM01}.
These tables were computed by \cite{W09} using the photo-ionization
package CLOUDY \citep{F13}, and are a function of
temperature,
density, redshift, and the abundances of individual elements.
The new tables also allow for
net heating that occurs at  very low densities
($\lesssim 10^{-4}$cm$^{-3}$) and
temperatures $\sim 10^4$ K.
The cooling tables cover a range in densities of  [$10^{-8}$, 1]cm$^{-3}$,
in temperatures of [100, 9.2$\times 10^8$]K, and are given for redshifts between 0 and 8.989
(for higher redshifts we use the $z=8.989$ cooling curve).

\subsection{Star formation and feedback model}\label{sec:sf-feed} 

As explained above, we use the same modules for star formation and
feedback of the standard S05 and  \cite{S06} model. Star formation
takes place stochastically, in particles that are eligible
to form stars (i.e.
denser than  $\rho_{\rm crit}=7\times 10^{-26}$g cm$^{-1}$ and in a converging flow).
For these, we assume a star formation rate (SFR) per unit volume equal to:
\begin{equation}
\dot{\rho_{\ast}} = c \frac{\rho}{\tau_{dyn}}
\end{equation}
where $\rho_{\ast}$ and $\rho$ are respectively the stellar
and gas densities, $c$ is the star formation efficiency
(in general we use $c=0.1$) and $\tau_{dyn} = 1/\sqrt{4\pi G \rho}$
is the local dynamical time of the gas. 
We do not examine the sensitivity of our results to the factor $c$ or this star formation scaling, which we take to approximate observed star formation scalings relatively well \citep{Schaye_2008}.

Each supernova explosion injects $0.7\times 10^{51}$ ergs of energy into the
interstellar medium, in equal proportions to the local hot and cold gas phases.
Energy is released to the hot phase at the time of the explosion, whilst
for the cold phase it is accumulated in a reservoir until it is sufficient
to thermalize the cold particle with the local hot phase.
Our feedback scheme works together with a multiphase model for the gas component,
which allows coexistence of cold and hot phases in the interstellar medium making
the deposition of supernova energy efficient. We have shown in previous
work \citep{S06,S08} that our model is able to regulate star formation
with a dependency on the total halo mass, and to drive galactic
winds that transport mass and metals from the inner to the outer regions
of galaxies.

We refer the interested reader to \cite{S06} for details on the implementation
of energy feedback from supernova, and to \cite{S08}, \cite{S09}, \cite{S10}, and \cite{S11} for previous works on the formation of disk galaxies in a cosmological context using this model.


\section{The effects of varying the assumptions of the chemical model} 
\label{sec:isolated}

In this section we study the effects of changing the various
assumptions of our chemical enrichment model, in particular
the IMF, the chemical yields of SNII, the DTDs for SNIa and the cooling tables.
In order to test the sensitivity of our
model to the different assumptions, we ran a set of isolated galaxy
simulations from idealized initial conditions (ICs).
The ICs consist of a  spherical grid of  superposed dark matter
and gas particles, that is radially perturbed to produce a
density profile of the form $\rho(r)\sim r^{-1}$ in solid
body rotation \citep[e.g.][]{Navarro93}. This halo has a total mass of $10^{12}$M$_\odot$,
initial radius of $100$ kpc and a spin parameter of $\lambda=0.1$.
The baryon fraction is 10$\%$, the assumed gravitational
softening is 100 pc for all particles, and the number of
particles is $N=2\times 64^3$.
In Table \ref{table:isolated} we summarize the characteristics of the different
simulations analyzed in this Section (see also Appendix~\ref{sec:resolution}
for a resolution study).

We note that, given the isolated nature of our test galaxy, evolution
is not realistic for long time periods, in particular as there is no resupply of
gas via accretion. All our runs exhibit a  strong star
formation burst after the collapse of the gas cloud, followed by an
inevitable  decay (regulated by gas return)
until the gas reservoir is exhausted or has been
expelled from the galaxy. We have run most of our simulations  up to 2 Gyrs, except
those involving AGB stars and experiments related to the DTDs, that
we have continued to 10 Gyrs to encompass their long characteristic time-scales.

\begin{table*}
\caption{List of the isolated galaxy simulations used in this work with their respective parameters used in the chemical enrichment model. The Table contains: the simulation name, the choice for the IMF, the type of yields for SNII, the incorporation of AGB stars, the DTD of SNIa and the assumed cooling tables. }
\begin{center}
\begin{tabular}{lccccccc}
\hline
\hline
 Name &  IMF$^1$ & SNII Yields$^2$ & SNIa DTD  & AGB &  Cooling$^3$
\\\hline
I01       & S       &  WW95     & uniform          & no    & SD93 \\
I02       & C       & WW95      & uniform          & no    & SD93 \\
I03       & K       & WW95      & uniform          & no    & SD93 \\
I04       & C       & P98       & uniform          & no    & SD93 \\
I05       & C       & P98       & uniform          & yes   & SD93 \\
I06       & C       & P98       & exponential      & yes   & SD93 \\
I07       & C       & P98       & wide Gaussian    & yes   & SD93 \\
I08       & C       & P98       & narrow Gaussian  & yes   & SD93 \\
I09       & C       & P98       & power-law        & yes   & SD93 \\
I10       & C       & P98       & bimodal          & yes   & SD93 \\
I11       & C       & P98       & exponential      & yes   & W09  \\

\hline
\end{tabular}
\end{center}
{{\sc notes:}\\
$^1$ S, C and K are abbreviations for Salpeter, Chabrier and Kroupa IMFs respectively.\\
$^2$ WW95 and P98 stand for \cite{WW95} and \cite{P98}, respectively.\\
$^3$ SD93 and W09 refer respectively to the use of the \cite{SD93} and \cite{W09} cooling tables.}
\label{table:isolated}
\end{table*}

\subsection{The effects of varying the Initial Mass Function} 
\label{sec:IMF}

\begin{figure*}
\begin{center}
{\includegraphics[width=5.75cm]{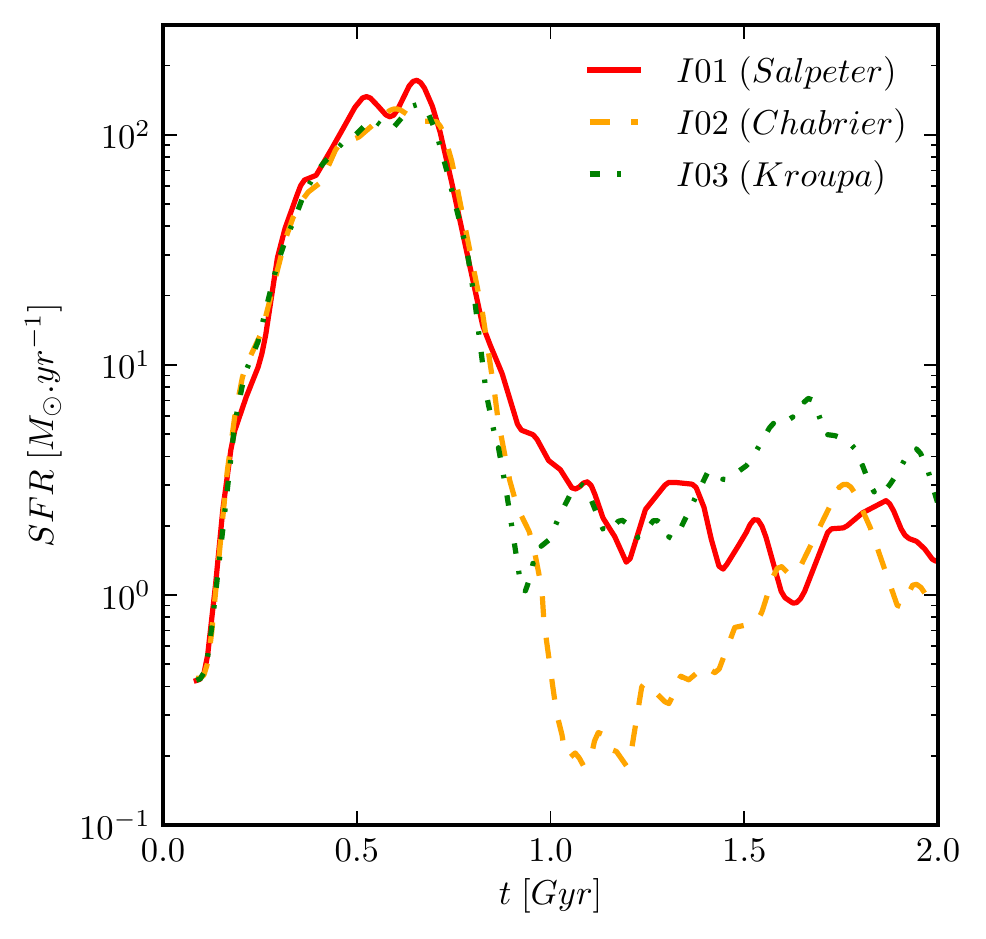}\includegraphics[width=5.75cm]{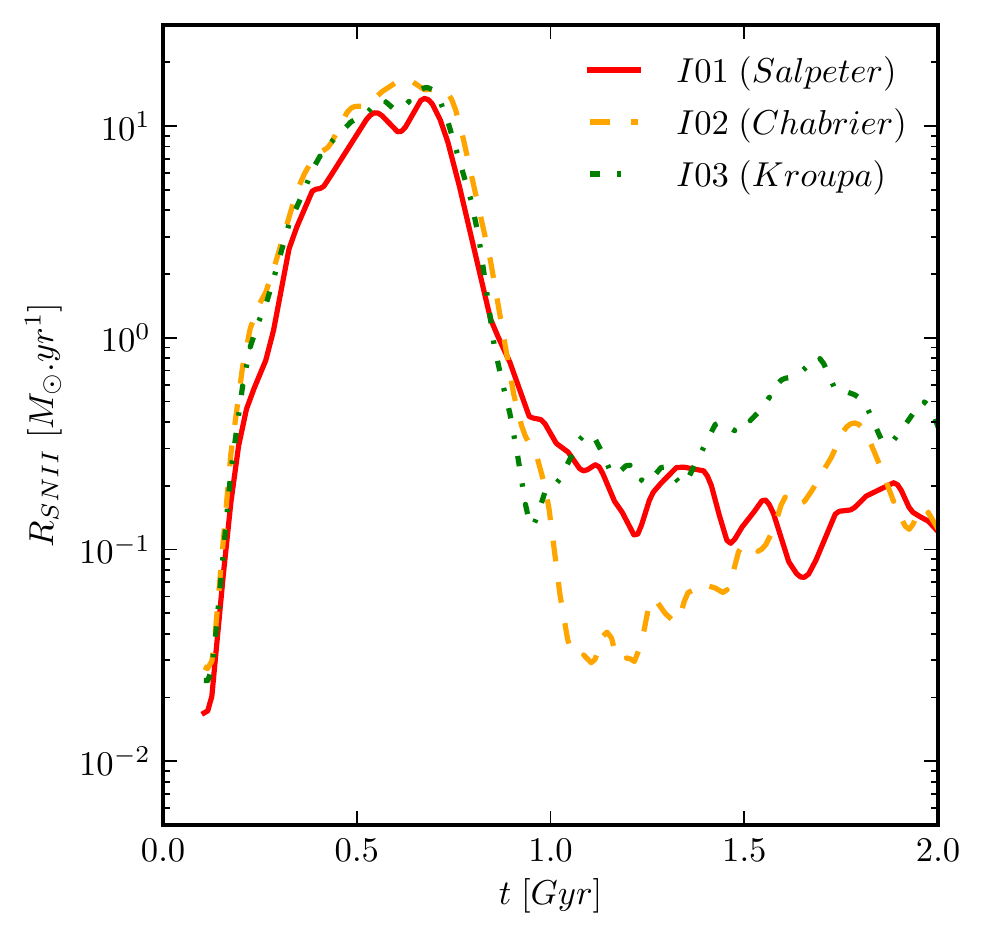}\includegraphics[width=5.75cm]{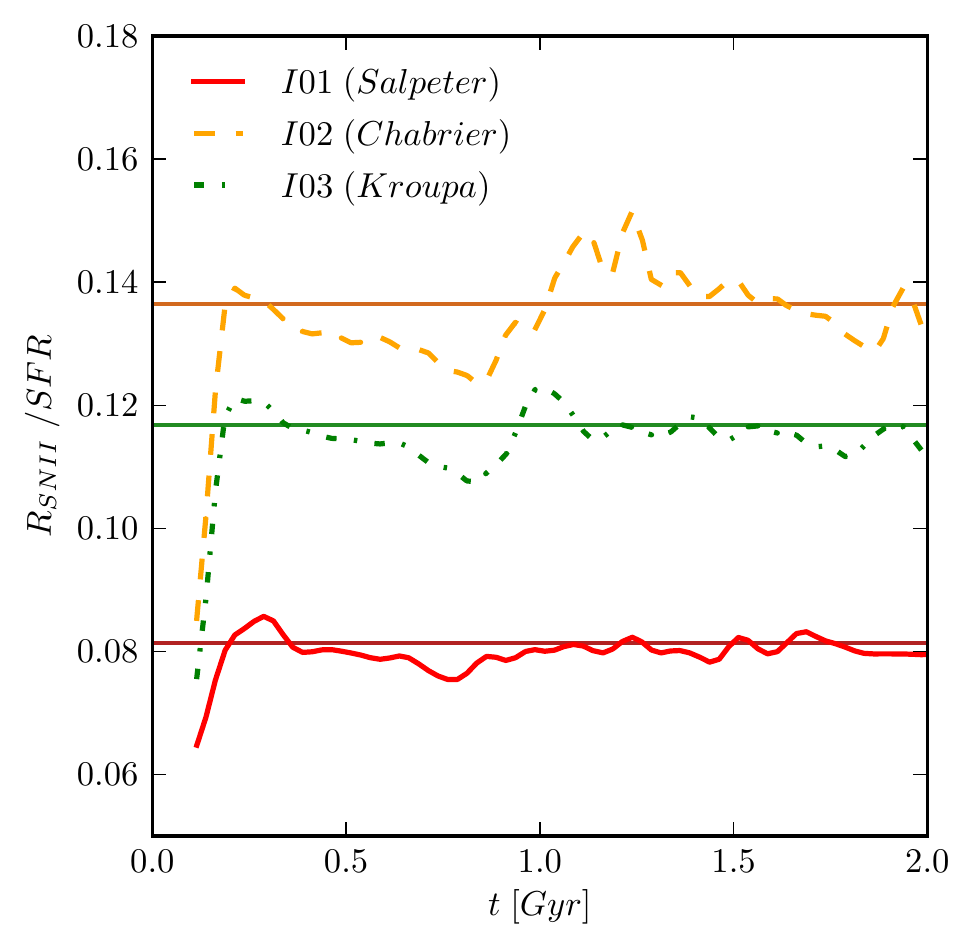}}
\caption{
Star formation (left-hand panel) and SNII (middle-hand panel) rates of our test runs I01, I02 and I03, with
the different IMFs. The right-hand panel shows
the ratio between the SNII and the star formation rates for the same tests. The horizontal lines represent the expected ratios given the choice of the IMF and the total
mass ejecta assumed when using the WW95 model. }
\label{fig:SFR_IMF}
\end{center}
\end{figure*}

\begin{figure}
\begin{center}
{\includegraphics[width=\linewidth]{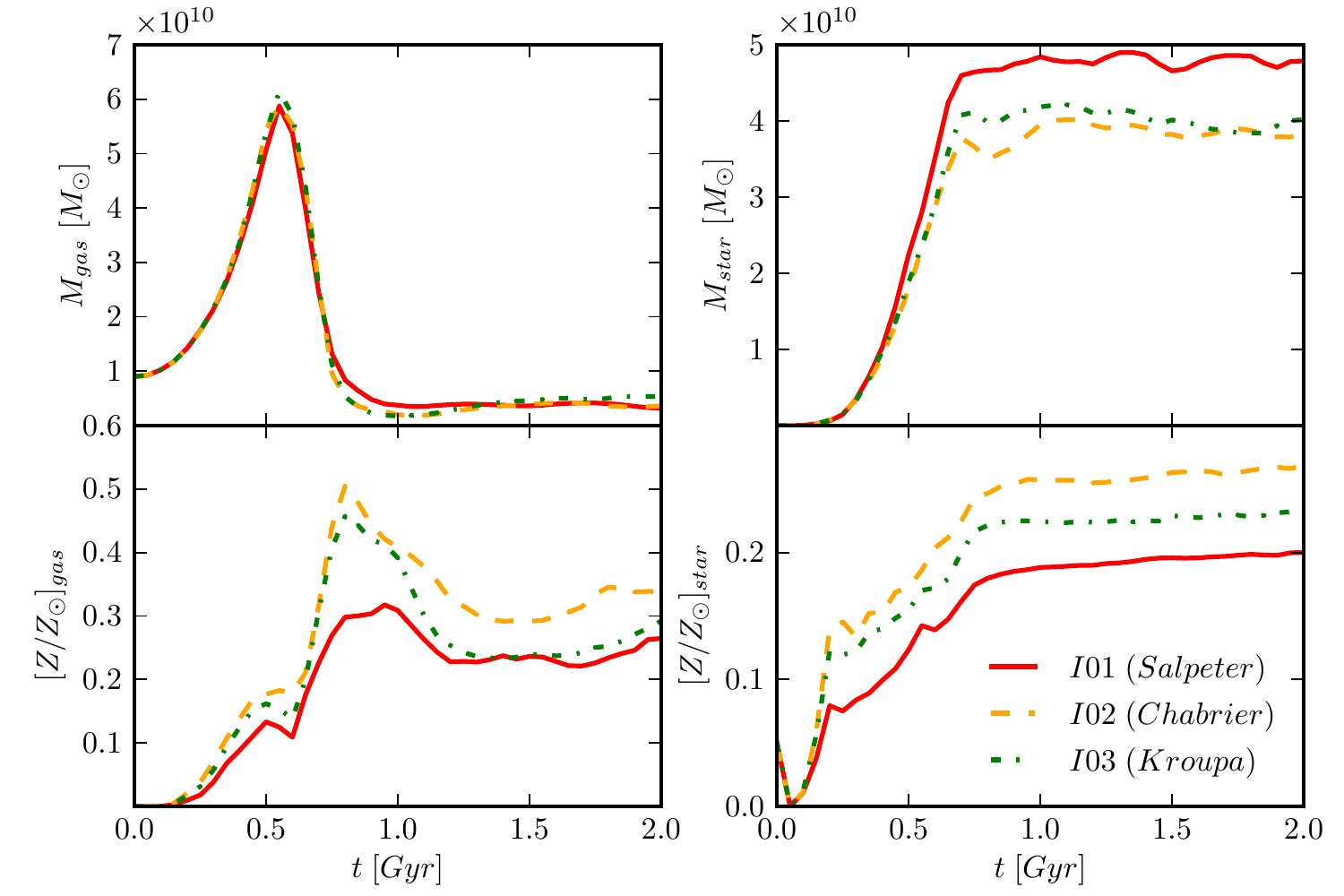}}
\caption{The evolution of the mass and metallicity for the gas (left panels) and stars
(right panels) for simulations I01, I02 and I03, which assume a Salpeter (1955, red), Kroupa (2001, green) and Chabrier (2003, yellow) IMFs, respectively. We consider here particles within the inner 30 kpc of the galaxy, in order to avoid gas particles that
have left the system due to feedback effects. }
\label{fig:Imfmetals}
\end{center}
\end{figure}

Different choices of the IMF will affect
the evolution of the simulated galaxies, as the
relative contribution of stars of different mass and their
resulting feedback will change, affecting the star formation
and supernova rates, and therefore the amount of metals
in the stars and the gas.
As explained in the previous section, in our model we assume
that stars with masses between 7 and 40M$_\odot$ end up as SNII,
stars with masses between 0.8 and 7M$_\odot$ will ultimately reach the AGB phase,
and progenitors of SNIa come from stars with masses between 1 and 7 M$_\odot$.
The Kroupa and Chabrier IMFs produce higher fractions of SNII progenitors
(0.14$\%$ and 0.18$\%$, respectively) compared to Salpeter ($0.11\%$), of stars
 that reach the AGB phase
($0.35\%$ and $0.41\%$, respectively, against $0.28\%$ for Salpeter),
and of SNIa progenitors
($0.30\%$ and $0.36\%$, respectively, against $0.24\%$ for Salpeter).
As AGB stars do not contribute to feedback, their effects on the overall evolution
of the galaxies are
expected to be moderate
(however, they will be important for the abundance of elements such
as C and N); and as the SNIa rate is much lower than that of SNII, SNIa effects
will be  sub-dominant in terms of their contribution to feedback compared to SNII.
For these reasons, and
in order to better isolate the effects of the various choices for the IMF,
the tests presented
in this Section do not include the modeling of AGBs (see Section ~\ref{sec:AGBs}).
We do however include SNIa, assuming a uniform DTD (see Section ~\ref{sec:DTD} for
the effects of varying the DTD).

Fig.~\ref{fig:SFR_IMF} shows the star formation and SNII\footnote{We quantify the SNII rate in terms of the total mass ejected per unit time.} rates, as well as their ratio, for our tests I01, I02 and I03, that are identical except for the
choice of the IMF (Table~\ref{table:isolated}). In all cases, the SFR peaks at
about $0.5$ Gyr, starting to decline at $\sim 0.8$ Gyr as the gas is
exhausted (note that no resupply of gas is
considered in these simulations).
The drop in SFR results from feedback effects, particularly from
SNII whose
progenitors are short-lived, massive stars, therefore
triggering a rapid effect after the starburst.
The SNII rate behaves similarly to the SFR but,
as quantified in terms of the ejected mass per time unit, has an additional
dependence with the chemical yields.
The test that assumes a Chabrier IMF produces the highest number of SNII
progenitors, and thus
the highest amount of feedback energy, followed by the tests that use a Kroupa and Salpeter IMF. For this reason, it reaches the lowest SFR level at the starburst, and the
most significant drop in both the SF and SNII rates.

The changes in the SFRs and the corresponding effects of the  feedback from
SNII of tests
I01-I03 translate into changes in the total stellar mass
produced and therefore in the metallicity distributions of the gas
and the stars in the simulated galaxies.
Fig.~\ref{fig:Imfmetals} shows the evolution of the mass and metallicity
for the gas (left panels) and stars (right panels) in these simulations.
The galaxy formed in test I01, which assumes a Salpeter IMF, has
the highest stellar mass, as it has the lowest SNII rate/SFR (Fig.~\ref{fig:SFR_IMF}, right-hand panel) and feedback effects are less important compared to
those in tests I02 and I03.
The higher stellar mass of I01 can not compensate the lower amount
of SNII events, resulting in galaxies with lower metal
content, both for the gas and for the stars.
In contrast, the Chabrier IMF is the one that produces the highest
enrichment levels, and the Kroupa IMF predicts a
final metallicity that is slightly lower.

\subsection{The effects of varying the SNII yields} 

\begin{figure*}
\begin{center}
{\includegraphics[width=8.5cm]{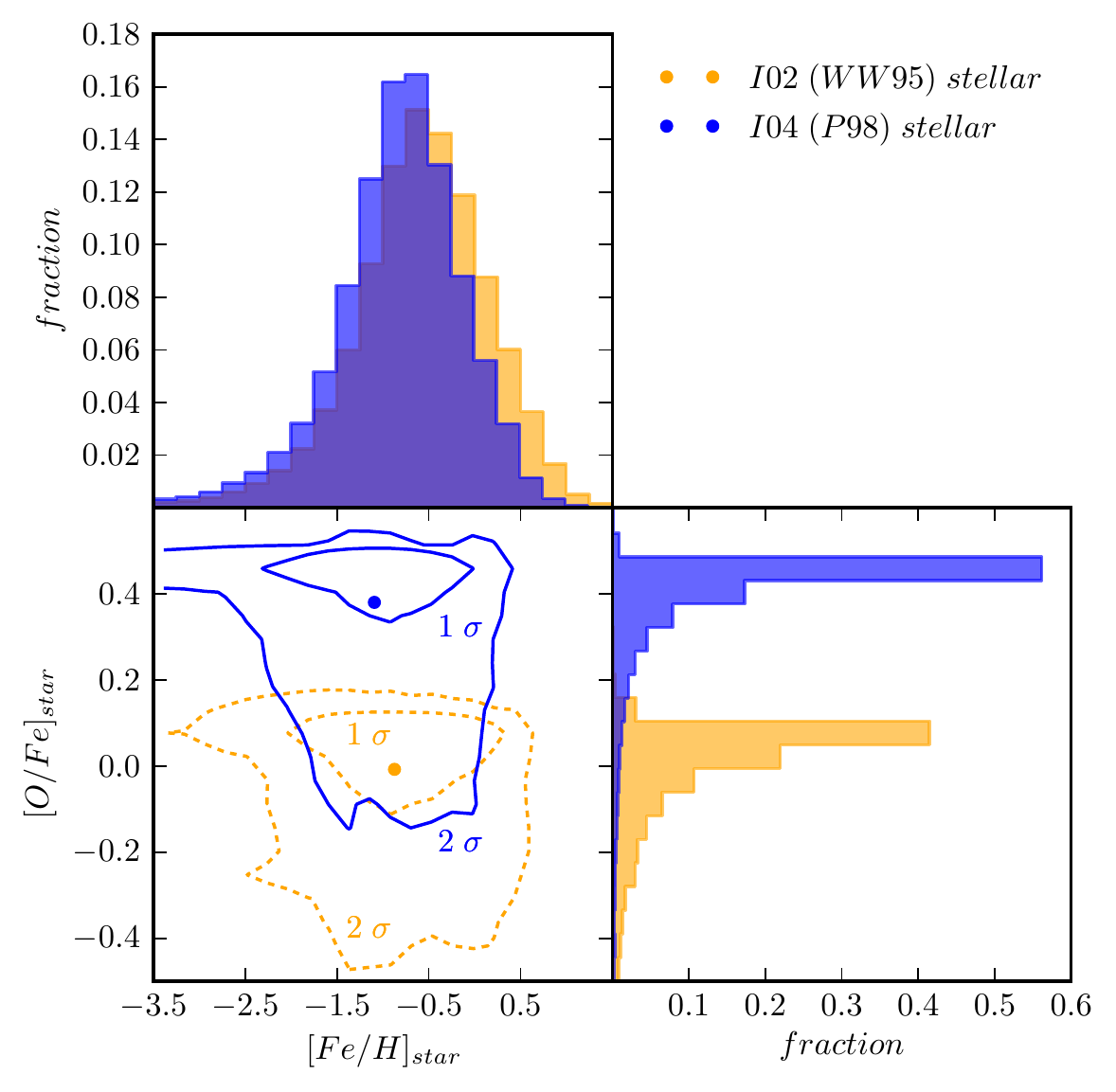}\includegraphics[width=8.5cm]{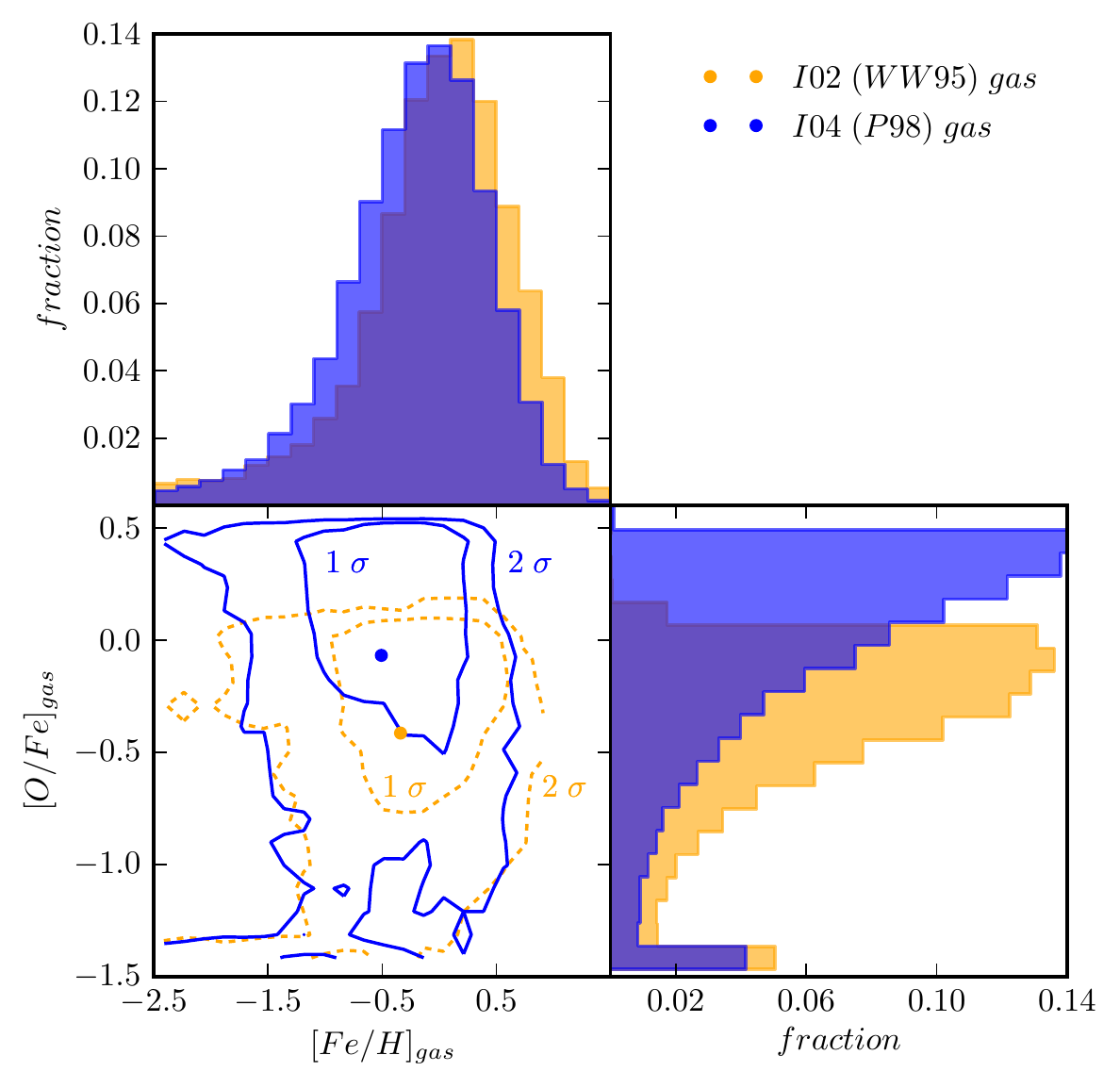}}
\caption{The stellar (left-hand panel) and gaseous (right-hand panel)
 distribution functions of [O/Fe] and [Fe/H] of runs I02 and I04, which differ
in the chemical yields adopted. In the case of the gas, we use particles
within 30 kpc from the center.}
\label{fig:Yofefeh_yields}
\end{center}
\end{figure*}

\begin{figure*}
\begin{center}
{\includegraphics[width=15cm]{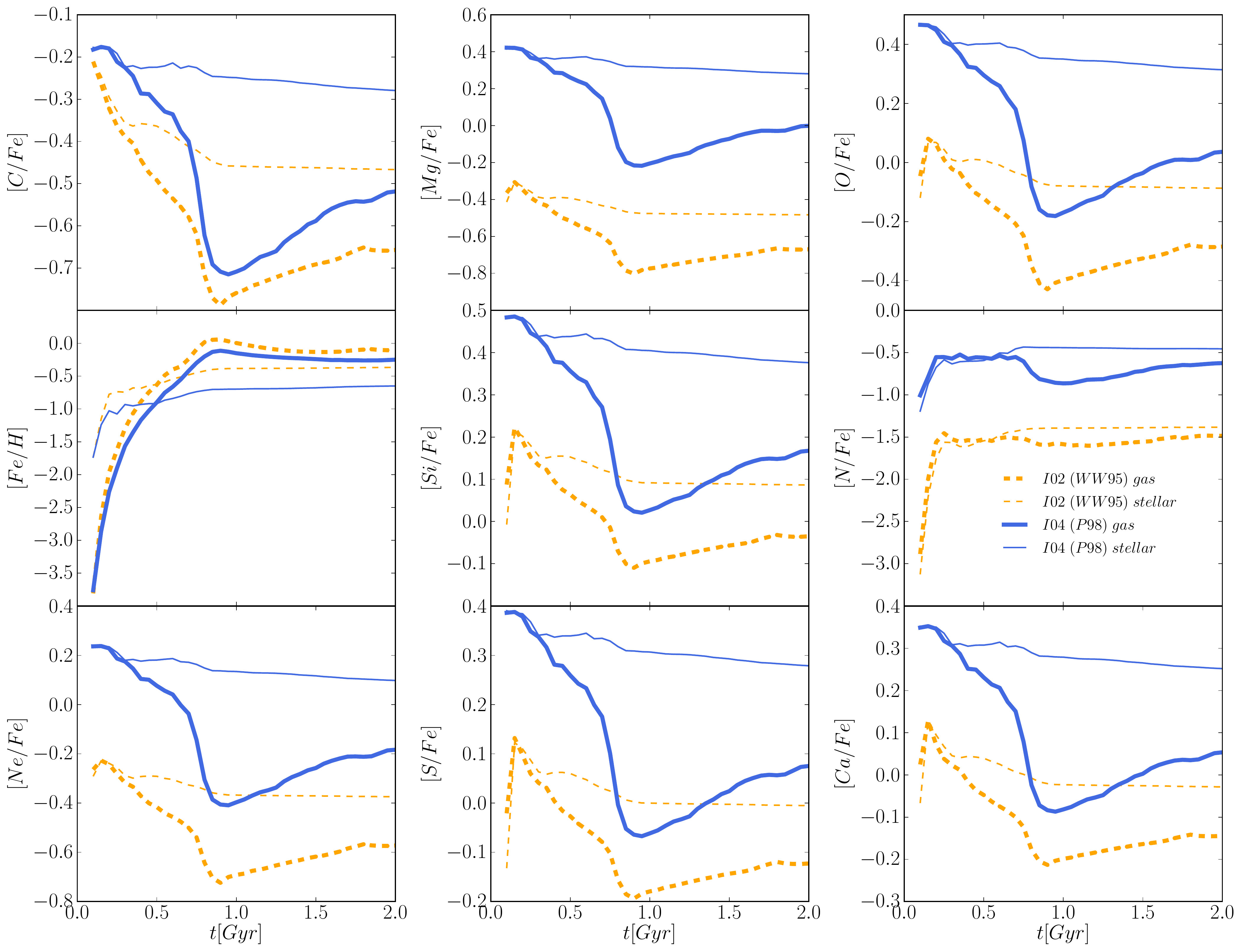}}
\caption{ Time evolution of the various stellar (thin lines)  and gaseous (thick
lines) element ratios
for simulations I02 and
I04 which assume, respectively, the WW95 and P98 SNII chemical yields.
The top row shows the relative abundances of Carbon, Magnesium and Oxygen
with respect to iron, in the middle row we show  [Fe/H] and the
relative abundances of Silicon and Nitrogen to Iron, and the
lower row shows Neon, Sulfur and Calcium, relative to Iron.
}
\label{fig:Yelems_yields}
\end{center}
\end{figure*}

\begin{figure*}
\begin{center}
{\includegraphics[height=6cm]{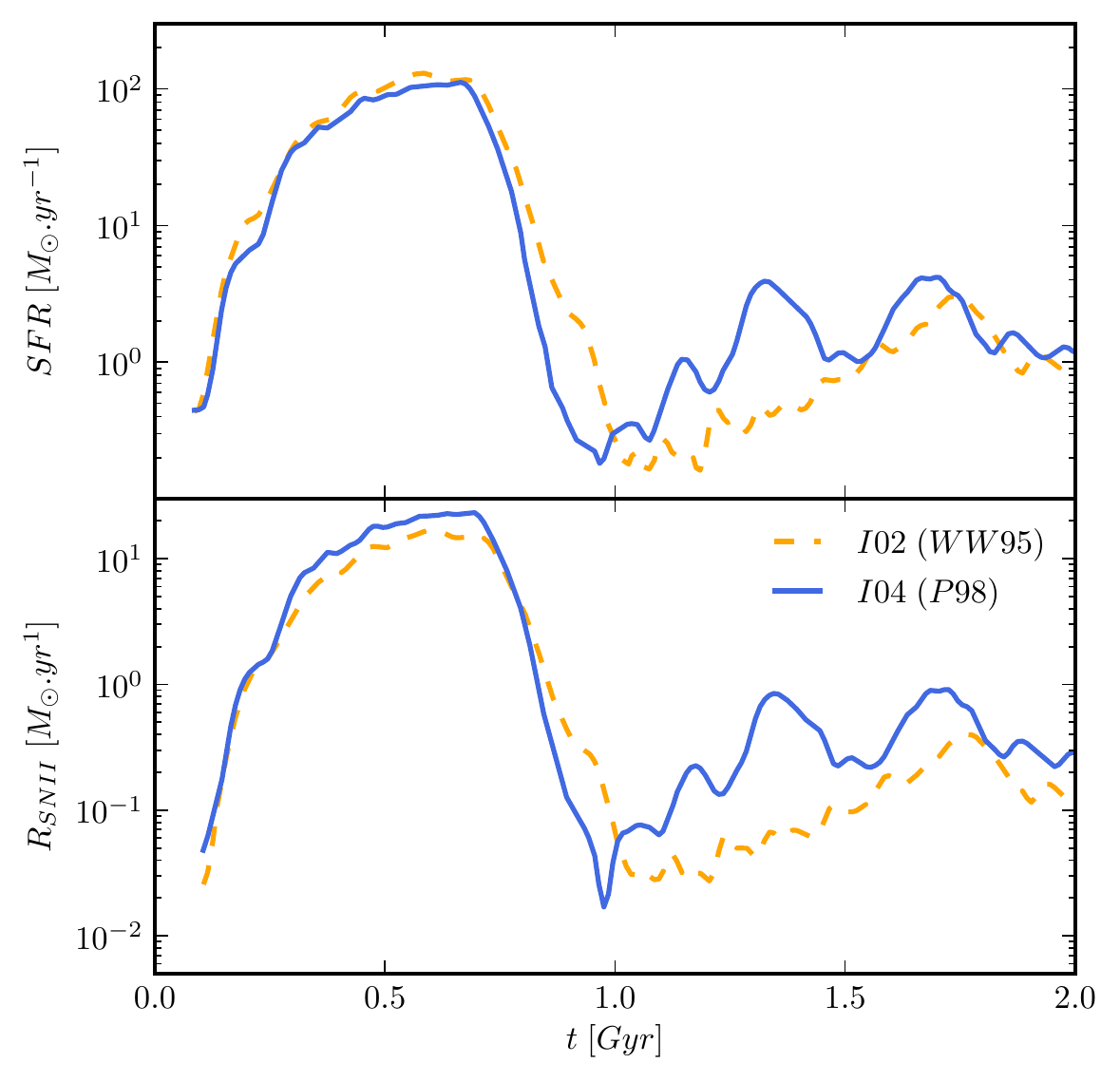}\includegraphics[height=6cm]{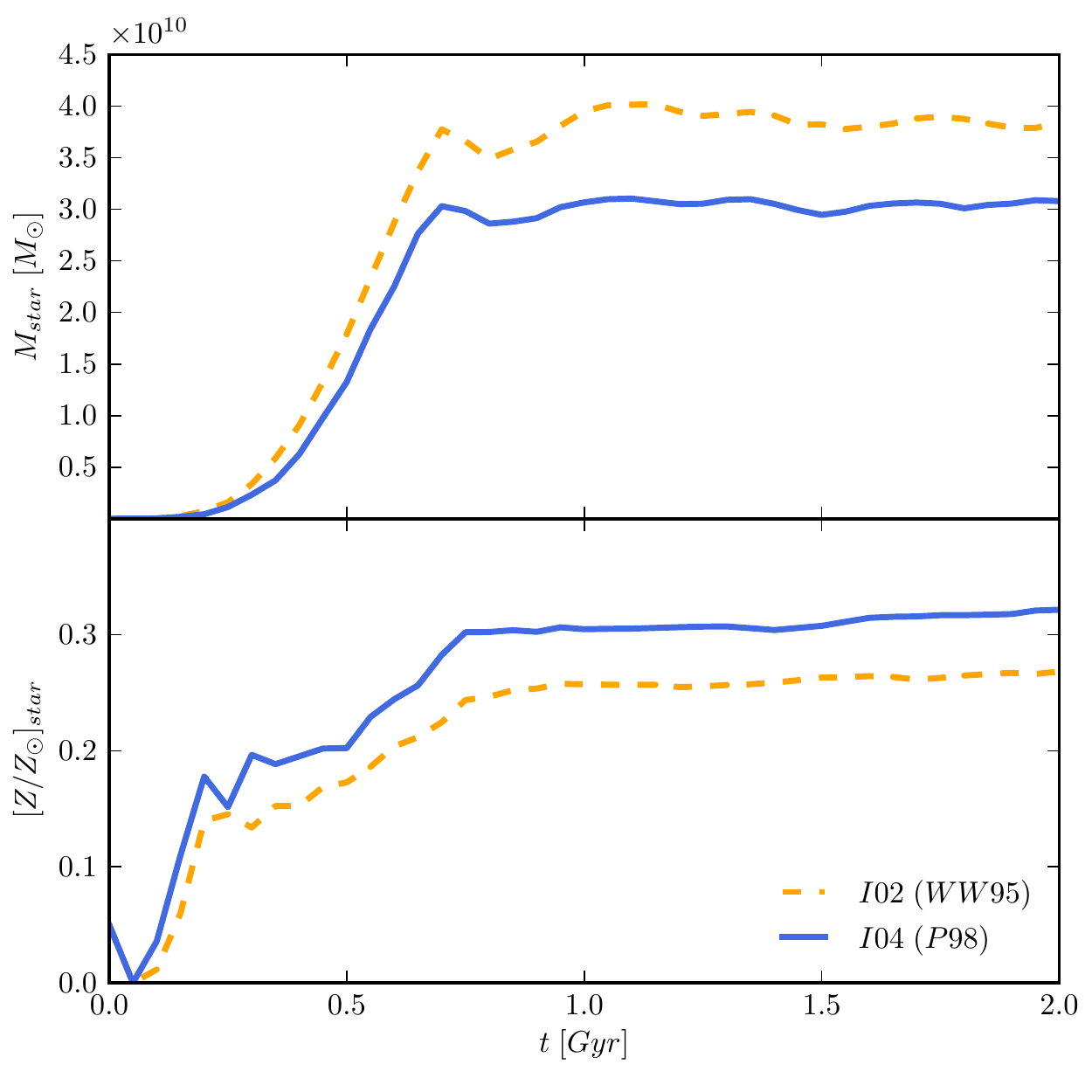}\includegraphics[height=6cm]{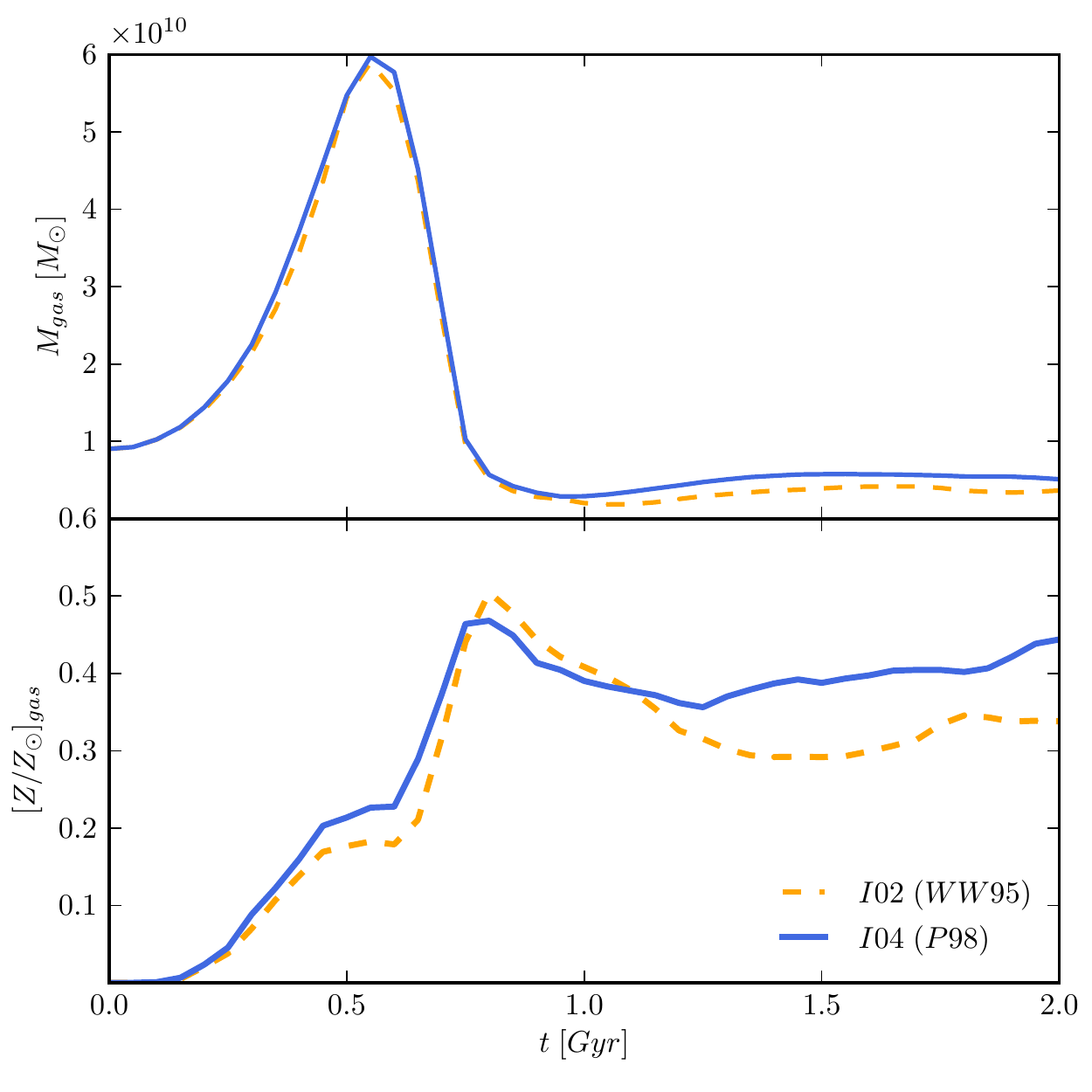}}
\caption{Star formation and SNII rates (left-hand panel), and the evolution of the mass
and metals in the stellar (middle-hand panel) and gaseous (right-hand panel) components in our test runs I02 and I04, which
are identical except for the choice of the SNII yields.
The masses and metallicities have been calculated using particles
in the inner 30 kpc,
 in order to facilitate the comparison between the runs, particularly for
the gas component, as a different fraction of the gaseous mass will be
ejected from the galaxy in the two simulations.}
\label{fig:Ysfr_yields}
\end{center}
\end{figure*}

In this section we compare our simulations I02 and  I04, which differ only
in the choice of SNII yields. Simulation I02 assumes the chemical yields
of WW95, while simulation I04 applies those of P98.
The main effect of varying the chemical yields is on the
metallicity evolution of the galaxies, in terms of total
metallicity and of the individual element ratios, as we describe below.

 Fig.~\ref{fig:Yofefeh_yields} shows
the stellar \& gaseous distribution functions of [O/Fe] and [Fe/H]
in runs I02 and I04, at the end of these simulations (2 Gyr).
In particular, we show the fraction of star \& gas particles per [O/Fe] and [Fe/H]
bin, as well as their distribution in this plane.
The main difference between the two runs comes from
a difference in the amounts of oxygen of the two simulations.
Simulation I04, that assumes the P98  yields,
predicts a much higher level of enrichment,
of the order of $0.4$ dex, compared to test I02, which is a consequence of a
higher value of the oxygen yield for stars less massive than $\sim 30$M$_\odot$ and of our choice of IMF upper limit ($40$M$_\odot$). 
This occurs both for the stars and for the
gas, the latter always presenting broader distributions (note the different
ranges shown in the plots).
The [Fe/H] distributions of the two tests are much
more similar than those found for the [O/Fe] ratio,
with differences of the order of $0.1-0.2$ dex.
In this case, the use of the P98 yields leads to a slightly lower
amount of iron, even though, at all masses, the yields
are higher in P98 compared to WW95. This results from
the different star formation/feedback levels of the two
simulations, as different levels of iron affect
the cooling and the star formation in the galaxy, therefore producing
non-trivial changes in the subsequent levels of feedback
and star formation activity.

This is clear from Fig.~\ref{fig:Ysfr_yields}, where we show
the star formation and SNII rates for tests I02 and I04,
together with
the evolution of the gaseous/stellar mass and metallicity (within the
inner 30 kpc).
Run I04 shows a lower maximum SFR, but higher SNII rates (recall
that the SNII rates are calculated as the ejected mass per time unit)
at the starburst.
Simulation I04 reaches a lower final stellar mass, and a slightly higher final
gas mass,  and exhibits a higher metallicity,
both in the gas and the stars, due to the higher chemical
yields.

Fig.~\ref{fig:Yelems_yields} shows the evolution of the ratio between the different
chemical elements and iron, for the stars and the gas in our simulations I02 and I04,
as well as the evolution of [Fe/H].
In the case of the stellar abundances,
the differences driven by the use of the two
sets of yields are significant, particularly for N and Mg,
with differences of the order of $\sim 1$ dex, and
at a lower level for Ne and O, with differences of about
$0.5$ dex. For the rest of the elements, we find differences of the order of
$0.25$ dex.
These differences are also partly driven by the differences
in the stellar [Fe/H] ratio (also shown in Fig.~\ref{fig:Yofefeh_yields}),
which is lower by about 0.2 dex when we
use the P98 yields.
In the case of the gas,  the abundances
show a stronger evolution, as the gas traces
the instantaneous chemical state of the simulated
galaxies. Note that there is a small amount of gas
left-over after the first starburst, and a more realistic
simulation that includes resupply
of gas from accretion would certainly change
the chemical abundances (see Section~\ref{sec:cosmo}).

\subsection{The effects of including AGB stars}  
\label{sec:AGBs}

Simulations that either ignore (I04) or include (I05)
the effects of stars in the AGB phase, that are otherwise
identical, are compared in this section.
As explained above, the typical time-scales of the release
of chemical elements by AGB stars is long; and we
have modeled AGB assuming three enrichment episodes per particle,
occurring at $\sim 10^8$ yr, $\sim 10^9$ yr and $\sim 8\times 10^9$ yr.
In order to see the effects of these three episodes, we
have run these simulations up to
10 Gyr although, as discussed previously,
we do not consider the resupply of gas that would occur
in real galaxies. Instead, these simulations allow us
to test and isolate the effects produced by the modeling
of this process in the code, before running more realistic
simulations in a cosmological context (see next section).

\begin{figure*}
\begin{center}
{\includegraphics[height=6cm]{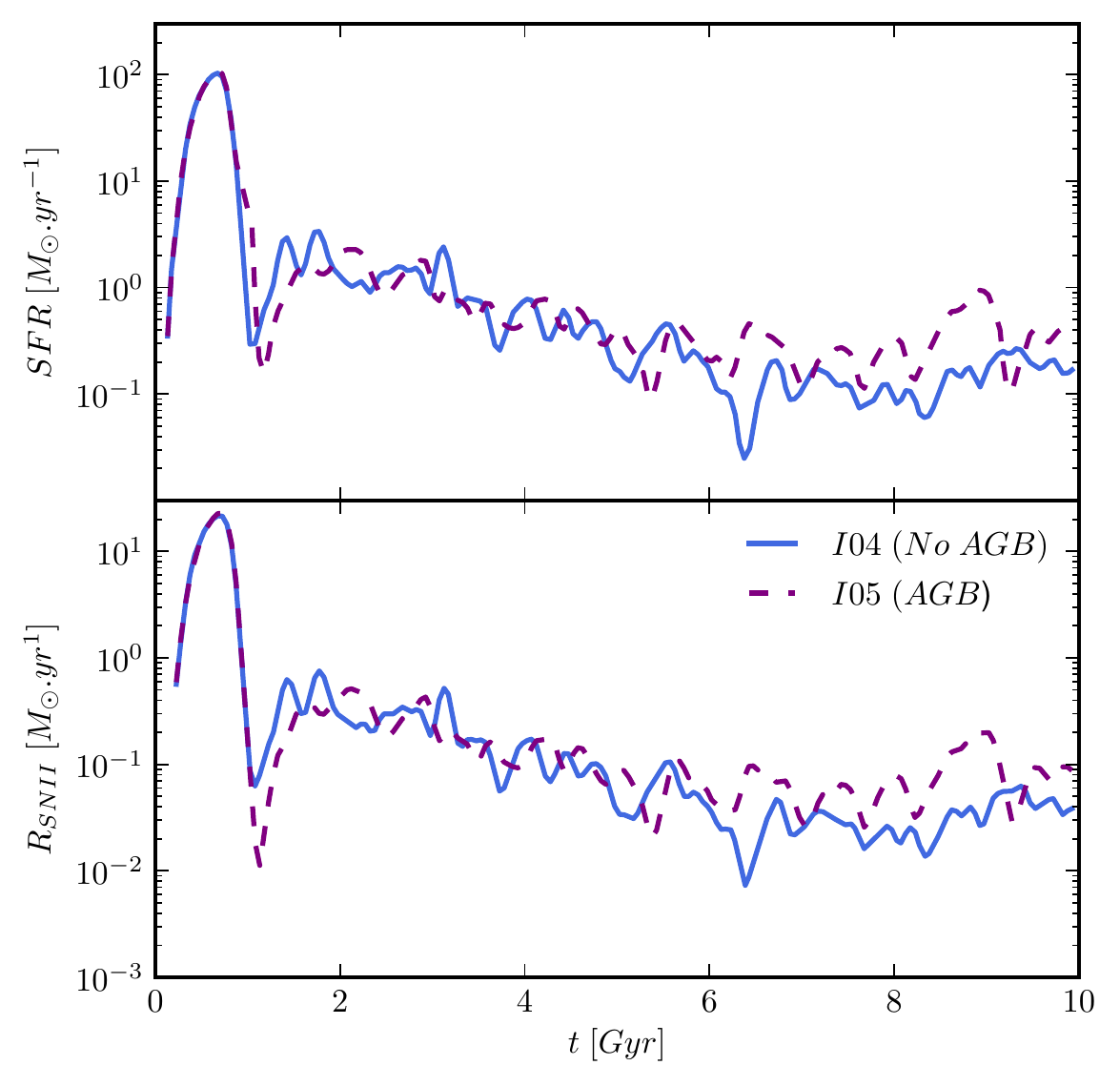}\includegraphics[height=6cm]{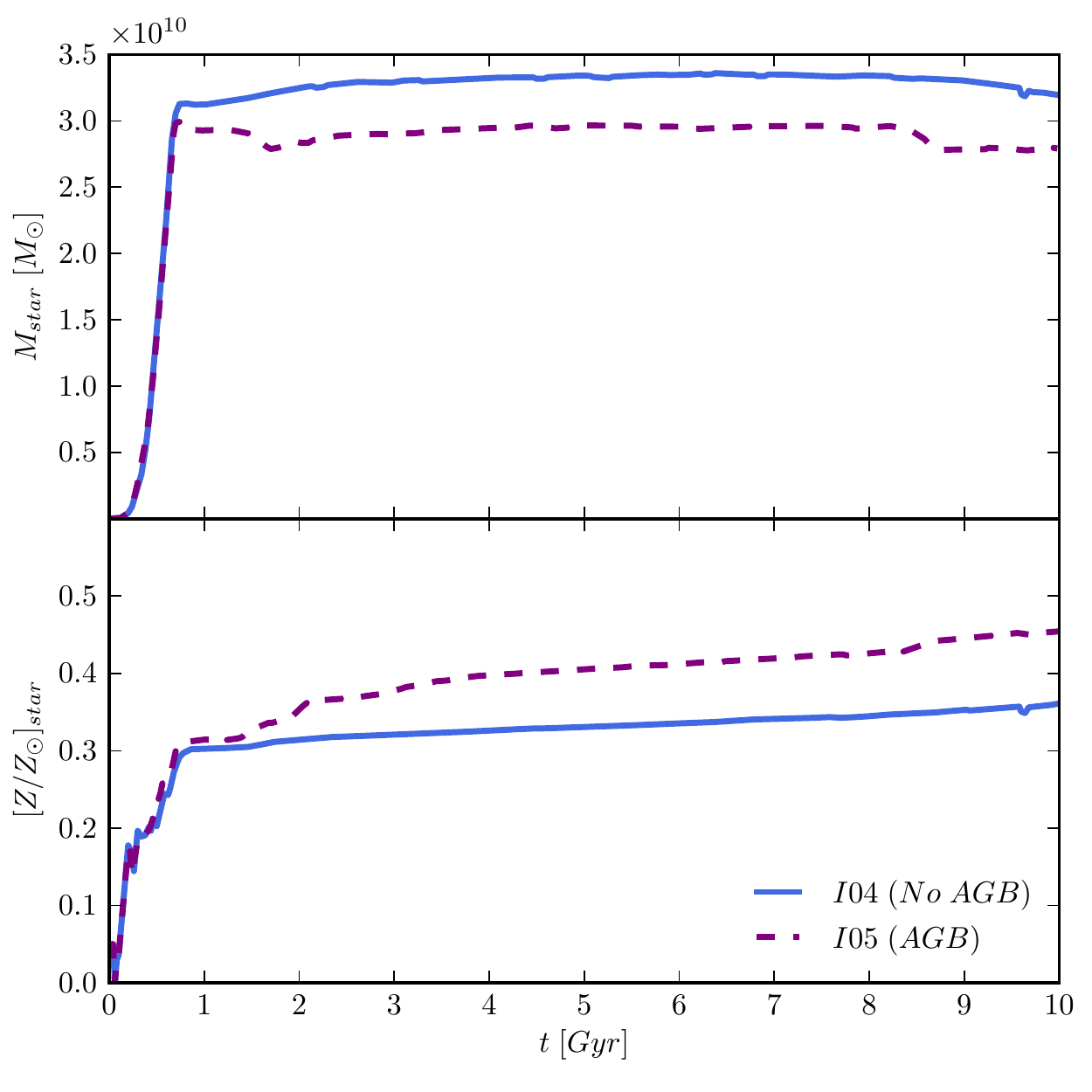}\includegraphics[height=6cm]{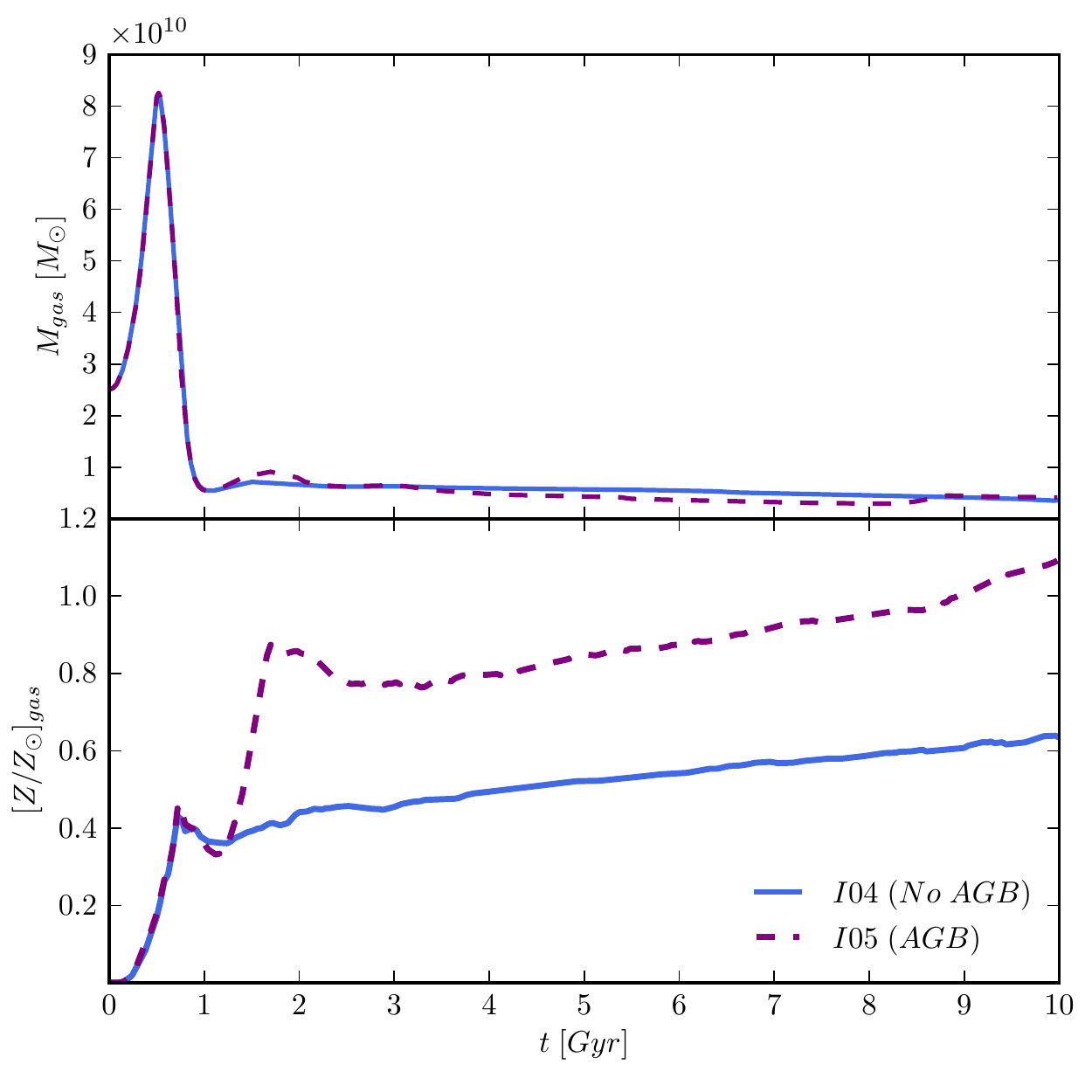}}
\caption{Star formation and SNII rates (left-hand panel), and the evolution of the mass
and metals in the stellar (middle-hand panel) and gaseous (right-hand panel) components in our test runs I04 and I05, which
are identical except for the inclusion of stars in the AGB phase.
The masses and metallicities have been calculated using particles
in the inner 30 kpc,
 in order to facilitate the comparison between the runs, particularly for
the gas component, as a different fraction of the gaseous mass will be
ejected from the galaxy in the two simulations.}
\label{fig:SFR_AGB}
\end{center}
\end{figure*}

As expected, the inclusion of AGB stars has a small effect
on the star formation and SNII rates, and a larger effect
on the metallicity distributions. This is clear from  Fig.~\ref{fig:SFR_AGB},
where we compare the star formation and SNII rates, as well as the evolution
of the gaseous and stellar mass and metallicity of these tests.
Note that AGB stars do not contribute feedback to the ISM,
and the only changes in SF/SN rates come from differences
in the metallicity distribution and corresponding cooling efficiency,
and from the different amounts of gas return.
The main effect of AGBs is to increase the overall metallicity
of the galaxies, both for the gas and for the stars.
At the end of the simulations, the run with AGB has twice the
gas metallicity of the run without AGBs, and also a higher
stellar metallicity, even though it has a lower  stellar mass. Note that
the gas metallicity in run I05 increases significantly right
after the starburst, and again, albeit more moderately, starting at around
1.5 Gyr, which corresponds to the first  and second AGB enrichment episodes
that follow the starburst at 0.5 Gyr.
The third AGB enrichment occurs at
$\sim 9$ Gyr, and also leads to an increase in the metallicity, although at a lower
level (see also next figure).
These changes are not so clearly seen in the stars, as the stellar
metallicity is the result of the cumulative effect of the
past star formation/enrichment history, which in these simulations
is dominated by the star-burst.

\begin{figure*}
\begin{center}
{\includegraphics[width=15cm]{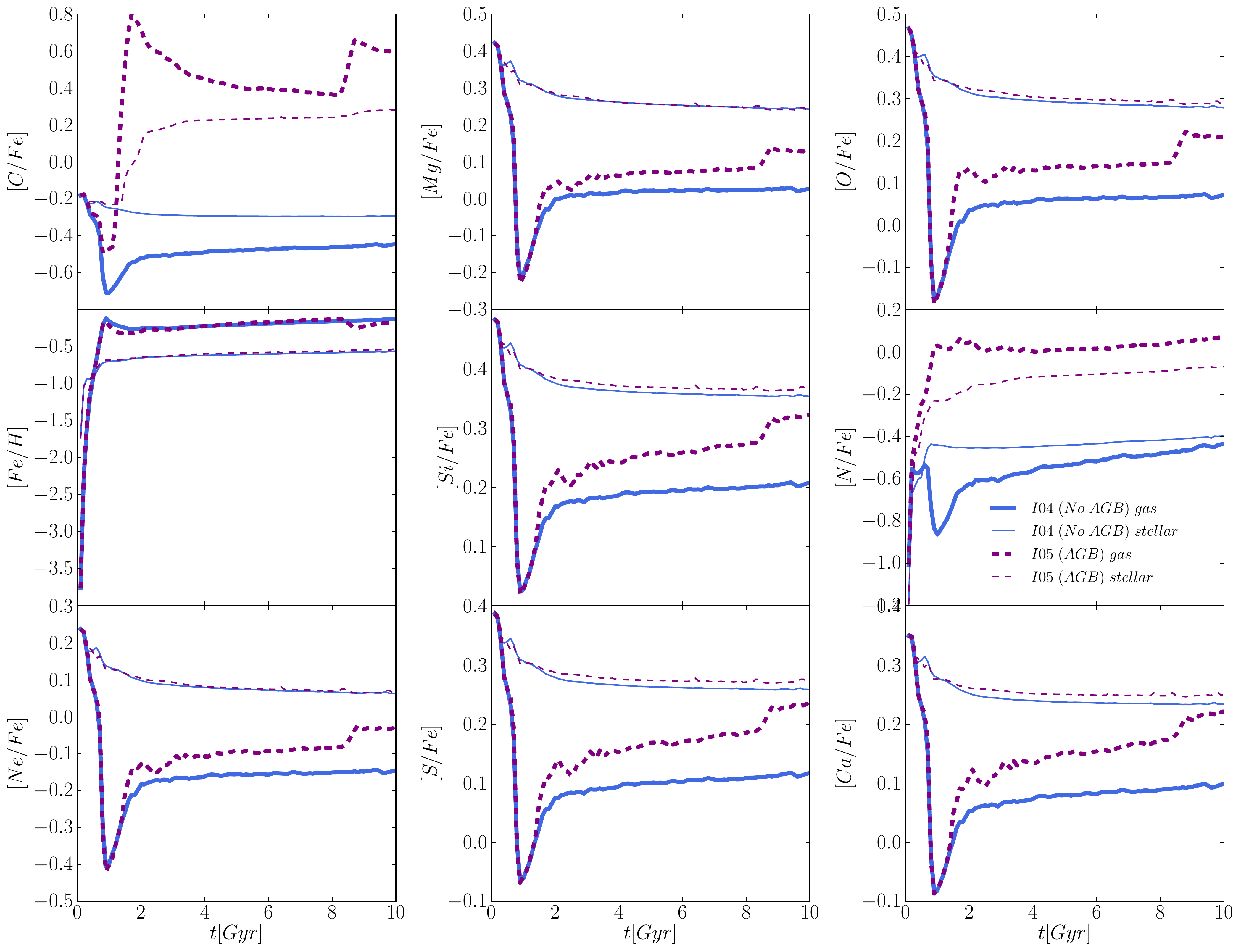}}
\caption{Comparison of the time evolution of stellar and gas abundance ratios of simulations I04 and I05 which differ in the inclusion or exclusion of AGB stars. }
\label{fig:Yelems_AGB}
\end{center}
\end{figure*}

Fig.~\ref{fig:Yelems_AGB} shows the evolution of the different
chemical elements with respect to iron, as well as the [Fe/H] ratio, for
stars (thin lines)
in our runs I04 and I05. As expected, significant differences are
found for the C and N  abundances. As AGB are main contributors
of these elements, the [C/Fe] and [N/Fe] stellar ratios are
higher by $0.6$ dex and $\sim 0.4$ dex respectively in I05 compared to I04.
It is important to note that, even though the release of elements
via AGB stars occurs in long time-scales, the prompt enrichment
phase produces the most important changes,
right after the first starburst. This can be different in a cosmological
context, particularly in galaxies with smoother SFRs.
Note also that, as we have seen in the previous figure,
our implementation of AGB enrichment in three episodes
can produce rapid increases in the
abundances.
This effect is exaggerated when looking at
an isolated galaxy simulation and could be reduced by
increasing the number of chemical-enrichment timesteps
at the expense of increasing the run-time of the simulation (see Appendix~\ref{sec:n_SNIa_episodes}).

Fig.~\ref{fig:Yelems_AGB} also shows the evolution of element ratios
 for the gas (thick lines) which,  unlike the stars,
 traces the instantaneous chemical properties of a galaxy.
In this case, we find more noticeable differences in all elements,
 and the three
enrichment episodes are clearly seen after
 $\sim 10^8$ yr, $\sim 10^9$ yr and $\sim 8\times 10^9$ yr of the star-burst.
In particular,
a significant change in the gas metallicities results from
the third enrichment episode at $\sim 8.5\times 10^9$.
 Note, however, that the changes seen in [Si/Fe], [S/Fe] and [Ca/Fe] are
not a result of changes in the abundances of these elements, as AGB stars do
not produce them, but rather the return of material locked away when the 
AGB stars formed (that is preferentially enriched in $\alpha$-elements from type II SNe)
into a relatively gas poor ISM. The primary element to be returned is hydrogen, and so one can
also see a reduction (dilution) of the [Fe/H] ratio.

\begin{figure*}
\begin{center}
{\includegraphics[width=8cm]{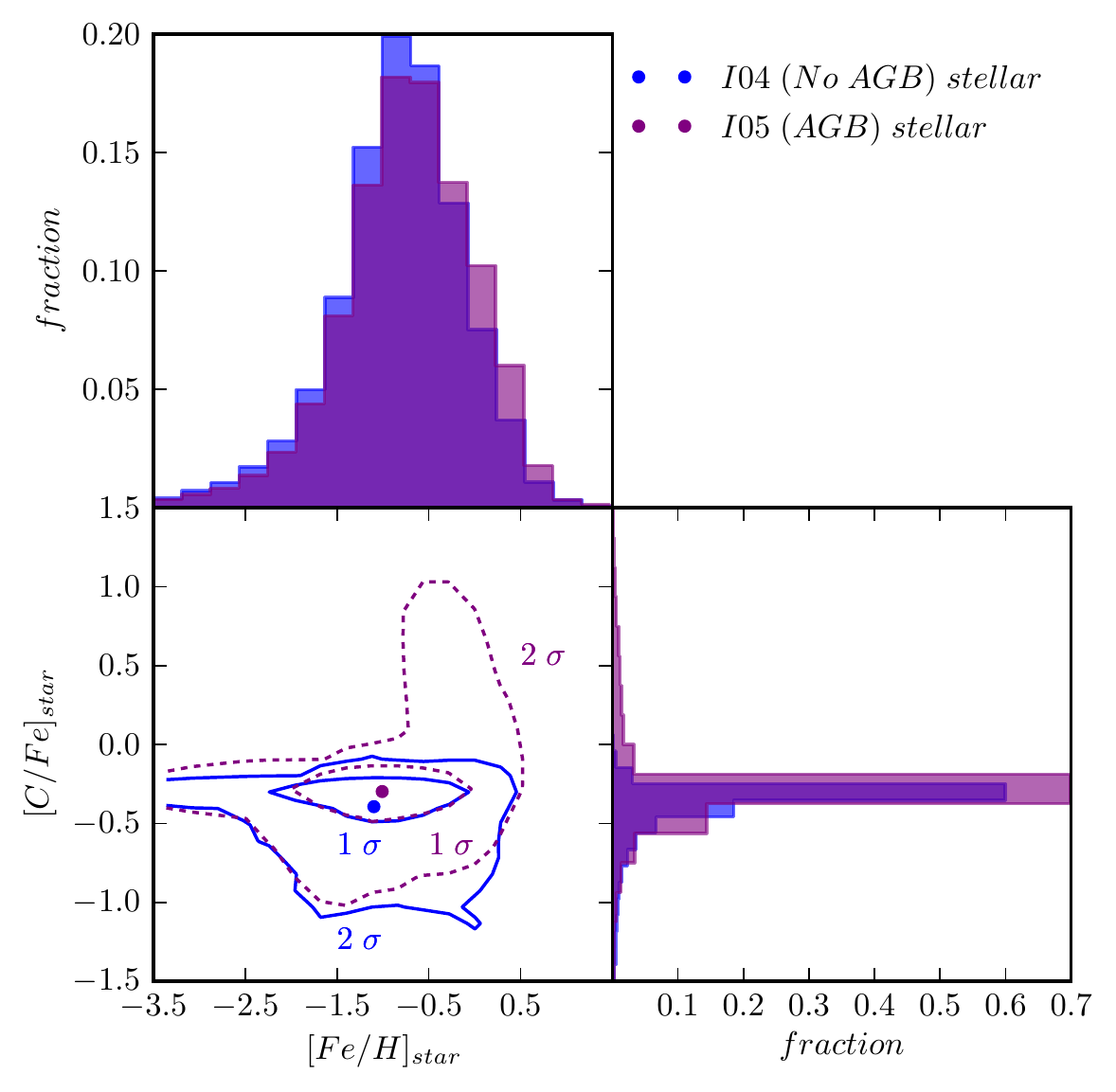}}
{\includegraphics[width=8cm]{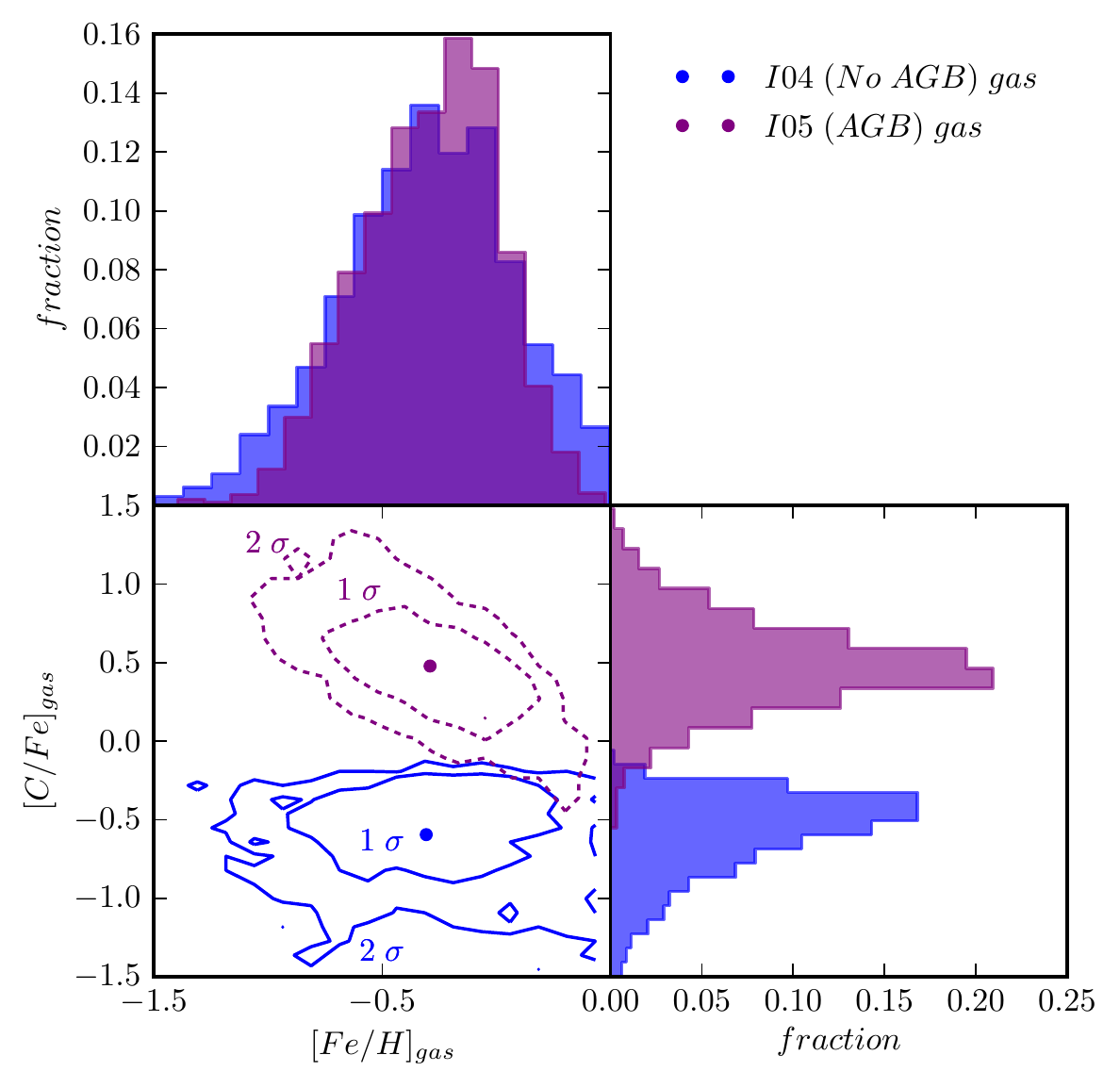}}
\caption{Stellar (left) and gaseous (right)
distribution functions of [C/Fe] and [Fe/H], and distributions
in the [C/Fe] vs [Fe/H] plane,  in simulations I04 and I05, at the
end of the simulation. }
\label{fig:Yofefeh_AGB}
\end{center}
\end{figure*}

Finally, we show in Fig.~\ref{fig:Yofefeh_AGB}
the stellar and gaseous distribution functions of [Fe/H] and
[C/Fe], as well as the position of particles in the [C/Fe]
vs [Fe/H] plane, for runs I04 and I05 at the end of
the simulations (10 Gyr).
In the case of the stars, we find very similar distributions
of [Fe/H] and [C/Fe], indicating that most of the stellar
mass has similar abundance ratios in both runs, although
there is a tail towards high [C/Fe] values in I05 that is absent in I04.
This tail originates from the
new stars that trace the instantaneous chemical state of the system.
These stars inherit the metallicities of the gas particles from
where they are formed, which are chemically enriched, as shown in
the right-hand panel of this figure. 
In fact, the
[C/Fe]  abundance ratios for the gas
particles in runs I04 and I05 are quite dissimilar, with
a much higher C abundance, of the order of 1 dex,  in the latter.
 Note that the uniform DTD assumed  ejects all its iron before 1 Gyr,
while carbon is ejected later on from AGB stars (after $\sim 1$ Gyr and $\sim 8$ Gyr of
the star formation event, in the second and third enrichment episodes)\footnote{We note that we have also studied the distribution of particles in the
[O/Fe] vs [Fe/H] plane, and found that, both for the gas and the stars, the distributions are very similar in the two runs. In this case, the most important difference is seen
in the oxygen gas abundances, that show a more peaky,
less broad distribution in run I05 compared to run I04. However, these differences are not significant.}. 
 Our results show that, after 10 Gyr of evolution, the gas is
much more chemically enriched with C as a result of stars that reach the AGB phase,
and demonstrates the significance of AGB stars in carbon production.
However, note that the
high [C/Fe] tail might be the result of considering a too idealized model for galaxy
formation, where the SFR has a dominant peak at early times in the galaxies evolution.

\subsection{The effects of varying the  Delay Time Distribution of SNIa}  
\label{sec:DTD}

\begin{figure*}
\begin{center}
{\includegraphics[width=8cm]{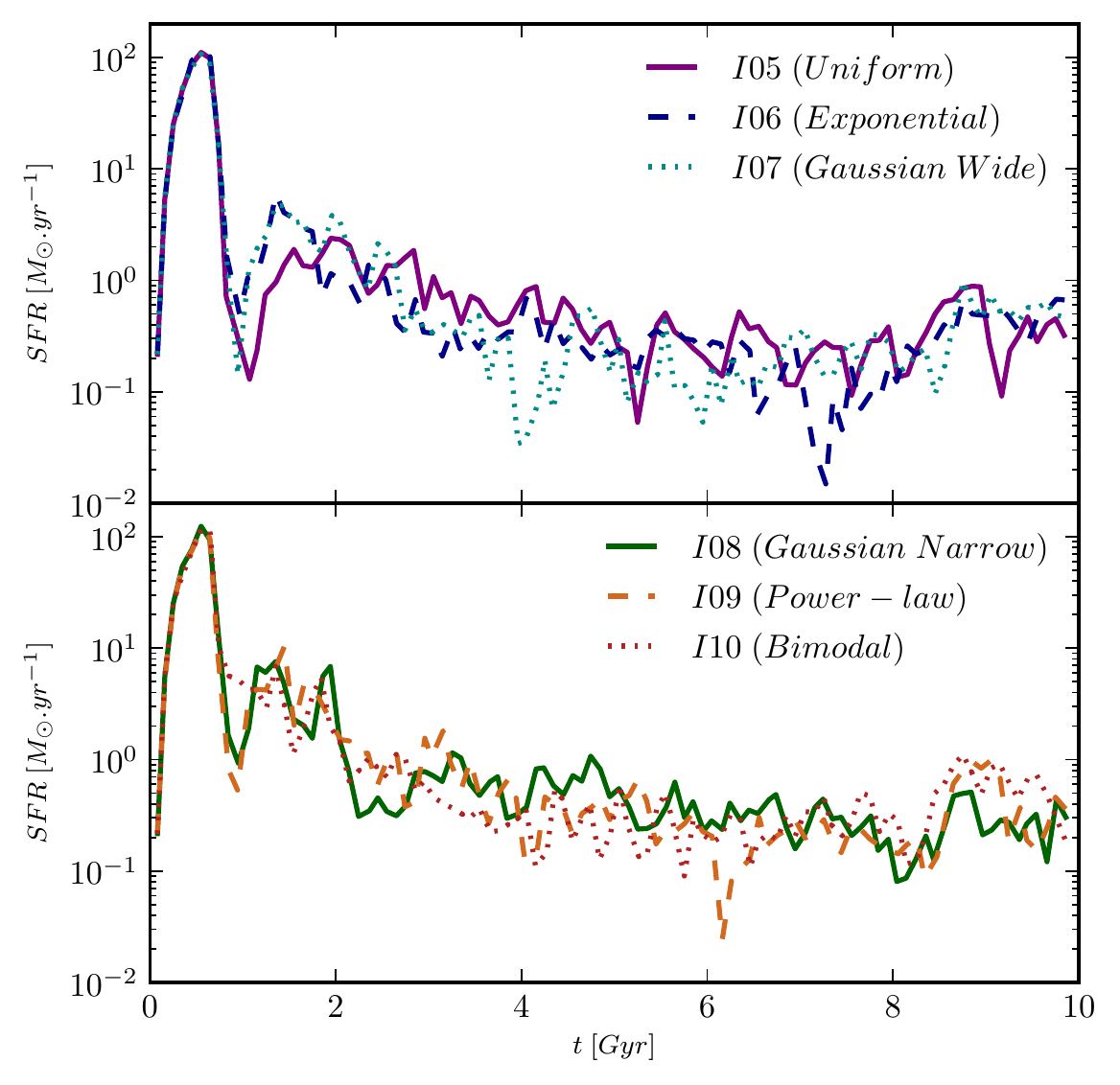}\includegraphics[width=8cm]{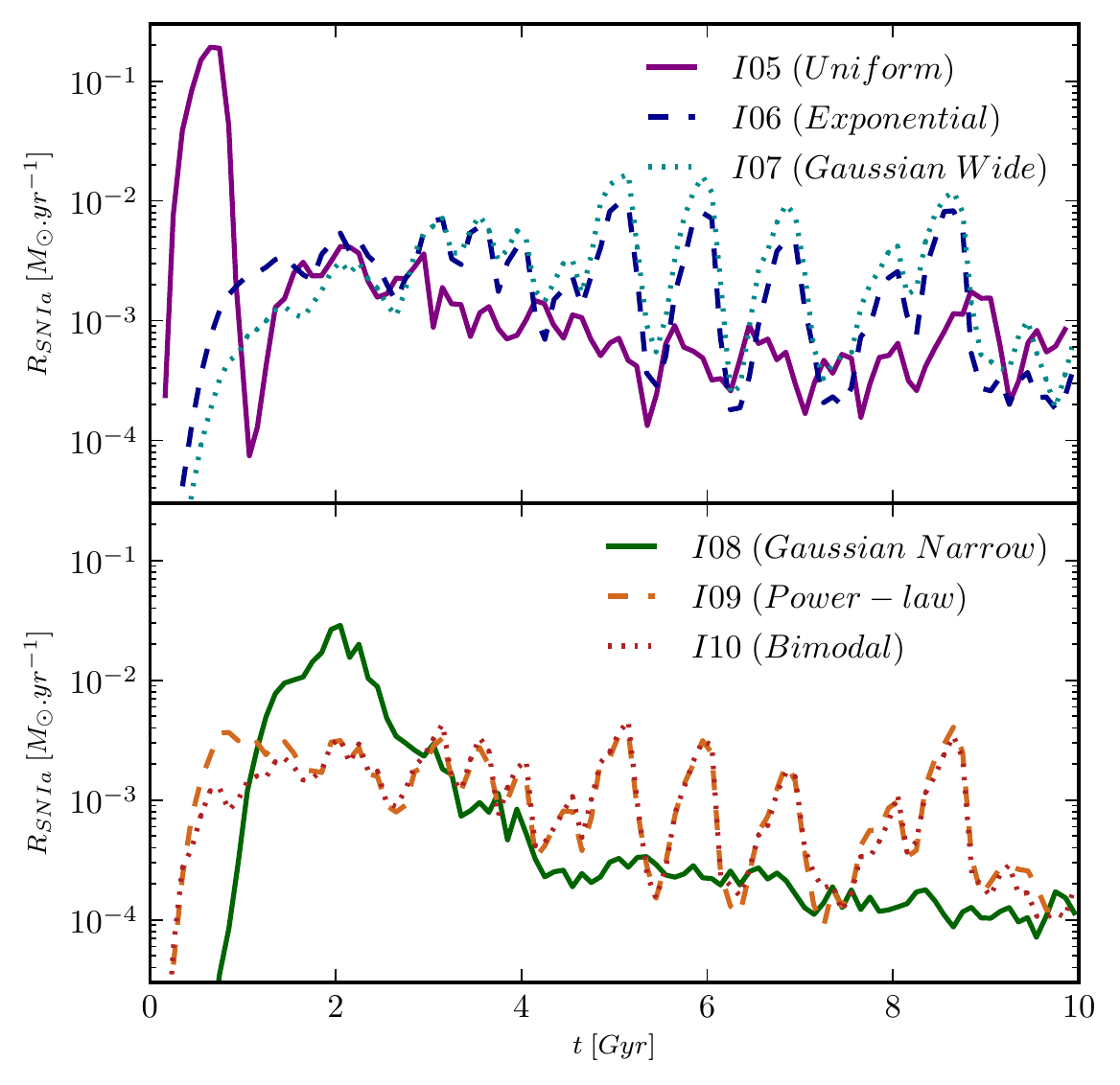}}
\caption{Star formation (left-hand panels) and SNIa rates (right-hand panels)
for tests I05-I10, which assume various DTDs for SNIa.}
\label{fig:SFRdtd}
\end{center}
\end{figure*}

\begin{figure*}
\begin{center}
{\includegraphics[width=8cm]{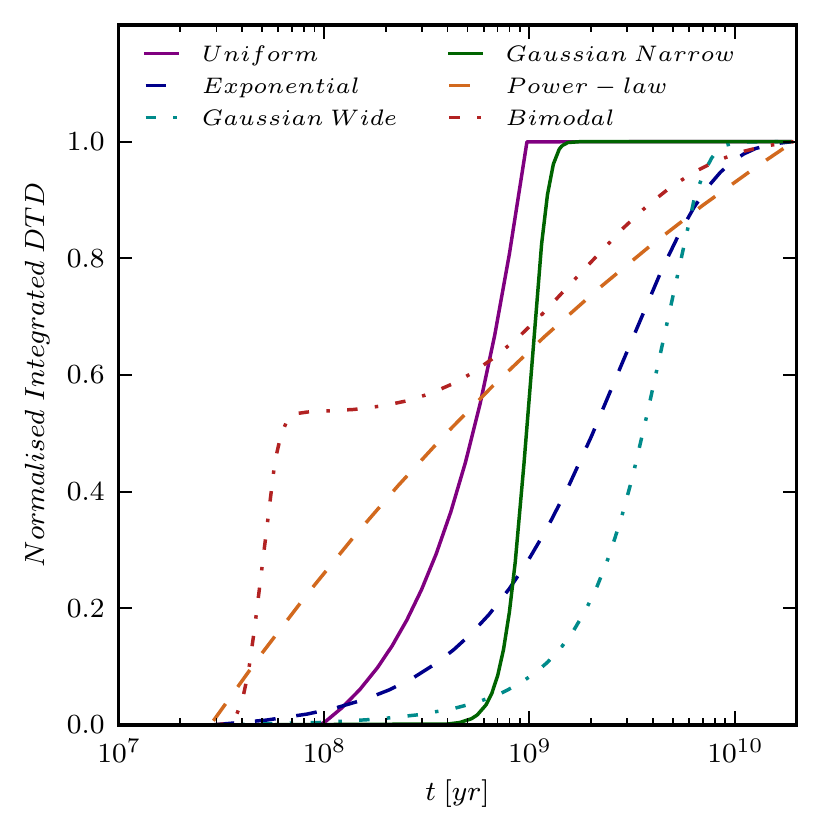}\includegraphics[width=8cm]{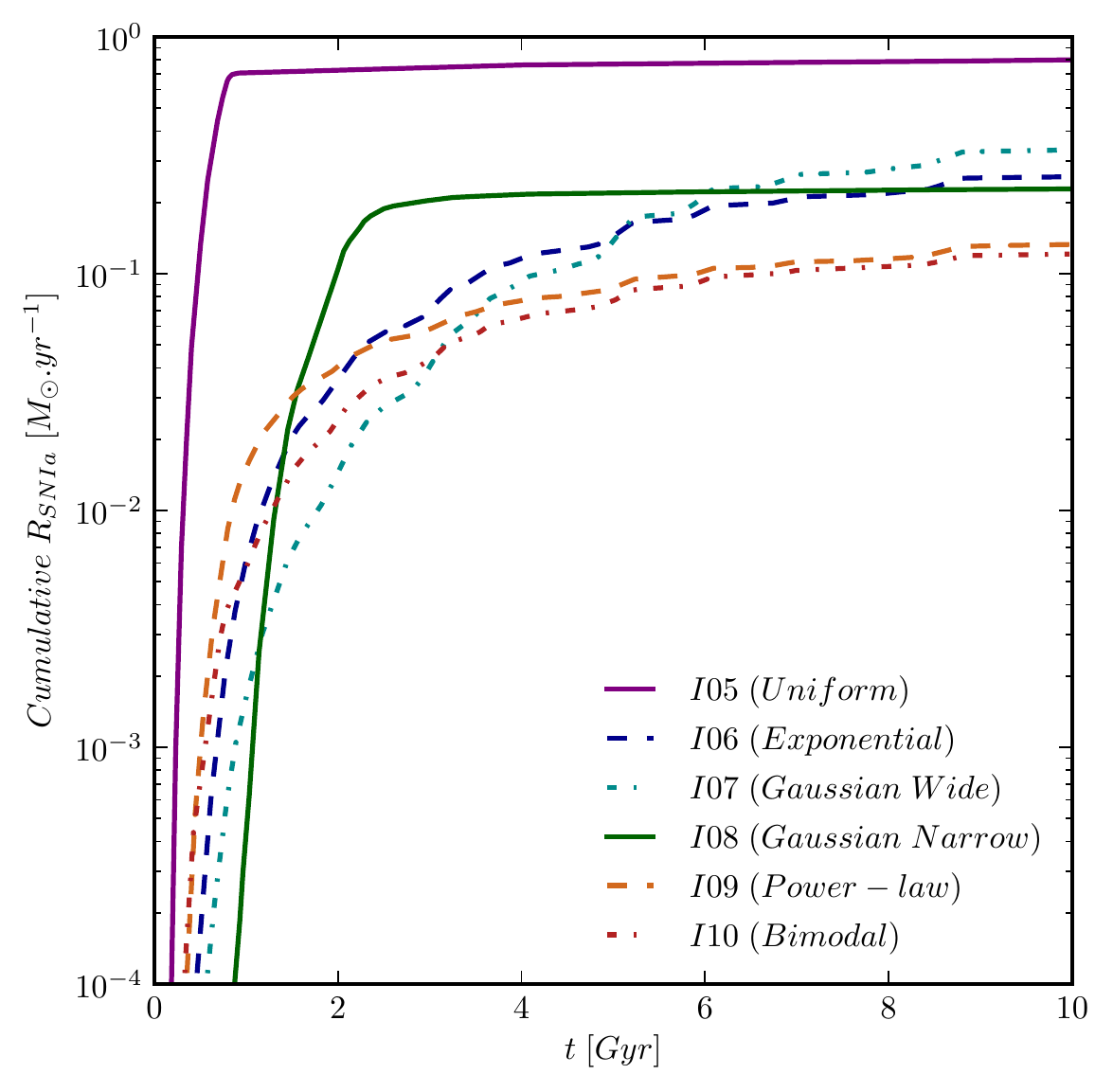}}
\caption{Integrated DTDs (left-hand panel) and cumulative SNIa rates for runs I05-I10 which assume various DTDs for SNIa.}
\label{fig:Dtdsni}
\end{center}
\end{figure*}

In this section we compare the results of our runs I05-I10, which
differ in the choice of DTD for SNIa (Table~\ref{table:isolated}).
(Note that these have run up to 10 Gyr as SNIa
typical time-scales can be this long.)
Fig.~\ref{fig:SFRdtd} shows the star formation (left-hand panels)
 and SNIa  (right-hand panels) rates for these tests.
As most of the feedback comes from SNII, the
SFR (and therefore the SNII rate) is not severely affected by the choice of
different SNIa
DTDs; note particularly the excellent agreement of all runs during  the star-burst.
As expected, more significant differences are found for the SNIa
rates. In particular, 
the use of  a fixed SNIa/SNII rate ratio (run I05)
naturally leads to a SNIa rate which roughly follows the SFR;
and a somewhat similar behavior is found for run I08 (narrow Gaussian
DTD), where the SNIa rate shows a peak at about
1 Gyr after the star-burst, followed by a smooth decline (see also
Fig.~\ref{fig:DTD})\footnote{Note that, due to the values assumed for the
normalization of the DTDs, the number of SNIa of the tests is different,
particularly in the case of a uniform DTD (see also Fig.~\ref{fig:Dtdsni}).}.
In contrast, the SNIa rates  of the remaining tests
show a much smoother increase at early times, followed
in general by a bursty behavior. The bursts happen in all the
tests at similar times, most notably around 4-9 Gyr.

As discussed in appendix~\ref{sec:n_SNIa_episodes}, this appears to be due to interplay between 
star formation timescales and the time discretisation of the SNIa and is likely a numerical artefact, that is 
exacerbated for the more extended DTDs. 
Furthermore, we find a very similar behavior of runs
I06 (exponential DTD) and  I07 (wide Gaussian DTD), and
of  I09 (power-law DTD) and I10 (bimodal DTD), which can be explained by their similar
DTDs at early and late times (Fig.~\ref{fig:DTD}).

The characteristics of the different SNIa rates when we assume
various DTDs can also be understood from
Fig.~\ref{fig:Dtdsni}, where we show the integrated DTDs and the cumulative
SNIa rates for the different tests. 
Runs I05 (uniform DTD) and I08 (narrow
Gaussian DTD) are the ones where the SNIa events appear earlier,
with most of them having occurred after $\sim 1-2$ Gyr after the star-burst.
In contrast, in the remaining tests only
after $\sim 10$ Gyr the majority of SNIa events coming from the first star-burst have
happened. Runs I06 (exponential DTD) and I07 (wide Gaussian DTD) appear
as the ones with larger delay between the formation of the stars and
the SNIa events, while runs I09 (power-law DTD) and I10 (bimodal DTD)
appear as intermediate cases, due to the presence of an early component in the corresponding DTDs. The similarities/differences in the DTDs then explain the behavior
seen in the previous figure.

\begin{figure*}
\begin{center}
{\includegraphics[width=17cm]{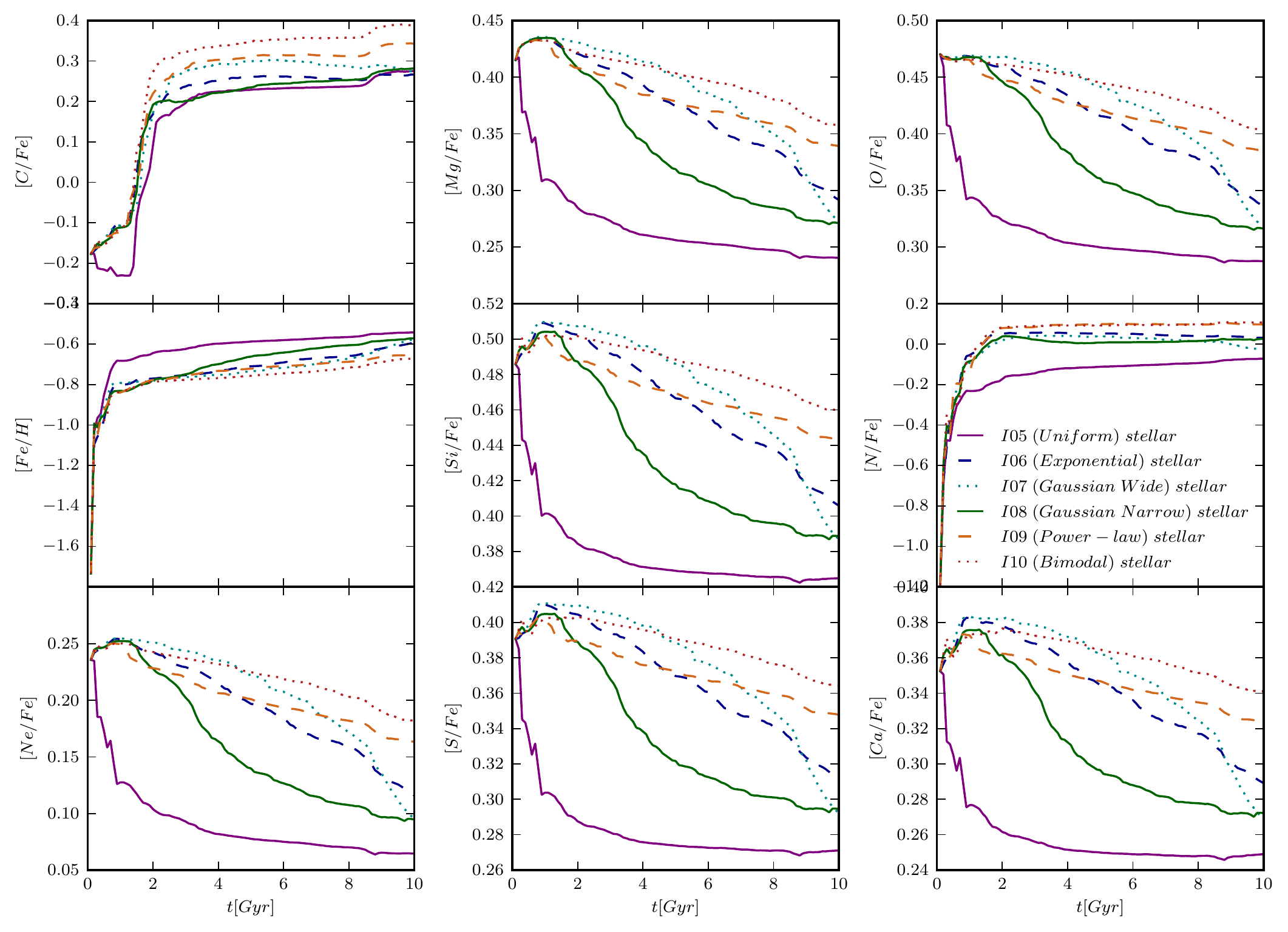}}
\caption{The evolution of the abundance ratios for the
stars in our simulations I05-I10, which assume various
DTDs for SNIa.
}
\label{fig:dtdmetals}
\end{center}
\end{figure*}

As a result of the different characteristic time-scales of the
release of chemical elements produced by SNIa (and the SNIa rate normalization),
we detect differences in the evolution of the stellar
abundances when different DTDs are assumed, as shown in Fig.~\ref{fig:dtdmetals},
although these are moderate compared to the differences when changing SNII yields.
In general, the variations in abundance ratios of the different runs
is of the order of $0.15$ dex for all elements (note the different scales
plotted in the y-axis for the different elements).
In general, the variations in abundance ratios increase with time, but note that this
might be, at some level,  an artifact of the idealized ICs.
Run I05 (uniform DTD) has the highest iron abundance and the lowest
 [X/Fe] for all other elements X, due to the higher
SNIa rates (which is due to the higher DTD normalisation adopted); 
while all other tests
show the opposite behavior, except for
I08 (narrow Gaussian) which lies at an intermediate position.

\begin{figure*}
\begin{center}
{\includegraphics[width=8cm]{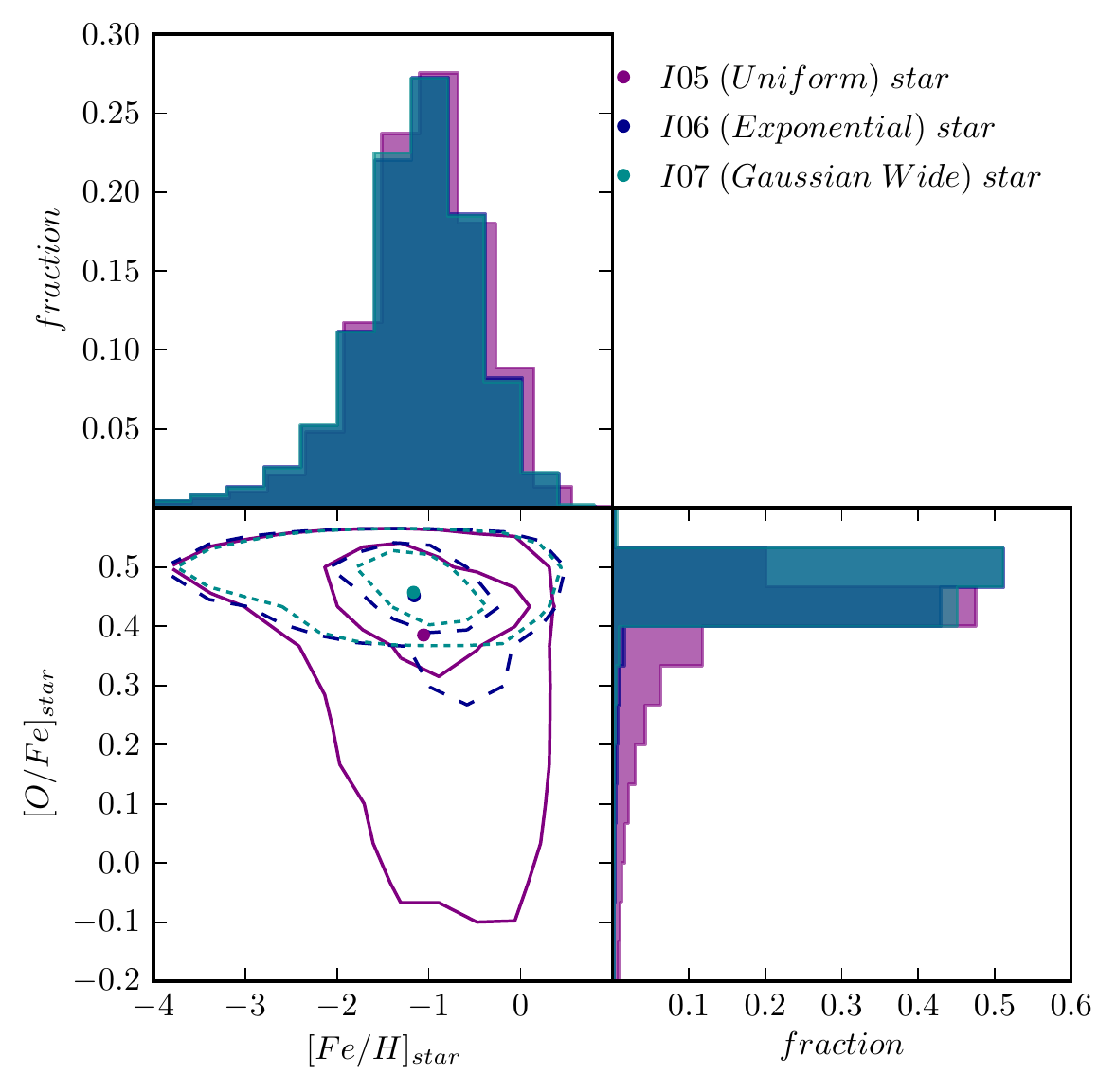}}
{\includegraphics[width=8cm]{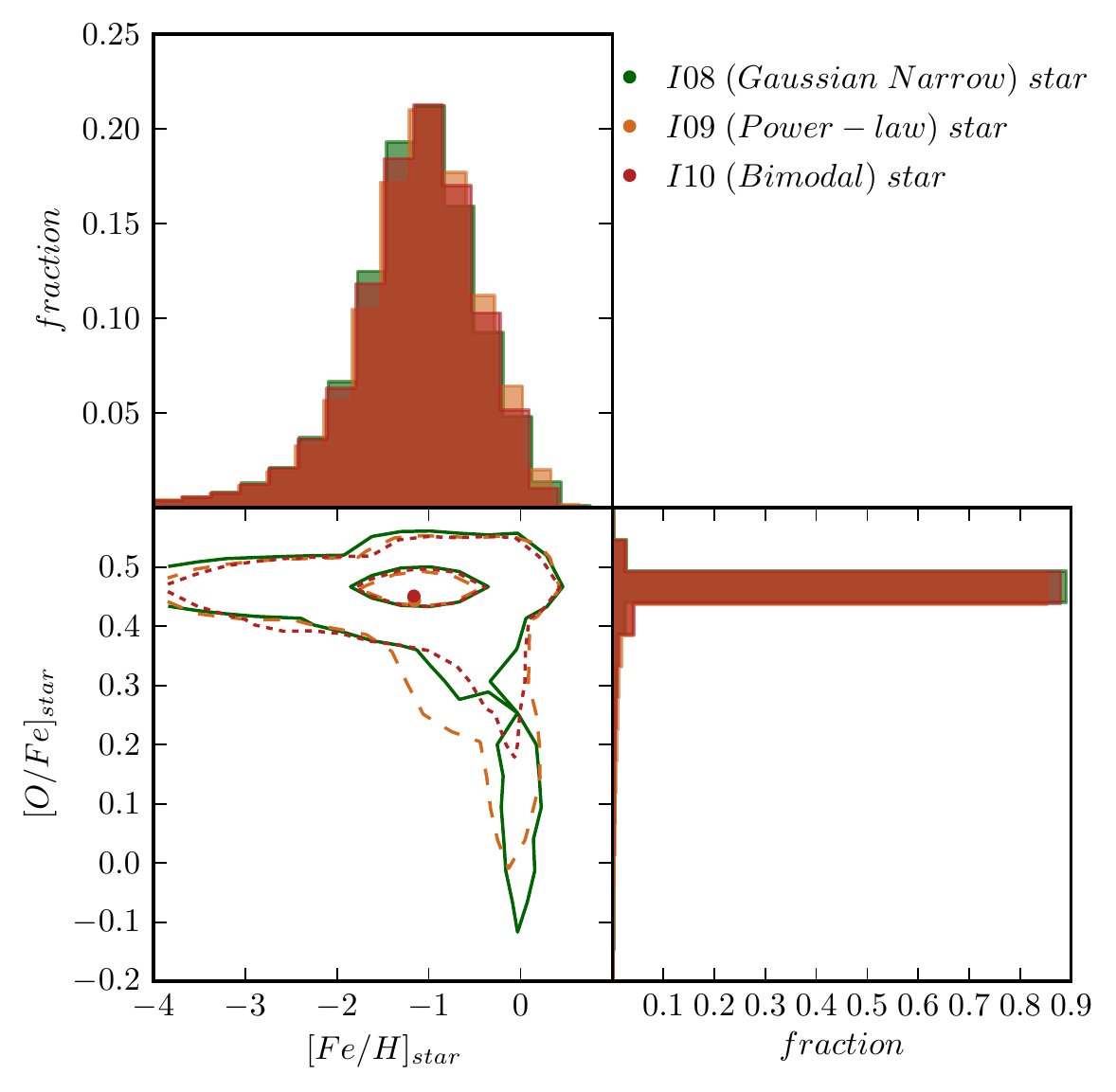}}
\caption{Stellar
distribution functions of [O/Fe] and [Fe/H], and distributions
in the [O/Fe] vs [Fe/H] plane,  for  simulations I05-I10 which assume
different DTDs for SNIa. }
\label{fig:Yofefeh_DTD}
\end{center}
\end{figure*}

Fig.~\ref{fig:Yofefeh_DTD} shows the stellar  distribution functions
of [O/Fe] and [Fe/H] for runs I05-I10.
The use of various DTDs produces similar
distributions of [Fe/H], which in all cases peak at [Fe/H]$\sim -1$.
In general the [Fe/H] distributions exhibit a long tail to low
[Fe/H] (a `G-dwarf problem', \citealp{Tinsley_1980}) which is typical
for simulations without gas inflow \citep{Edmunds_1990,
Creasey_2015}. 
The [O/Fe] distributions for the different runs
are more dissimilar, and although they all peak at a similar
value of [O/Fe]$\sim 0.45$, we find that a uniform
DTD predicts a much broader [O/Fe] distribution, extending to lower
oxygen abundances.
 The rest of the simulations show narrow [O/Fe] distributions,
and only small differences at the highest metallicities.
Note that the differences between I05 and the other tests comes
not only from the DTD itself, but from the different normalization
which produces a significantly higher number of SNIa events.

 \cite{Matteucci_2009} (M09 hereafter) studied the effects of assuming different
DTDs on the galactic chemical evolution, focusing on DTDs with various
fractions of prompt SNIa components. They found that models with
a non-negligible prompt component produce narrower [Fe/H] distributions
compared to models with no prompt component, and the [O/Fe] distributions
in better agreement with observational results.
These results seem in line with our findings, although we note that our
simulations and the galactic chemical evolution models differ in various
aspects. First, we include in our distribution functions and SFRs
 all stars in the galaxy (i.e. disk and bulge),
unlike  M09 who focus on the disk stars in the solar vicinity. 
Second, and perhaps more important, is the fact that 
the SFRs of the two studies  are different in terms of the relative
importance of the early and late SF levels, which will affect
the distributions of chemical abundances in the galaxies.

\begin{figure*}
\begin{center}
{\includegraphics[height=9cm]{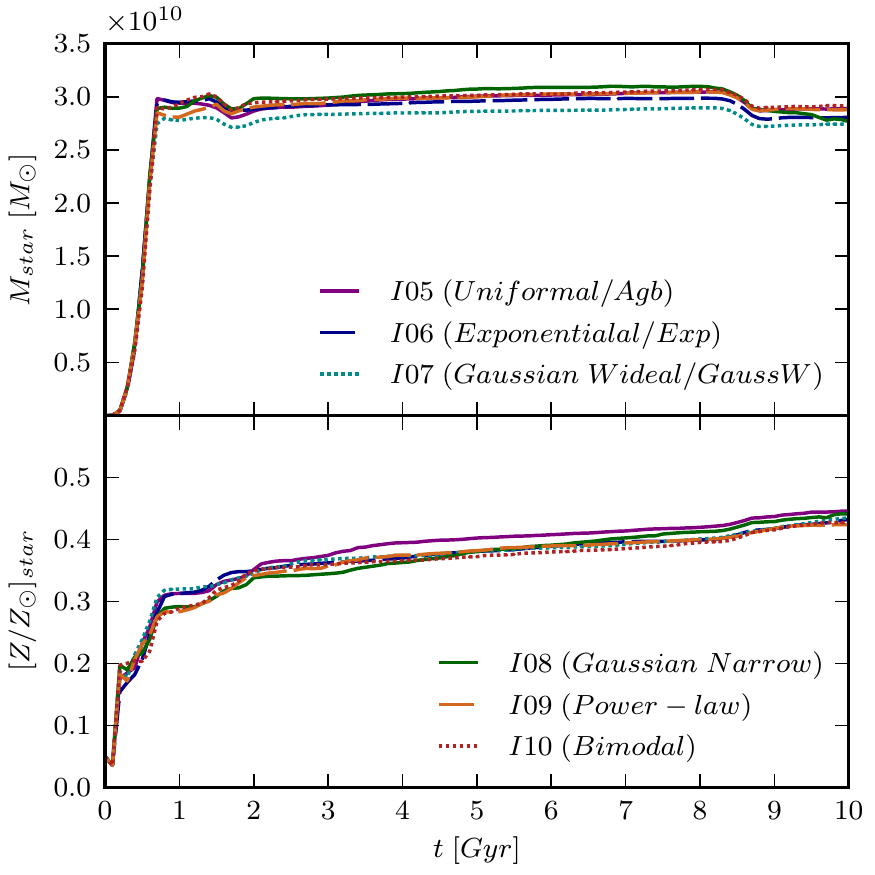}\includegraphics[height=9cm]{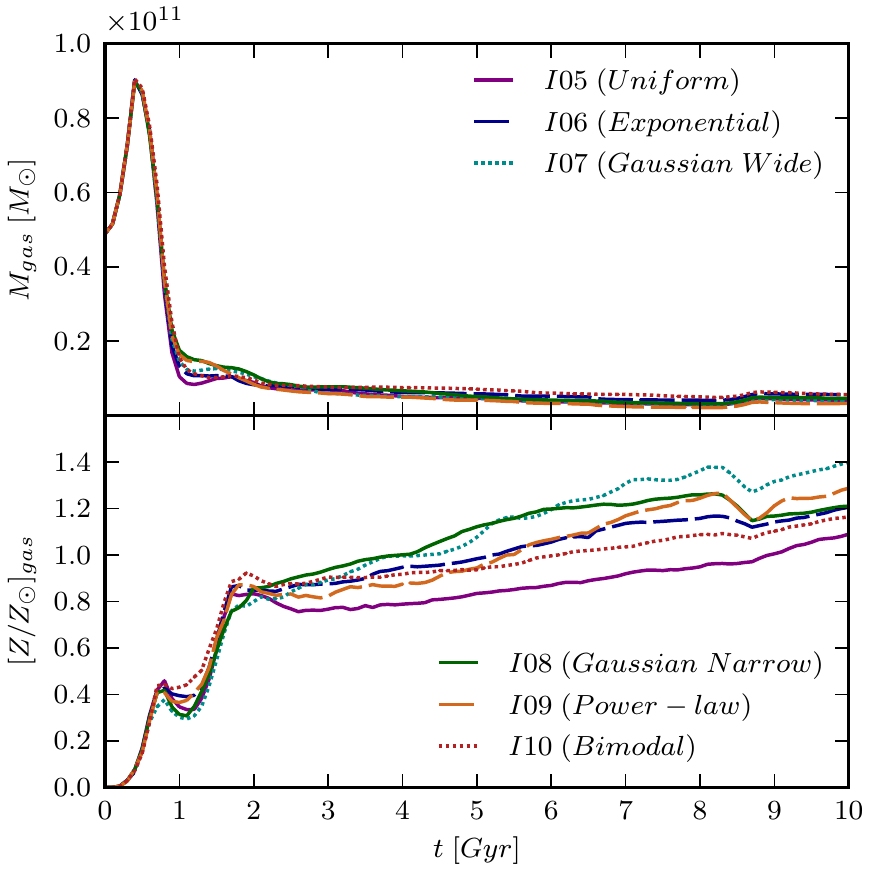}}
 \caption{The evolution of the mass and metals in the stellar (left-hand panel) and gaseous (right-hand panel) components in our test runs I05-I10, which are identical except for the DTD assumed. The masses and metallicities of the gas have been calculated using particles in the inner 30 kpc, in order to facilitate the comparison between the runs (note that a different fraction of the gaseous mass could be ejected from the galaxy in the different simulations).}
\label{fig:SFR_DTD}
\end{center}
\end{figure*}

As evident from the previous plots, the runs assuming various
DTDs for SNIa are similar, in terms of stellar/gas masses
and metallicities, as seen also from Fig.~\ref{fig:SFR_DTD}.
All runs produce a similar amount
of stellar mass and show a similar metallicity evolution
of the stars. More
notable differences are found for the gas metallicities,
with variations of the order of $0.2$ dex  at the end of the simulations. Note, however, that the lack of accretion of gas might affect
our results, particularly in the long-term. We investigate
the effects of varying the DTD in a more realistic, cosmological
set-up in the next section.

\subsection{The effects of varying the cooling}\label{sec:cool}

\begin{figure}
\begin{center}
\includegraphics[width=\linewidth]{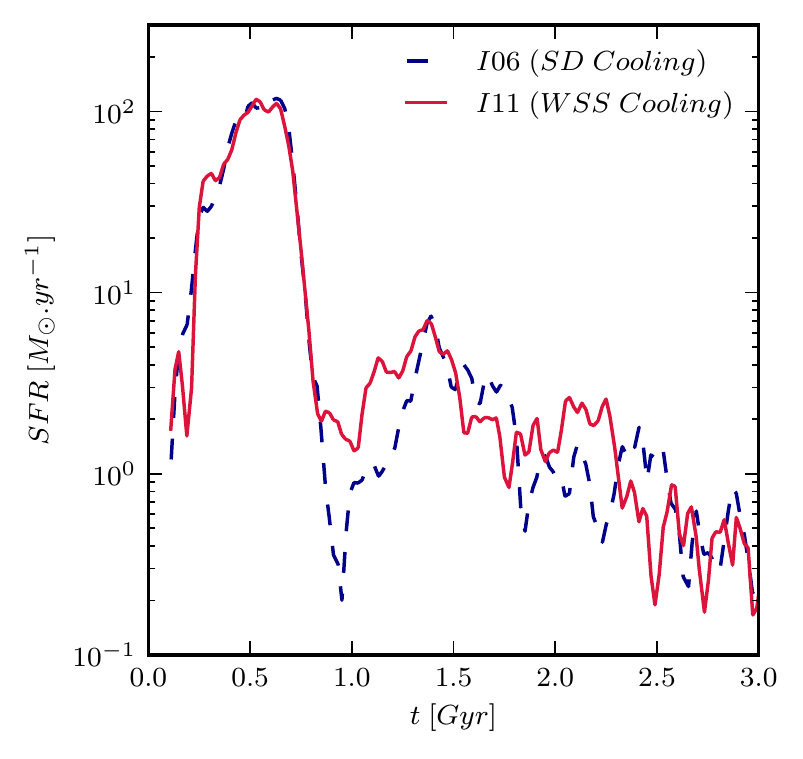}
\caption{ A comparison of the star formation rates of our runs I06 and I11 which differ only in the cooling function assumed.}
\label{fig:sfrcool}
\end{center}
\end{figure}

As explained in the previous Section, we have updated the cooling 
routines of our standard model, that previously used
the metallicity-dependent cooling functions of \cite{SD93} that assume 
gas in collisional ionization equilibrium and are tabulated as a function
of the [Fe/H] content. The new cooling tables are instead
based on the method described in \cite{W09}, which provide cooling rates
sensitive to the relative abundances of the 12 studied elements
and consider photo-ionisation due to the UV background.
In the case of a non-cosmological simulation, 
such as the case of the
idealized runs of this section, however, differences produced by
changing the cooling are not expected to be significant.
In fact, we find similar star formation/supernova
rates and  metallicity content for  runs
I06 and I11 (see Table~\ref{table:isolated}) that only differ in the cooling functions
adopted. Additionally, the phase-space of gas particles is similar in the
two runs. As an example, we show in Fig.~\ref{fig:sfrcool} a comparison of
the SFRs of these two tests, and further discuss the effects of varying the
cooling in the next section where we present results for cosmological simulations.


\section{Cosmological simulations}  
\label{sec:cosmo}

In this section we describe a series of zoom-in simulations of the formation of a halo in full cosmological context, and explore the effects of varying the DTD for SNIa -- a distribution that is poorly constrained by current observations (see e.g. \citealt{Bonaparte13,Hilldebrandt13}) -- on the chemical properties of the gas and stars, as well as the effects of varying the cooling function. 
All tests of this Section include  AGB stars, assume a Chabrier IMF and adopt the P98 life-times and chemical yields for SNII. We have also run a simulation with no AGB stars, a Salpeter IMF and WW95 yields (referred to as the 'standard model') which allows comparison with previous works which used the S05 model. Table~\ref{tab:cosmo} summarizes the characteristics of our cosmological simulations.

\begin{table*}
\caption{List of the cosmological simulations used for this work.}
	\begin{small}
		\begin{center}
	\begin{tabular}{lccccccc}
	\hline
	\hline
	Name      & IMF$^1$    & SNII yields$^2$ & SNIa rates      & AGB & Cooling$^3$ \\\hline
	AqC1$^4$  & S          & WW95            & uniform         & no & SD93  \\
	AqC2      & C          & P98             & uniform         & yes & SD93  \\
	AqC3      & C          & P98             & exponential     & yes & SD93  \\
	AqC4      & C          & P98             & wide Gaussian   & yes & SD93  \\
	AqC5      & C          & P98             & narrow Gaussian & yes & SD93  \\
	AqC6      & C          & P98             & power-law       & yes & SD93  \\
	AqC7      & C          & P98             & bimodal         & yes & SD93  \\
        AqC8      & C          & P98             & exponential     & yes & W09  \\
	\hline
	\end{tabular}
	\end{center}
	\end{small}
	\label{tab:cosmo}
{{\sc notes:}\\
$^1$ S and C are abbreviations for the Salpeter and Chabrier IMFs.\\
$^2$ WW95 and P98 stand for \cite{WW95} and \cite{P98}, respectively.\\
$^3$ SD93 and W09 refer respectively to the use of the \cite{SD93} and \cite{W09} cooling tables.\\
$^4$ This simulation will be referred to as 'standard'.}
\end{table*}

The ICs used in this section are those of halo C (AqC for short) of the Aquarius Project  \citep{Springel08}, in its hydrodynamical version \citep[see][]{S09,S12}. The ICs use the zoom-in technique, which allows to describe the formation of a galaxy and its surroundings with very high resolution, at the same time allowing the description of the matter distribution at larger scales. AqC has  been selected from the parent dark-matter only simulation with the  conditions to have a similar mass to the Milky Way and to be mildly isolated at the present time (no neighbor exceeding half its mass within 1.4 Mpc). The virial mass of AqC at $z=0$ is $1.2\times 10^{12}$M$_\odot$, and its virial radius  167 kpc (in our standard run, these values are  $1.3\times 10^{12}$M$_\odot$ and 173 kpc).

The ICs of AqC use a $\Lambda$CDM cosmology with $\Omega_\Lambda=0.75$, $\Omega_{\rm m}=0.25$, $\Omega_{\rm b}=0.04$, $\sigma_8=0.9$ and $H_0=73$ km s$^{-1}$ Mpc$^{-1}$ consistent with the WMAP-1 cosmology \citep{Spergel_2003}.   The mass resolution is $2.2 \times 10^6$ M$_\odot$ and $4\times 10^5$ M$_\odot$, for dark matter and gas particles, respectively, and  we have used a gravitational softening of $0.7$ kpc, which is the same for gas, stars and dark matter particles. The AqC halo has been extensively analyzed using various simulations codes and set-ups, and has been used in the Aquila code comparison Project \citep{S12}. For details on its formation history, disc/bulge evolution, numerical effects and resolution, in particular using the standard \cite{S05,S06} code, we refer the reader to \cite{S09,S10,S11}.

We emphasize that these simulations have been designed to test the effects  of assuming different DTDs on the chemical properties of the galaxies. For this reason, we
have not `tuned' these models, choosing instead to keep
 the input parameters as similar as possible
to those used in previous studies with this code, to allow for direct comparison. 
Our run AqC1 is then fixed to the original prescription (we set the SNIa to SNII rate
to 0.0015 as in \citealt{S09}), while the rest of the runs assume a fixed
normalization
for the DTD (Section~\ref{sec:rates}). We note that differences between AqC1 and the other runs are expected
(see next sections),
particularly due to the change of IMF and the differences in the SNIa assumptions.
In order to keep the same $z=0$
cosmological stellar density and SNIa density, a different normalization of the
various DTDs would be required.  A detailed analysis of input parameters, as well
as comparison with available observations\footnote{Note that  in order
to select the best input parameters of the model, a detailed comparison with observations is needed. This is a complex task, as different results can be obtained depending on the way used to calculate the observables, as shown, e.g.  in \cite{S10} and \cite{Guidi15,Guidi16}.},  are out of the scope of
this paper and will be presented elsewhere.

\subsection{Star formation and SN rates} 

\begin{figure*}
	\begin{center}
	{\includegraphics[width=8cm]{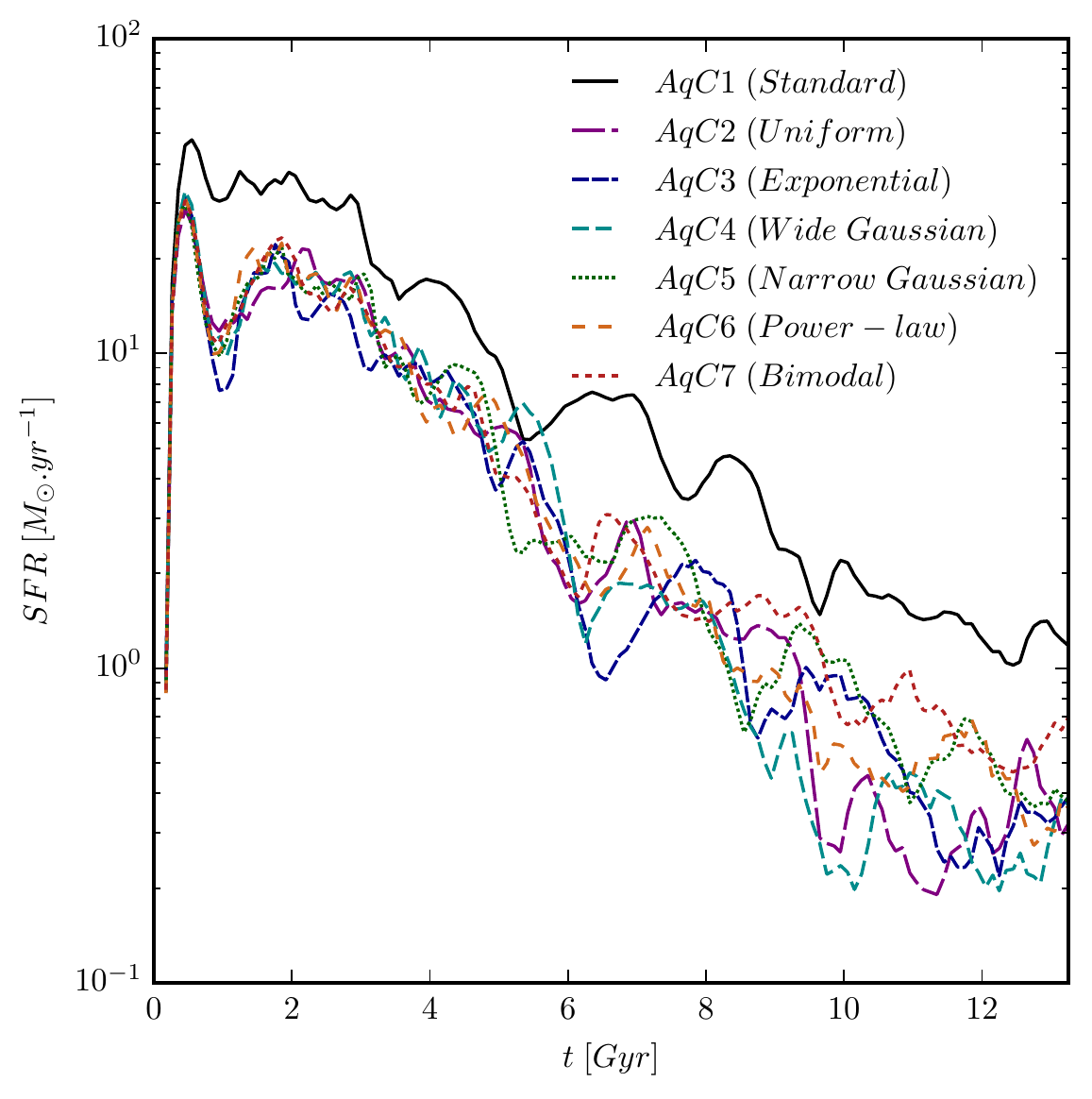}\includegraphics[width=8cm]{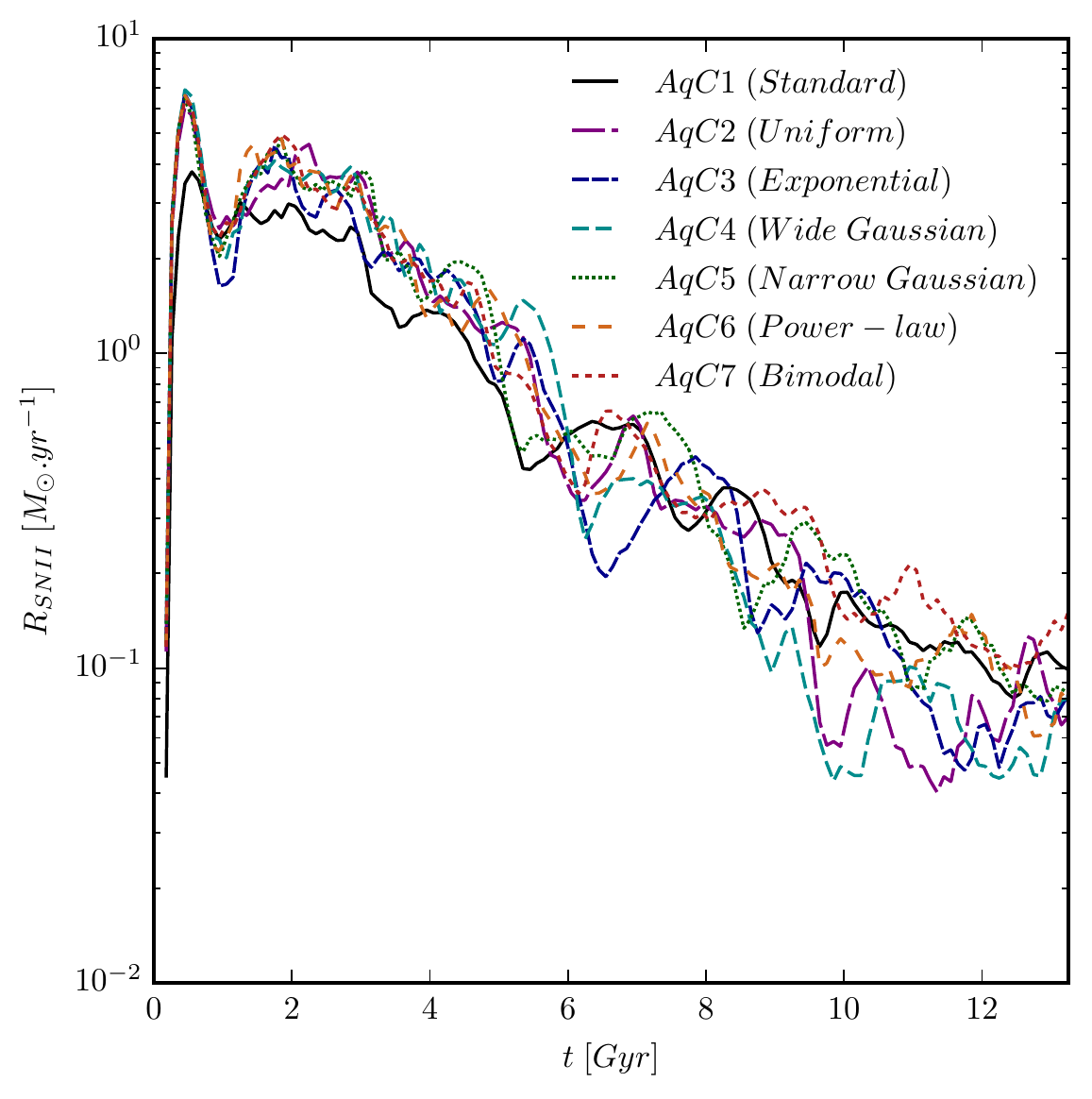}}

	\caption{Star formation  and SNII rates for our simulations AqC1-AqC7.
We note that  in the case of the SNII rates, the figure includes all SN events in the
simulations, i.e. those originated in the main halo and also those produced in
other structures.}
	\label{fig:cosmosfr}
	\end{center}
\end{figure*}

Various choices of DTDs for SNIa are not expected to have a significant role
in the formation of galaxies in terms of star formation/SNII rates, which are
mainly affected by the feedback produced by SNII and therefore by the choice
of IMF. Fig.~\ref{fig:cosmosfr} shows the SFRs and SNII rates for our
simulations AqC1-AqC7.
Note that the SFRs of these galaxies peak at early times, as a consequence
of the star formation/feedback routine used in this work. 
More recent models have shown that it is possible to shift
star formation to later times, e.g. invoking additional feedback
such as that coming from radiation pressure \citep{Aumer_2013}, 
producing  SFRs with a much more moderate early peak and an approximately
constant SF level at later times, more similar to e.g. the SFR of our
Milky Way and similar galaxies. In this work we used our standard
routines which have been extensively tested and discussed
in previous works, to allow for a cleaner comparison, while we will discuss updates to
the feedback model in future work.

Fig.~\ref{fig:cosmosfr} shows that the shape of the SF and SN rates are similar in all runs,
reaching a maximum  after $\sim 0.5$ Gyr of evolution
and decreasing until $z=0$ in a constant but bursty manner.
Short-term, strong variations appear as a result of mergers or accretion events
which are more evident in the SF evolution that only includes the star formation
originated in the main halo.
Our standard model (AqC1), which assumes a Salpeter IMF, has a significantly
higher SFR compared to AqC2-AqC7, that results from a lower
SNII rate -- particularly at the starburst -- and  a reduced amount of feedback.
Variations in the SF/SNII rates due to the different choices of SNIa
DTDs appear, as expected, later on when SNIa have a larger effect through the
changes in the metallicities and cooling functions, in  particular
after 6-8 Gyr  of evolution, as explained in the next sections.

\begin{figure*}
	\begin{center}
	{\includegraphics[width=8cm]{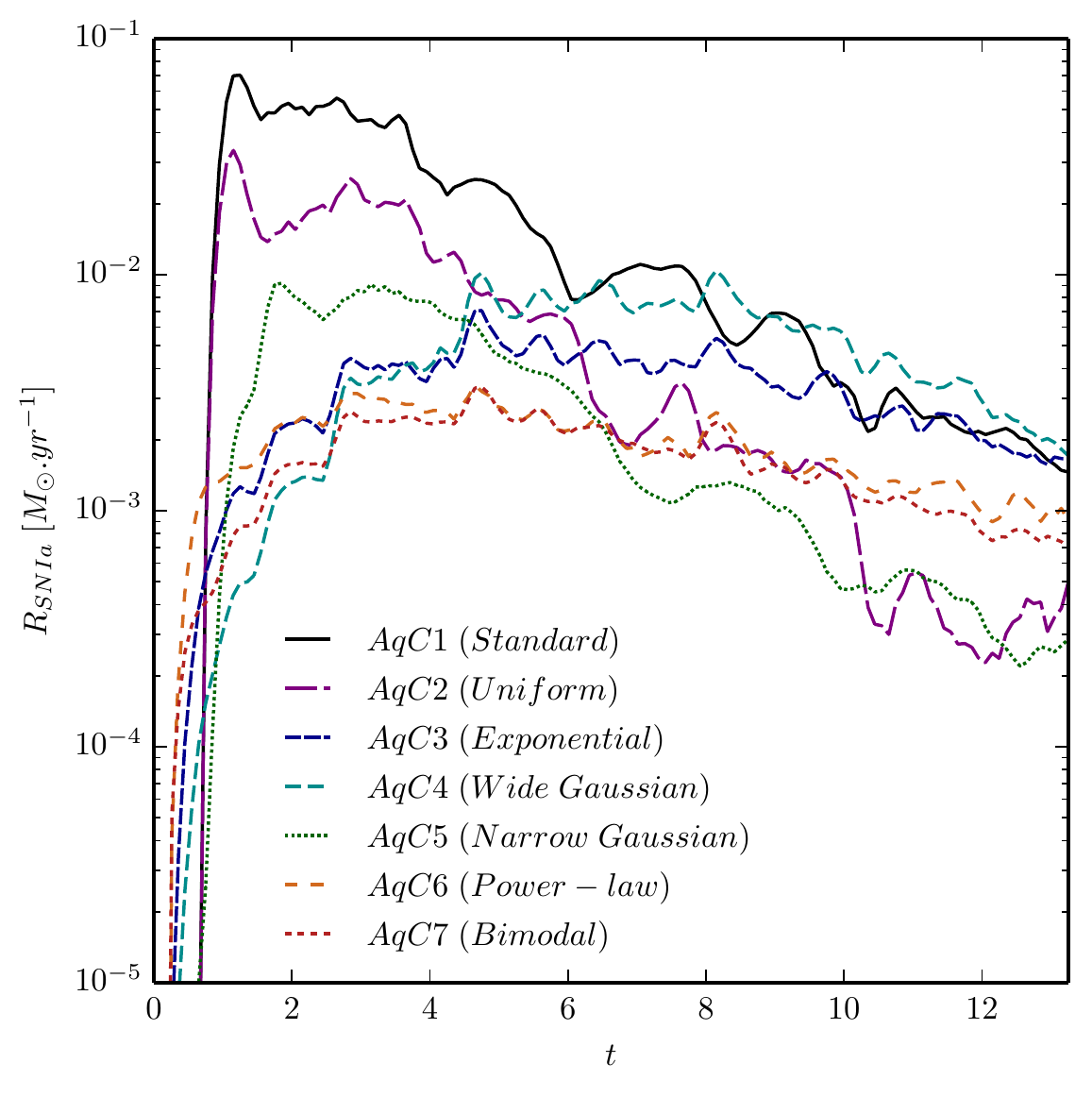}{\includegraphics[width=8cm]{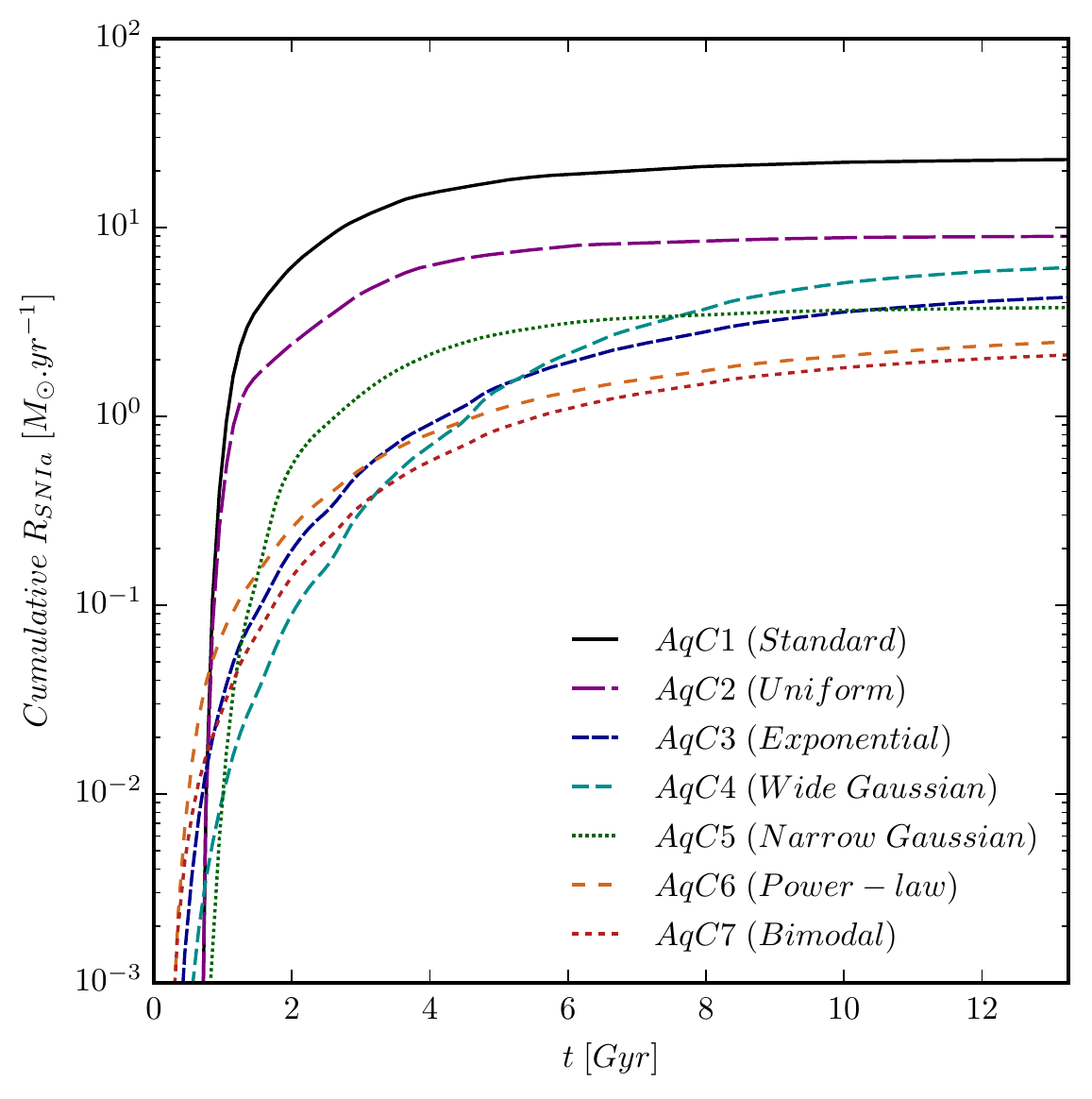}}}
	\caption{Instantaneous (left-hand panel) and cumulative (right-hand panel)
SNIa  rates of our cosmological runs AqC1-AqC7.}
	\label{fig:cosmosn}
	\end{center}
\end{figure*}

The instantaneous and cumulative SNIa rates of runs AqC1-AqC7 are presented
in Fig.~\ref{fig:cosmosn}.
AqC1 and AqC2, which assume a uniform DTD and a fixed SNIa/SNII rate,
exhibit the highest SNIa rates due to the higher normalization, and a behavior
similar to the SF and SNII rates, with a dominant component at early times.
A similar evolution, albeit at a later time, is found for AqC5 (narrow Gaussian)
which exhibits a high SNIa
rate between 2 and 5 Gyr and the lowest SNIa rate of all runs after 6 Gyr of evolution.
The similarity of these runs originates in the similar characteristics of the DTD
(Fig.~\ref{fig:DTD} and left-hand panel of Fig.~\ref{fig:Dtdsni}): the uniform and narrow Gaussian DTDs have the
strongest late-time cutoffs of all distributions, with all SNIa events  
occurring within 2 Gyr after the formation of the progenitor
stars, i.e. all the SNIa are concentrated in a shorter time interval and
have a correspondingly higher rate.
From the other runs, AqC6 (power-law) and AqC7 (bimodal) show very
similar SNIa rates during the whole evolution, exhibiting the highest
SNIa rates of all runs up to 0.5 Gyr due to the high contribution of
a prompt component (Fig.~\ref{fig:DTD} and left-hand panel of Fig.~\ref{fig:Dtdsni}).
Finally,  runs AqC3 (exponential) and AqC4 (wide Gaussian) have
 the least number of SNIa before 5 Gyr, 
 and experience the highest
SNIa rates of all runs after $\sim 9$ Gyr.
Note that, even assuming the same normalization in Eq.~\ref{eq:number}, the
different characteristics of the DTDs result in different final values
for the cumulative SNIa rates in the various runs.

The choice of DTD distribution and resulting SNIa rates of our runs,
although having almost no effect on the global SFR\footnote{Note also that,
due to the similarities in SF and SNII rates,
the galaxies formed in all cosmological
simulations present similar morphologies, with an extended stellar disk and
a bulge.
Furthermore, AqC1, AqC4 and AqC7
show clear signatures of a bar at $z=0$, and if
one compares this with the cumulative DTDs in Fig.~\ref{fig:Dtdsni},
this may be due to the uniform (AqC1), wide-Gaussian (AqC4) and bimodal (AqC7) models  all having the fewest SNIa after $\sim 6 \, \rm Gyr$. In
the other cases there will be increased feedback in the bulge,
which may have some impact on the bar formation, though the effect
must be relatively modest.}, will
affect the chemistry of the simulated galaxies, the imprints of which we discuss in the next section.

\subsection{Evolution of chemical abundances}  

\begin{figure}
\begin{center}
{\includegraphics[width=\linewidth]{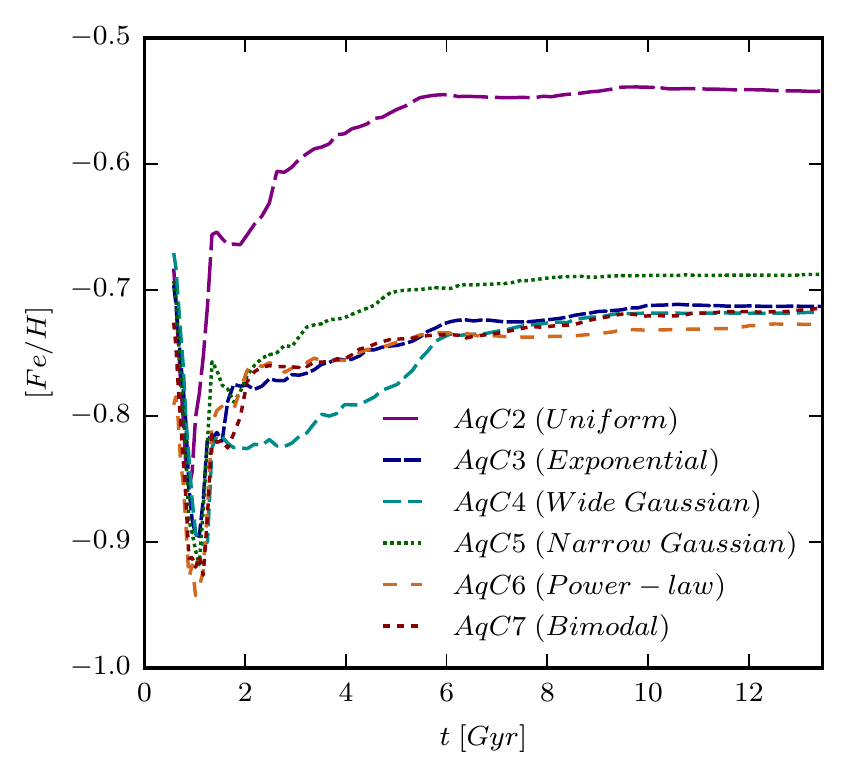}}
\caption{The evolution of [Fe/H] in our cosmological simulations AqC2-AqC7, which
only differ in the DTD assumed for SNIa.}
\label{fig:FeHEvolution}
\end{center}
\end{figure}

In this Section we focus on the evolution of the chemical abundances of galaxies
formed in our
runs AqC2-AqC7. As shown above, these have a similar SFR and are identical
except for the DTD assumed, allowing us to isolate the effects of this choice
on the simulated galaxies.
In Fig.~\ref{fig:FeHEvolution} we show the evolution of [Fe/H] for the galaxies
formed in AqC2-AqC7.  Due to its higher SNIa rate, the run with a uniform DTD, AqC2,
has incorporated more iron than the other simulations and is at the present day a factor $0.12$~dex above the galaxies  with other DTDs (i.e. $\approx 30\%$ more iron).
Furthermore, the enrichment is fast, as AqC2 has produced almost all of its SNe Ia within $1$~Gyr of the star formation events. The galaxy with the next highest [Fe/H] level is
AqC5 (narrow Gaussian DTD) which has essentially completed its DTD by $2$~Gyr after the main star formation event.
The DTD that produces the least [Fe/H] at $z=0$ is the power-law (AqC6),
which is the one with the most SNIa in the $>7$~Gyr tail.
Finally, it is interesting to see that at intermediate times of $2$-$6$~Gyr,
the simulation with the lowest [Fe/H] in the main halo is the
wide Gaussian DTD (AqC4), that has the least number of SNIa before $5$~Gyr.
Note that the enrichment level of the galaxies at each time is determined
by the SNIa rate and the amount of un-exploded SNIa whose ejecta has yet to be incorporated into the stellar populations.

\begin{figure*}
\begin{center}
{\includegraphics[width=16cm]{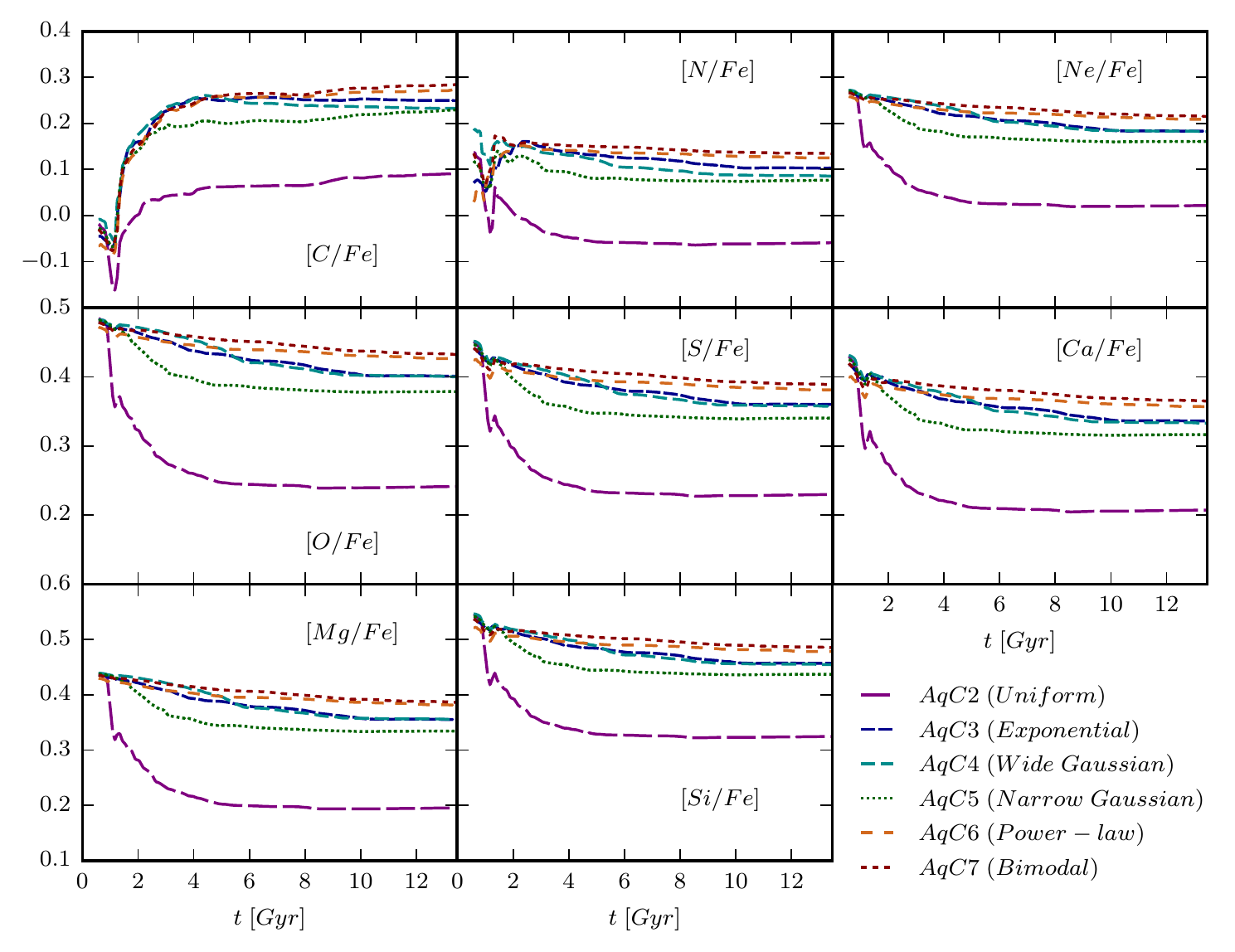}}
\caption{The evolution of the stellar chemical elements in our cosmological runs
AqC2-AqC7 which assume various DTDs for SNIa.}
\label{fig:XFeEvolution}
\end{center}
\end{figure*}

Fig.~\ref{fig:XFeEvolution} shows the evolution of the stellar heavy element enhancements for the main halo when assuming different DTDs. These approximately correspond to our expectations from Fig.~\ref{fig:FeHEvolution}, the $\approx 30\%$ higher iron ratios produced
in AqC2 (uniform DTD) causing correspondingly lower X/Fe ratios compared to the other
simulations. The quick response of Fe to star formation (i.e. prompt SNIa component) also causes the X/Fe ratios to be more stochastic, which can be observed for the
troughs in the first 2 Gyr for the uniform (AqC2) and bimodal (AqC7) DTDs, particularly for the N/Fe ratio. The X/Fe ratios for AqC5 (narrow Gaussian DTD) are intermediate
between those in AqC2 (uniform DTD) and the rest of the simulations and,
as a consequence of the [Fe/H] evolution of this simulation, has the strongest
changes at early times.
Finally, the galaxies with the highest X/Fe ratios are those produced
with a power-law (AqC6) and bimodal (AqC7) DTDs, due to the significant contribution
of  prompt SNIa events. (Note that even though the galaxies formed in
our cosmological simulations have experienced a more complex evolution compared
to the idealized tests of the previous Section, the most important features
found in the latter and produced by the use of various DTDs are still visible here.)

\subsection{[O/Fe] vs. [Fe/H] diagrams} 

\begin{figure*}
\includegraphics[width=16cm]{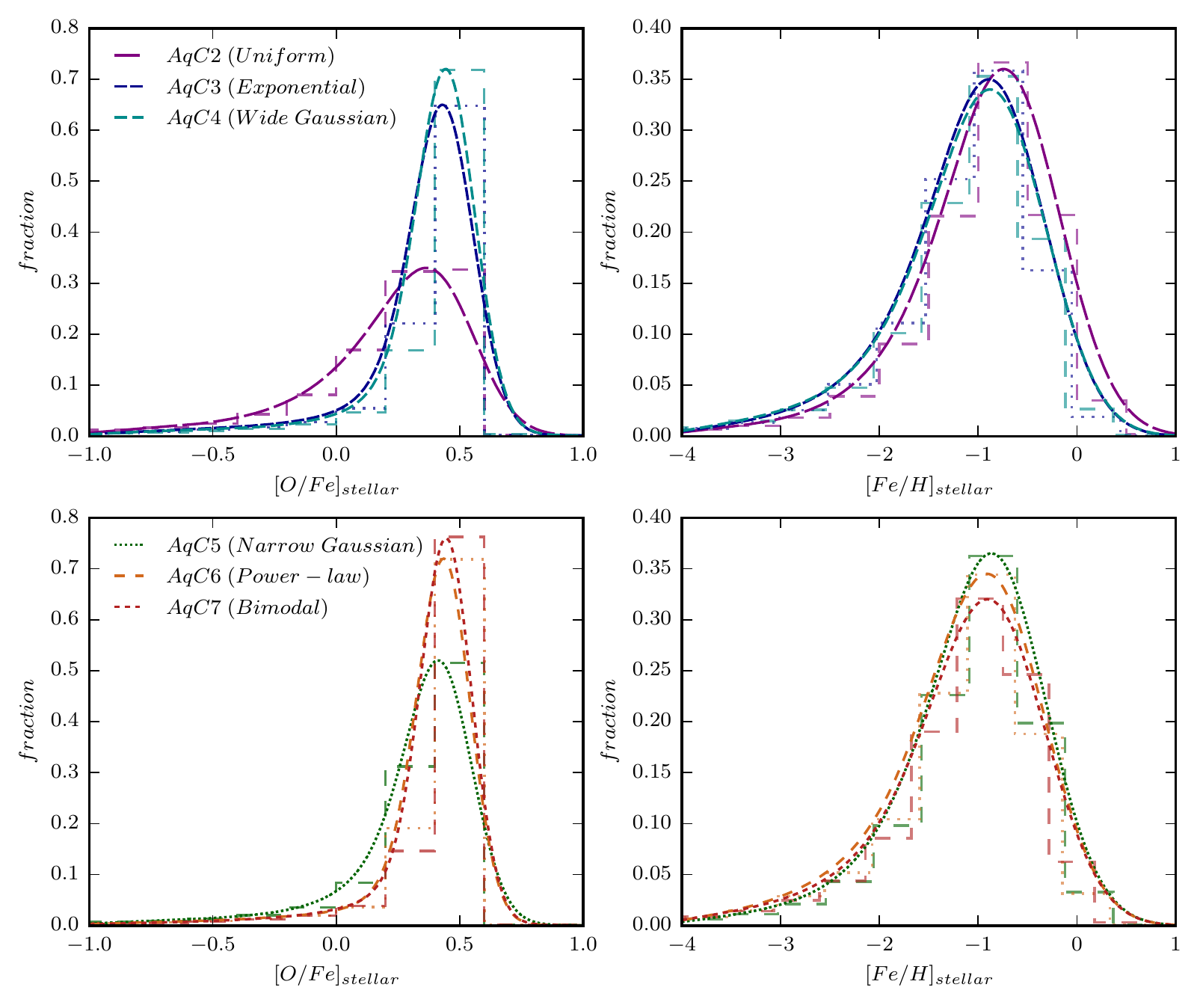}
\caption{Stellar distribution functions of [O/Fe] (left panels) and [Fe/H] (right panels) in simulations AqC2, AqC3 and AqC4 (upper panels) and in simulations AqC5, AqC6 and AqC7 (lower panels)}
\label{fig:Ofestar}
\end{figure*}

In this Section we investigate the effects of the different DTDs on the
final chemical properties of our galaxies. First, we focus on the
distributions of [O/Fe] and [Fe/H] to later discuss the [O/Fe] vs
[Fe/H] relations.
Fig.~\ref{fig:Ofestar} shows histograms of these ratios for our
simulations AqC2 to AqC7. While the shape of the [Fe/H] distributions
is similar for all runs, although with the galaxy in AqC2 (uniform DTD)
exhibiting a higher overall iron abundance as seen previously, more
significant differences are found for the [O/Fe] ratios.
Note that while the iron abundances are mainly determined
by SNIa and therefore affected by the choice of DTD, most of the oxygen
is produced by SNII, so that the O/Fe ratio is influenced
by the two types of SNe.
The most prominent feature here is the broader distributions of [O/Fe]
for the runs with uniform (AqC2) and narrow Gaussian (AqC5) DTDs. As explained
above, this is a consequence  of the shorter time-scales in which iron
is ejected of these DTDs. As a result, compared to the other DTDs, more iron 
is injected into the interstellar gas within $\sim 2$ Gyr after the peak of
star formation, leading to the formation of stars with lower 
[$\alpha$/Fe] thereafter. 
In contrast, the power-law (AqC6) and bimodal (AqC7) DTDs produce the narrower distributions, suppressing the long tail to low
stellar [O/Fe]. Finally, at an intermediate position, are
the exponential (AqC3) and wide Gaussian (AqC4) DTDs. In these
cases, the contribution of iron from SNIa is delayed with
respect to the starburst and consequent to the oxygen enrichment that follows
SNII events. 

According to our results, 
models with significant prompt component
(AqC6/AqC7) produce narrower
[O/Fe] distributions and suppress the long tail of
low [O/Fe] ratios. This is consistent with the results
of Galactic Chemical Evolution models (e.g. \citealt{Matteucci_2009,Maoz2012,Yates2013,Walcher16})
which seem to require a prompt component in order to reproduce
observational results on chemical abundances.
 Note however that, as explained above, the SFRs of our simulated galaxies
are dominated by an early starburst, and have a small fraction of young stars. A
SFR with more dominant late star formation would certainly 
affect our [Fe/H] distributions in terms of their peak values. 
In future work we will present
an update to our feedback model, and discuss the consequences of this on the chemical
properties of simulated galaxies.

\begin{figure}
\begin{center}
{\includegraphics{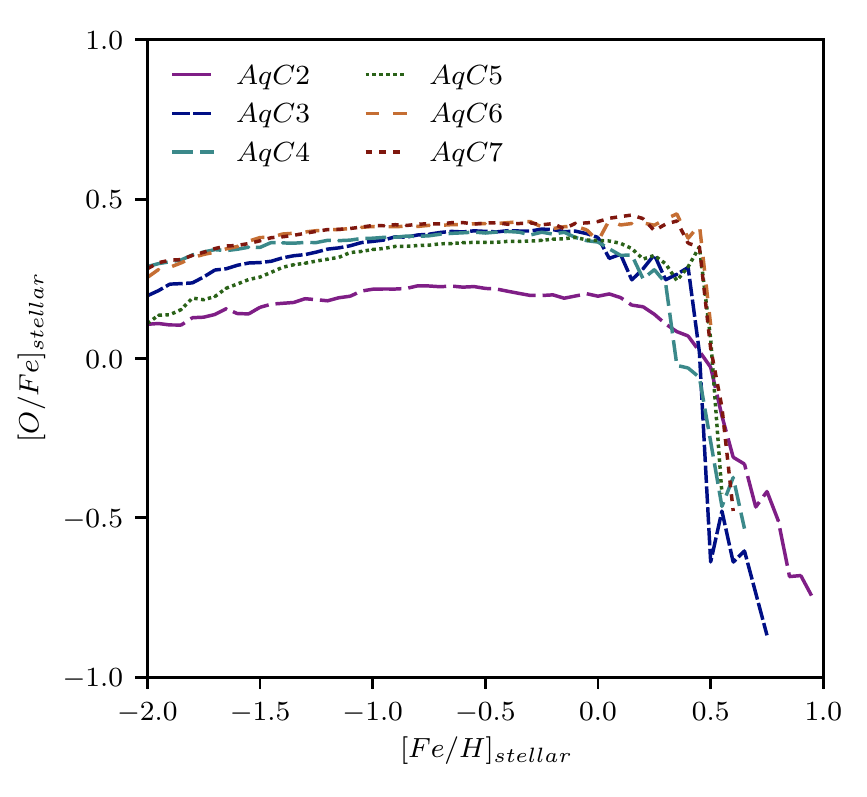}}
\caption{[O/Fe] vs. [Fe/H] relation at z = 0 for the stars  of the main halo
of our cosmological simulations simulations AqC2-AqC7 with different DTDs for SNIa.
}
\label{fig:OFeFeH_bulge}
\end{center}
\end{figure}

 Observations of the stellar [O/Fe] ratio as a function of [Fe/H] also
provide constrains that could help to decide on the DTD of SNIa. 
As explained before, $\alpha$-elements (such as oxygen) are produced by SNII while iron is mainly produced by SNIa, and so
the relative importance of  SNII and SNIa during evolution will shape the element
ratios and particularly the [O/Fe] abundances. As
the oxygen abundance remains high and only drops when
the number of SNIa events become important, the [O/Fe] vs. [Fe/H] diagram
will have a change in the slope  --  known as the `knee' -- which measures
the timescale of the SNIa enrichment (i.e. the maximum in the SNIa rate)\footnote{
 We note, however, that when interpreting these distributions for a
single halo one should be careful as the features in the [O/Fe] vs [Fe/H]
diagrams are controlled not
only by the DTD, but can also be
modified by inflow (which changes the [Fe/H] ratio)
and more extended star formation, and as such one could make
the features shown more or less pronounced by choosing a
halo with a different accretion and star formation history.}.
In Fig.~\ref{fig:OFeFeH_bulge} we show the average stellar abundances
[O/Fe] vs. [Fe/H] for all stars 
of the main halo for our simulations AqC2-AqC7\footnote{We find very similar results
if we consider only stars in the inner 5 kpc of the simulated galaxies, i.e. the bulge region. Only small differences are detected for the most metal-rich populations, which in any case do not affect the position/behaviour of the knee.}.

All models present a plateau for [Fe/H]
$\leq$ 0 followed by a decrease, although the shape of the ``knee'' is
distinctive of each simulation (and thus of each DTD assumed).
The model AqC2 (uniform DTD) lies at an extreme, as
the many SNIa events produced early on allow the model to reach
a low level of [O/Fe]$\sim 0.2$ at the plateau. As shown in
 Fig.~\ref{fig:FeHEvolution}, AqC2 reaches significantly higher
 iron abundances at all times, with most of the iron
of the galaxy being in place within the first 6 Gyrs. All other
 models reach a plateau with a higher [O/Fe] level of about 0.4.
 Models AqC6 (power-law) and AqC7 (bimodal) lie at an extreme,
as they produce so many early SNIa events and exhibit the most abrupt
decrease in [O/Fe], and at the highest [Fe/H] values. All other
simulations show a less abrupt behavior, with 
the wide Gaussian DTD (AqC4)  producing the softer knee.

The results shown in Fig.~\ref{fig:OFeFeH_bulge}
seem in contradiction to those of M09, in particular in that M09
finds that DTDs with a significant prompt fraction present shallower
slopes after the knee, opposite to DTDs with negligible prompt fractions.
In our case, we find that the bi-modal and power-law DTDs exhibit the
most abrupt decrease of all runs. Note however that a detailed comparison
between our results and those of M09 is not possible, in particular
because of the different SFRs and resulting  morphologies (as explained above,
the simulated galaxies have a dominant early starburst, which leads
to systems where the bulk of the stars are old and located in the bulge
regions).

\subsection{The effect of the cooling}

\begin{figure*}
	\begin{center}
		{\includegraphics[width=7cm]{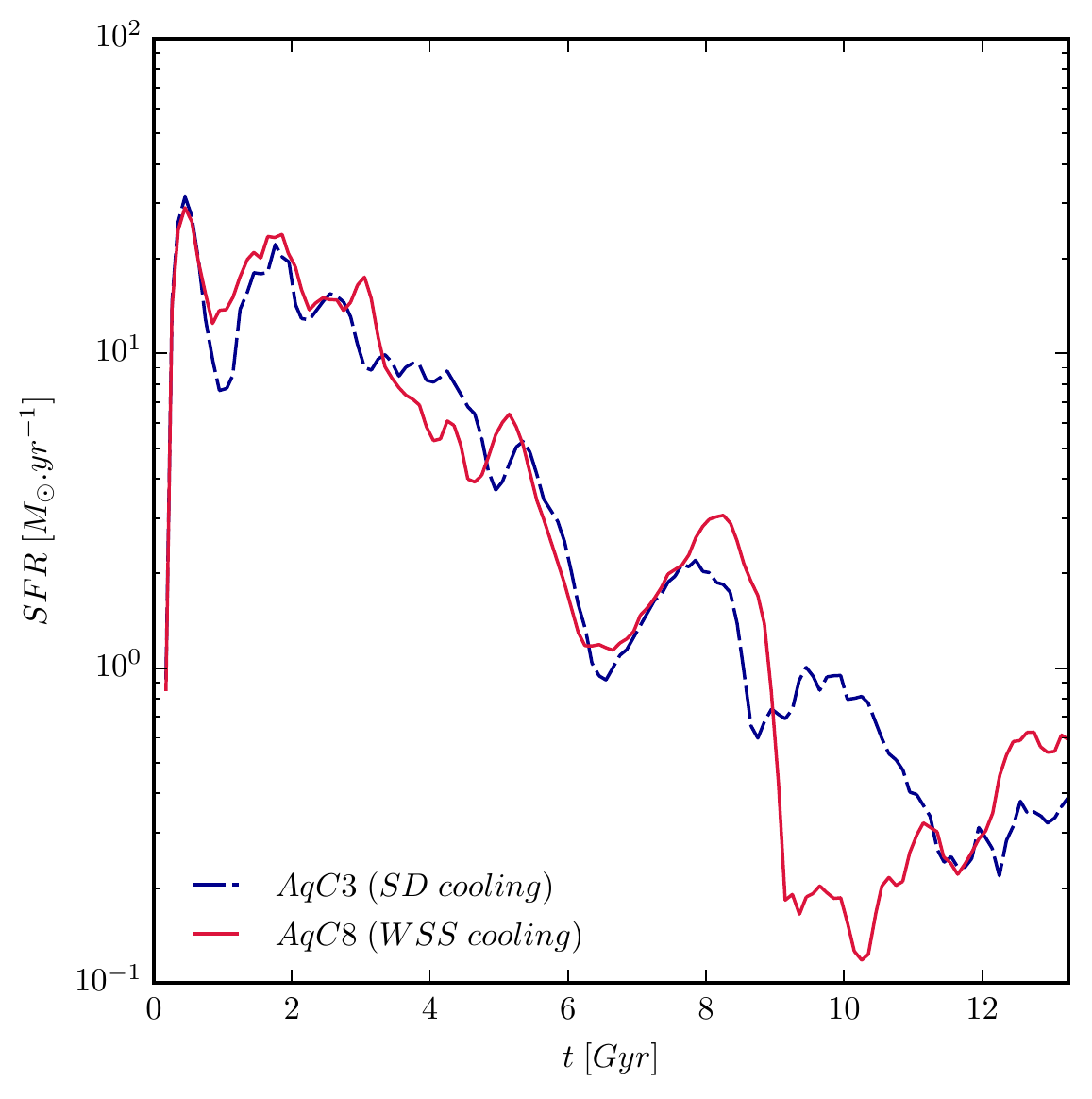}\includegraphics[width=7cm]{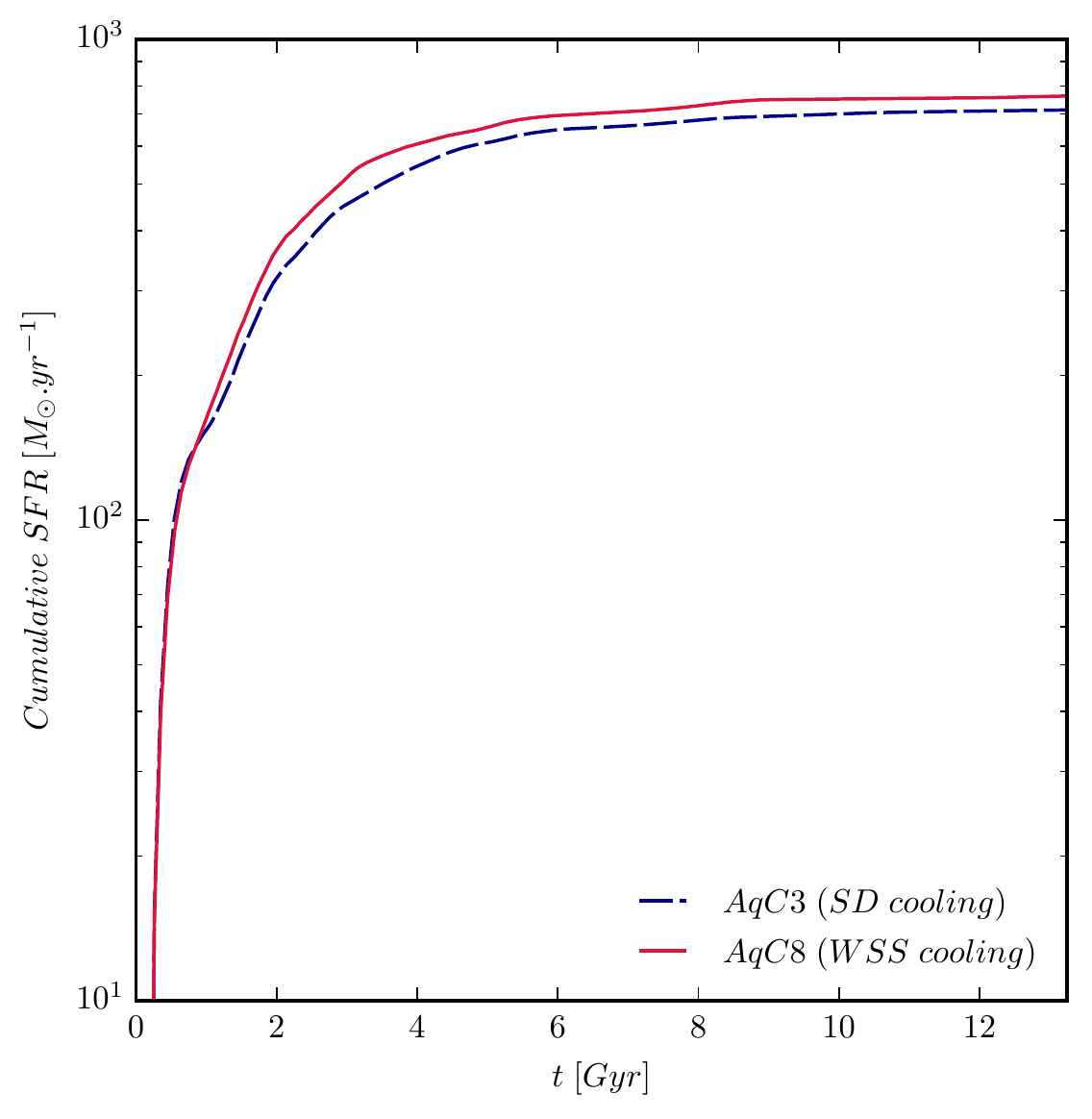}}
		\caption{Instantaneous (left-hand panel) and cumulative (right-hand
panel) star formation rate 
for our  cosmological simulations AqC3 (dashed blue) and AqC8 (solid purple) 
which are identical except for the choice of cooling tables. 
AqC3 and AqC8 use, correspondingly, the cooling tables of SD and WWS.
}
		\label{fig:cosmocoolsfr}
	\end{center}
\end{figure*}

As explained in Section~\ref{sec:cool}, our updated model uses
the cooling tables from \cite{W09}, instead of those given
by \cite{SD93} of our standard code.
In this Section we compare the results of cosmological simulations
AqC3 and AqC8 which only differ in the choice of the cooling
functions (Table~\ref{tab:cosmo}). 
Similarly to our results using idealized initial conditions, we do not 
detect significant differences in the two runs, in terms of the
global properties of the simulated galaxies. This is shown in
Fig.~\ref{fig:cosmocoolsfr}, which compares the SFR (instantaneous and cumulative) of
 simulations AqC3 (dashed blue) and AqC8 (solid purple).
Although the effects of photo-ionization would be visible
in some regions of the phase-space, the similarity in the SFRs of our two
runs indicates that differences are in any case moderate.
The galaxies formed in runs AqC3 and AqC8
look very similar in their morphologies and dynamical properties, 
and in their metallicity distributions. Note that our standard model
already included metal cooling, although in terms of the [Fe/H] abundance.

\section{Summary and Conclusions}\label{sec:conclu} 

In this work we present several updates to the \cite{S05} model of galaxy formation focusing on the chemical enrichment mechanisms, and perform a series of controlled experiments to disentangle their interacting effects. We implement additional models for the IMF, the type II SNe yields, included ejecta from asymptotic-giant branch stars, updated the cooling tables and added several models of the delay-time distribution of type Ia SNe. In order to validate these models 
without performing an overwhelming series of calculations 
we utilised two different sets of ICs. The first was an isolated galaxy which had the advantages of simplicity both in the computational sense and in the star formation history which allowed us to disentangle the effects of the various processes we introduced. For the second we used the cosmological zoom of the very well studied Aquarius-C halo, which has a much more realistic formation history and subsequent star formation episodes, with the increased complexity in computation and analysis that involves, and as such we simulated only the latest combination of IMF, AGB and SNII yields, and varied the delay time distribution. With this cosmological initial condition we have also tested the effects of varying the cooling function which also considers the coupling to the
UV background field. Our primary results are as follows.

\begin{enumerate}
\item{The ascending fraction of high mass stars in the Salpeter, Kroupa and Chabrier IMF respectively drives corresponding trends in stellar mass and metallicity, with the Kroupa IMF almost universally being the intermediate case. The Chabrier IMF has the most feedback and forms the least mass of stars, a trend that is seen in both the isolated and cosmological simulations. This higher feedback is, however, not sufficient to prevent it also producing the highest metallicities in both gas and stars.}

\item{The effect of WW95 vs. P98 SNII yields varies greatly between elements. Most significant is for N and Mg, with the P98 model producing excesses of the order of 1 dex in the stellar abundances, and at a lower level for Ne and O, with differences of about 0.5 dex in the isolated simulations. These correspond to large changes in the [O/Fe] ratios of about 0.5 dex with P98.}

\item{AGB stars return chemicals to the ISM in a time distribution with both a significant prompt fraction and relatively heavy tails. 
We settled on a default implementation of 3 enrichment periods of 100 Myr, 1 Gyr and 8 Gyr per star particle, the latter two capturing the extended phases  and the earliest for the material that is quickly recycled. The AGB stars are particularly effective at polluting with C and N, and our isolated simulations exhibit increases in [C/Fe] and [N/Fe] stellar ratios by $0.6$~dex and $0.4$~dex respectively.
}
\item{The effects of switching from \citet{SD93} to \citet{W09} cooling functions can be observed in the phases of star formation in isolated and cosmological halos, however the net effect is rather modest, likely due to the tight self-regulation of star formation via stellar feedback. }
\end{enumerate}

The delay time distribution of type Ia SNe is of particular interest as
it affects the typical time-scales of the iron release, and this effect might still
be imprinted in the properties of the chemical properties of the stars and gas in galaxies. 
In our various implementations of SNIa models, the distribution of SNIa takes the form of a very extended process along with the presence or absence of a prompt component. Since this can interplay with hierarchical formation and gas accretion, we tested these both with isolated and cosmological simulations.
In the following we summarize our main results. 

\begin{enumerate}
\item{The prompt component is maximized in our power-law and bimodal models, and at the other extreme the narrow Gaussian (centered on 1 Gyr) has the least prompt SNIa. 
The models with the prompt component produce the highest stellar element ratios   at late times, in both the isolated and cosmological simulations.}
\item{The models with the prompt component also  exhibit the narrower [O/Fe] distributions at the present time, suppressing the long tail to low stellar [O/Fe] ratios. Conversely at the opposite extremes the narrow Gaussian and uniform delay-time  distributions create the lowest overall [O/Fe] ratios, also having a  low [O/Fe] tail.}

\end{enumerate}

This work marks an essential step in linking the observed abundance patterns of stars in our own galaxy to its star formation, gas evolution and enrichment history over cosmic time. The most immediate application of this work is for more detailed studies of the abundances of individual elements and their distribution, both spatially and in terms of age in our galaxy. Having a validated cosmological model also allows the examination of the statistics of galaxies, for example the analysis of dispersions and gradients of $\alpha$ and Fe, and the correlations with star formation. This will require some additional simulation effort to provide a significant sample of galaxies evolved in a $\Lambda$CDM cosmology to capture the effects of diverse formation histories and environments, and we leave this to a future paper.

\section*{Acknowledgments}
The authors would like to thank the anonymous referee for his/her detailed reading of our work and the many suggestions which improved this manuscript considerably.
 The authors gratefully acknowledge the Gauss Centre for Supercomputing e.V. (www.gauss-centre.eu) for funding this project by providing computing time on the GCS Supercomputer SuperMUC at Leibniz Supercomputing Centre (www.lrz.de) through project pr49zo, and
the  Leibniz Gemeinschaft for funding this
project through grant SAW-2012-AIP-5 129.

\bibliographystyle{mnras}
\bibliography{biblio}

\newpage

\appendix

\section{Varying the resolution} 
\label{sec:resolution}

\begin{figure*}
	\begin{center}
		\includegraphics[width=6cm]{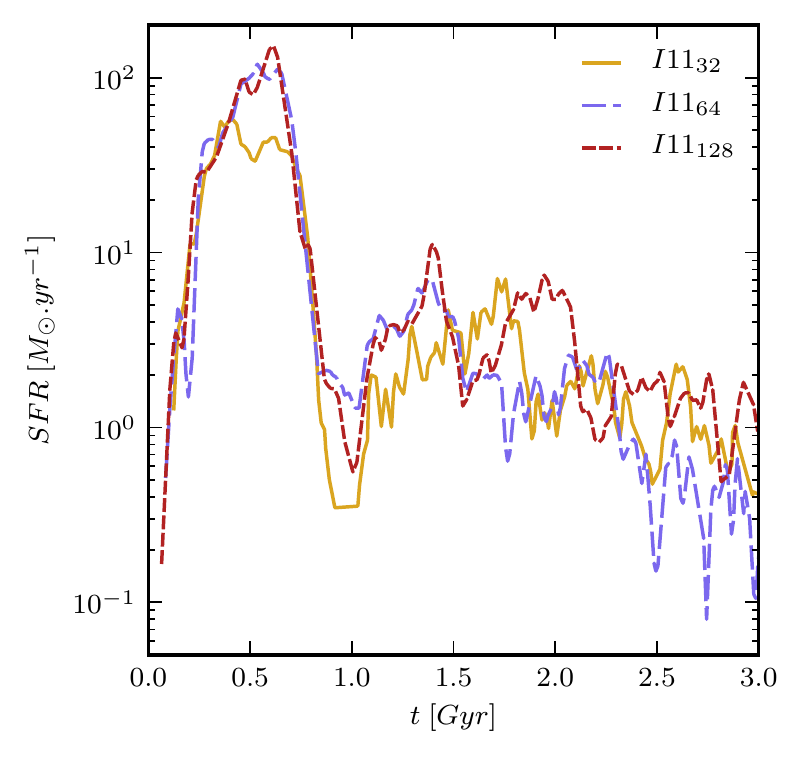}\includegraphics[width=6cm]{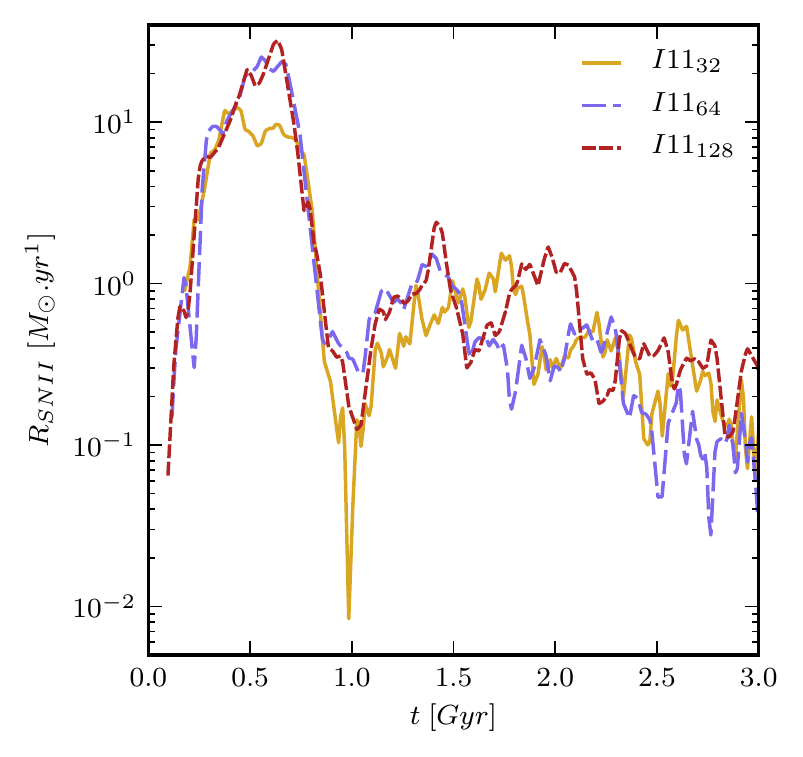}\includegraphics[width=6cm]{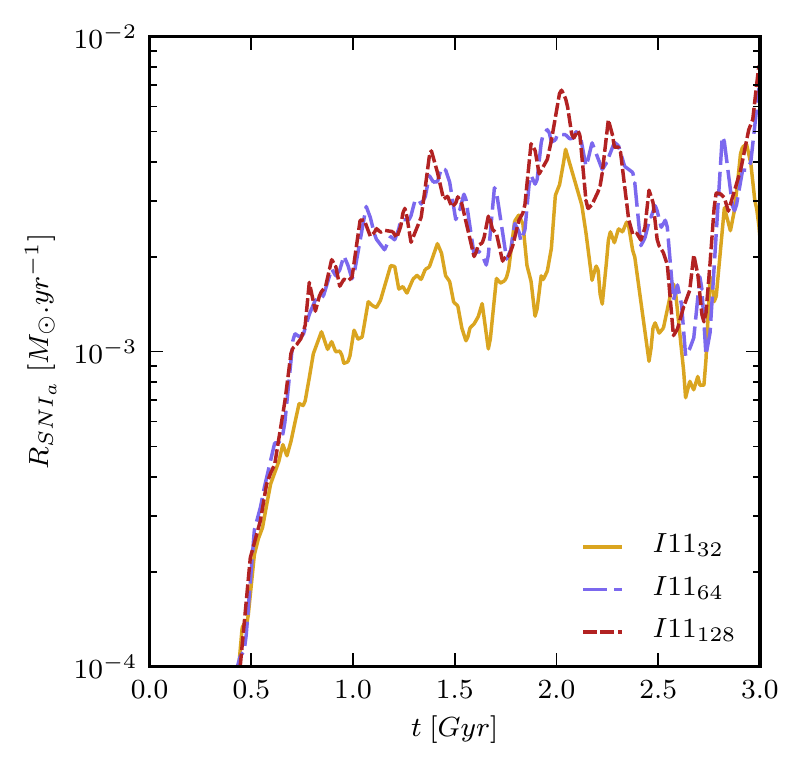}
		\caption{Star formation (left-hand panel), SNIa (middle-hand panel) and SNII (right-hand panel) rates for simulations I11$_{32}$ (solid yellow), I11$_{64}$ (dashed blue) and I11$_{128}$ (dotted red), which assume various resolutions.}
		\label{fig:Resolution_Rates}
	\end{center}
\end{figure*}

\begin{figure}
	\begin{center}
		\includegraphics[width=\linewidth]{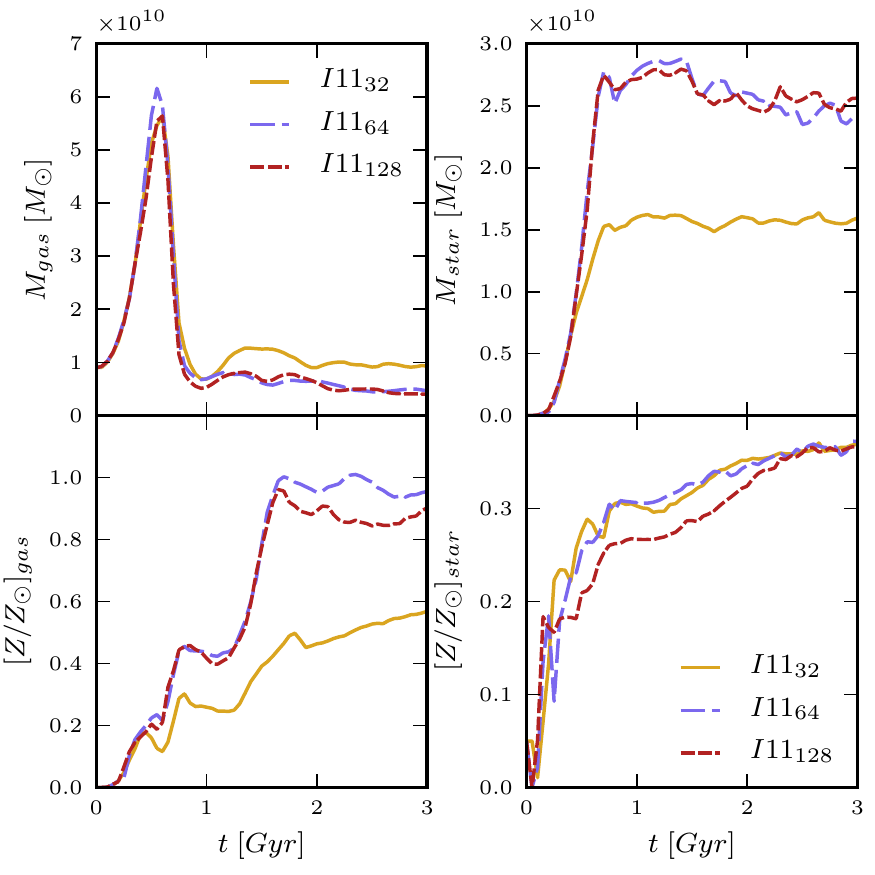}
		\caption{Evolution of the mass and metallicity for the gas (left panels) and stellar (right panels) components for simulations I11$_{32}$ (solid yellow), I11$_{64}$ (dashed blue) and I11$_{128}$ (dotted red). We consider only particles within the inner 30 $kpc$ of the galaxy.}
		\label{fig:Resolution_Metal}
	\end{center}
\end{figure}

In this section we compare the results of  simulations with different number of particles, that use the idealized initial conditions of Section~\ref{sec:isolated}. They correspond to various versions of our run I11 and use $2 \times N^3$ particles with $N=32$, $64$ (I11) and $128$. 

Figs.~\ref{fig:Resolution_Rates} and ~\ref{fig:Resolution_Metal}  
compare the star formation, SNIa and SNII rates, and the
evolution of the gaseous/stellar masses and metallicities for
the tests with different resolutions. 
Higher star formation levels are found when the number of particles
in the simulation increases; our highest resolution run produces a galaxy with
twice the stellar mass compared to that of the lowest resolution simulation. 
As $N$ increases, gas particles can achieve higher densities and therefore
higher star formation levels (note that the
density threshold for star formation is fixed in the  different runs).
The amount of metals in stars is therefore significantly higher in
the run with $N=128$ compared to $N=32$. In the case of the gas, the
metallicity evolution is more complex as a fraction of the gas is expelled
from the galaxy, and this fraction depends strongly on the amount of stars
formed. In Fig.~\ref{fig:Resolution_Metal} we show the mean metallicity
of the gas which stays within 30 kpc from the center of the galaxy and
which is different in the different runs, particularly if we compare
our highest and lowest resolutions.

We note that in these idealized simulations, the accretion of gas from the intergalactic medium is not considered, which can dilute the metallicity of the interstellar
gas affecting the metallicities of future generations of stars. Also, note
that after $\sim 1.5$ Gyr of evolution, the number of gas particles in our idealized
simulations has decreased significantly, either due to consumption into stars or the blow out of gas in winds, so that discrete effects can become noticeable. 
Our idealized initial conditions, however, allow us to
test the dependence on resolution without the complexity of the cosmological evolution,
where star formation, accretion and outflows act together in a non-trivial manner. 
Our results show that the differences between the runs 
with $N=64$ and $N=128$ for the SFRs, stellar/gas masses and metallicities remain
reasonably small, suggesting that
 $N\gtrsim 64$ is sufficient to reach reasonably convergent results.
In terms of gas particle mass, run I11$_{64}$ corresponds to $M_{\rm gas} = 1.3\times 10^5$M$_\odot$ and $M_{\rm dark matter} =  1.2\times 10^6$M$_\odot$, which are similar to the gas/dark matter particles in our cosmological simulations.

Resolution effects,  in terms of number of particles but
also other input parameters such as the density threshold
for star formation,  have been extensively discussed in previous works
using our simulation code \citep{S05,S11,S12}
as well as adopting other implementations (e.g. \citealt{Guedes11})
where we refer the interested reader.

\section{Effects of varying the number of assumed SNIa episodes} 
\label{sec:n_SNIa_episodes}

\begin{figure*}
	\begin{center}
		\includegraphics[width=17cm]{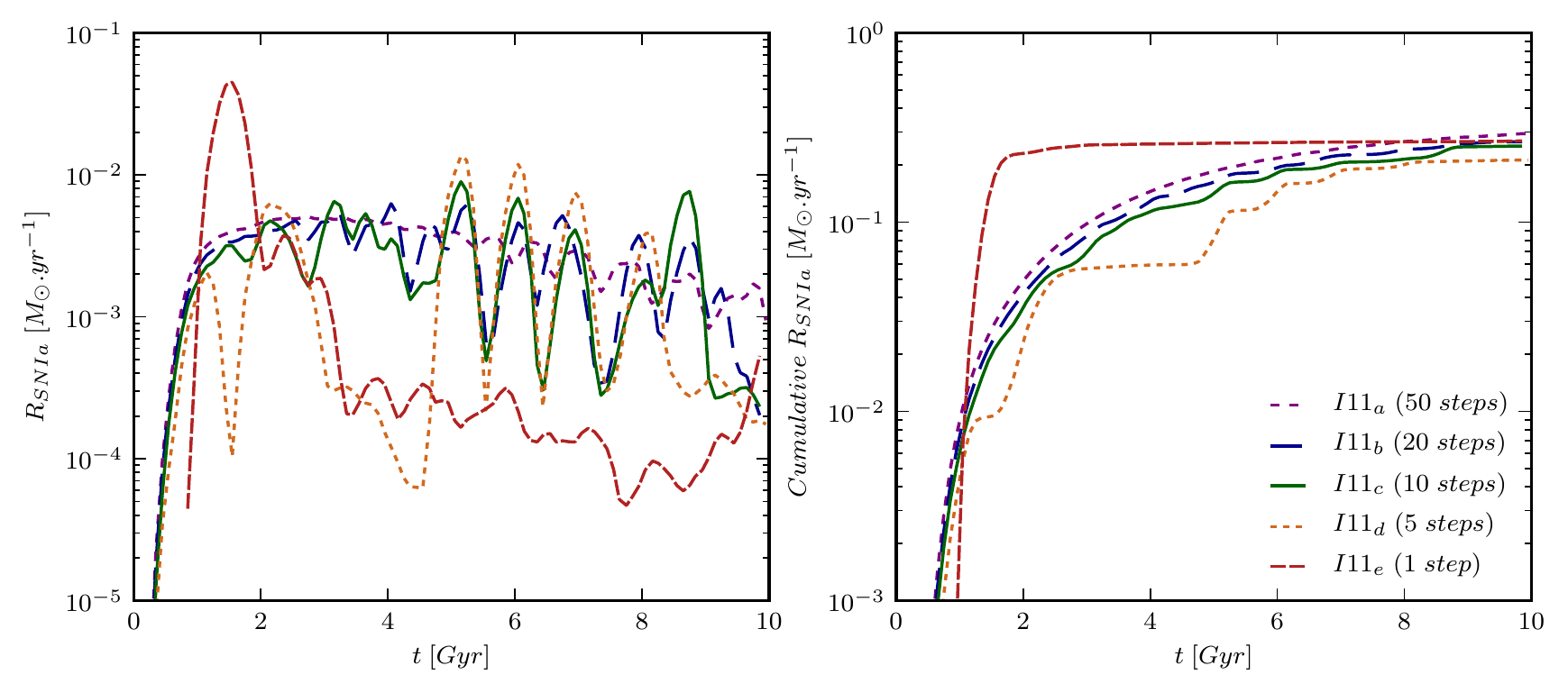}
		\caption{Instantaneous and cumulative SNIa rates for our tests assuming different number of SNIa episodes.}
		\label{fig:nssni}
	\end{center}
\end{figure*}

\begin{figure}
	\begin{center}
		\includegraphics[width=\linewidth]{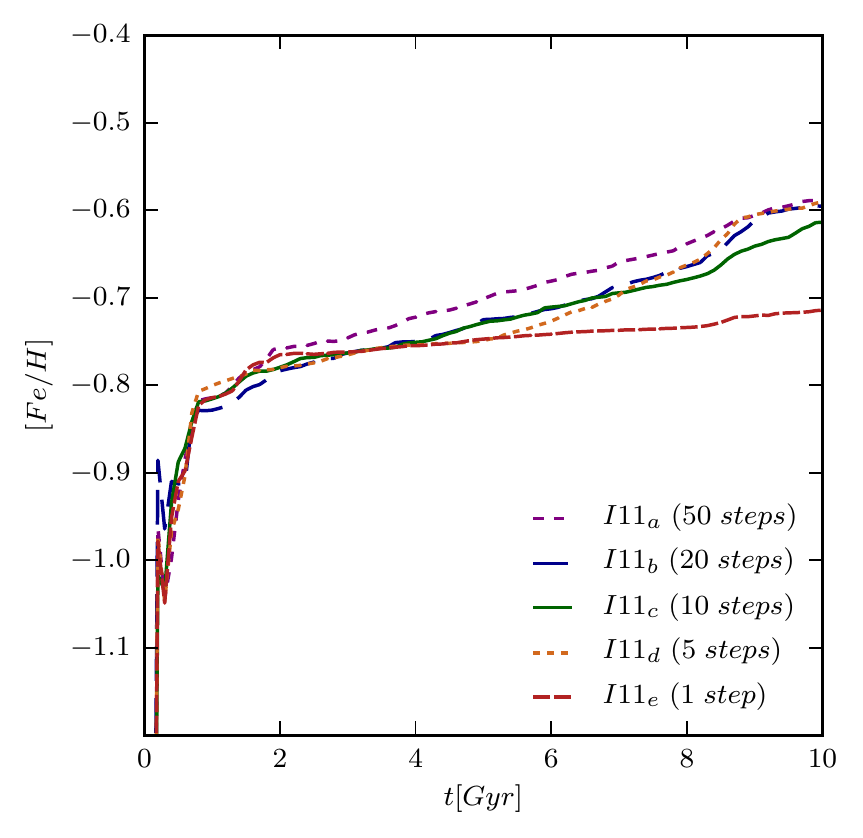}
		\caption{The evolution of the [Fe/H] for our tests with different numbers of SNIa episodes. }
		\label{fig:nselem}
	\end{center}
\end{figure}

In this section we describe the dependence of our results to our choice for
the number of enrichment steps assumed per SNIa event, $N_{\rm SNIa}$. 
For this purpose, we have
run $4$ additional simulations, that are identical to our idealized simulation
 I11 of Section~\ref{sec:isolated}, using different numbers of SNIa episodes:
from $1$ (I11$_a$) to $50$ (I11$_e$). Our standard I11 test which assumes
$10$ enrichment episodes is here referred to as I11$_c$.

In Fig.~\ref{fig:nssni} we plot the instantaneous and cumulative SNIa rates for these models. 
As one might expect, with coarse discretisations of 1 and 5 events we see some `bursty' behaviour, with more fine time discretisation the cumulative SNIa distribution seems converged. 
Interestingly some oscillations in the rates still appear in the \emph{rates} (Fig.~\ref{fig:nssni} left panel). At first sight one might attribute to the time discretisation, however this is unlikely to be the case since we discretise logarithmically in time and the period seems independent of the number of steps, $N_{\rm SNIa}$,  so this is likely due to the interplay/feedback between the star formation and the prompt SNIa events.

It is also important that our choice
of $N_{\rm SNIa}$ does not strongly affect the evolution of the different
chemical elements which might affect the efficiency of cooling. 
Indeed, we find a good convergence for the evolution of [Fe/H]
(and therefore for the rest of the elements produced during SNIa)
as long as   $N_{\rm SNIa}\gtrsim 10$, as shown in Fig.~\ref{fig:nselem}.
The results of this Section show that our choice of 
$N_{\rm SNIa}=10$ is a good  compromise between accuracy
in the calculations and computational cost.

\end{document}